\documentclass[fleqn,usenatbib]{mnras}

\usepackage{newtxtext,newtxmath}
\usepackage[T1]{fontenc}
\usepackage{ae,aecompl}

\DeclareRobustCommand{\VAN}[3]{#2}
\let\VANthebibliography\thebibliography
\def\thebibliography{\DeclareRobustCommand{\VAN}[3]{##3}\VANthebibliography}

\usepackage{graphicx}	% Including figure files
\usepackage{amsmath}	% Advanced maths commands
\title[ASAS-SN Catalog of Variable Stars IX]{The ASAS-SN Catalog of Variable Stars IX: \textit{The Spectroscopic Properties of Galactic Variable Stars}}

\author[T. Jayasinghe et al.]{T. Jayasinghe$^{1,2}$\thanks{E-mail: jayasinghearachchilage.1@osu.edu},
C. S. Kochanek$^{1,2}$,
K. Z. Stanek$^{1,2}$,
B. J. Shappee$^{3}$,
\newauthor 
T. W. -S. Holoien$^{4}$,
Todd A. Thompson$^{1,2}$,
J. L. Prieto$^{5,6}$,
Subo Dong$^{7}$,
M. Pawlak$^{8}$,
\newauthor 
O. Pejcha$^{8}$,
G. Pojmanski$^{9}$,
S. Otero$^{10}$,
N. Hurst$^{11}$,
D. Will$^{1,11}$
\\
$^{1}$Department of Astronomy, The Ohio State University, 140 West 18th Avenue, Columbus, OH 43210, USA\\
$^{2}$Center for Cosmology and Astroparticle Physics, The Ohio State University, 191 W. Woodruff Avenue, Columbus, OH 43210, USA\\
$^{3}$Institute for Astronomy, University of Hawaii, 2680 Woodlawn Drive, Honolulu, HI 96822,USA\\
$^{4}$Carnegie Observatories, 813 Santa Barbara Street, Pasadena, CA 91101, USA\\
$^{5}$N\'ucleo de Astronom\'ia de la Facultad de Ingenier\'ia y Ciencias, Universidad Diego Portales, Av. Ej\'ercito 441, Santiago, Chile\\
$^{6}$Millennium Institute of Astrophysics, Santiago, Chile\\
$^{7}$Kavli Institute for Astronomy and Astrophysics, Peking University, Yi He Yuan Road 5, Hai Dian District, China\\
$^{8}$Institute of Theoretical Physics, Faculty of Mathematics and Physics, Charles University, Czech Republic\\
$^{9}$Warsaw University Observatory, Al Ujazdowskie 4, 00-478 Warsaw, Poland\\
$^{10}$The American Association of Variable Star Observers, 49 Bay State Road, Cambridge, MA 02138, USA\\
$^{11}$ASC Technology Services, 433 Mendenhall Laboratory 125 South Oval Mall Columbus OH, 43210, USA\\
}

\date{Accepted XXX. Received YYY; in original form ZZZ}

\pubyear{2020}

\begin{document}
\label{firstpage}
\pagerange{\pageref{firstpage}--\pageref{lastpage}}
\maketitle

\begin{abstract}

The All-Sky Automated Survey for Supernovae (ASAS-SN) provides long baseline (${\sim}4$ yrs) $V-$band light curves for sources brighter than V$\lesssim17$ mag across the whole sky.
We produced V-band light curves for a total of ${\sim}61.5$ million sources and systematically searched these sources for variability. We identified ${\sim} 426,000$ variables, including ${\sim} 219,000$ new discoveries. Most (${\sim}74\%$) of our discoveries are in the Southern hemisphere. Here we use spectroscopic information from LAMOST, GALAH, RAVE, and APOGEE to study the physical and chemical properties of these variables. We find that metal-poor eclipsing binaries have orbital periods that are shorter than metal-rich systems at fixed temperature. We identified rotational variables on the main-sequence, red giant branch and the red clump. A substantial fraction (${\gtrsim}80\%$) of the rotating giants have large $v_{\rm rot}$ or large NUV excesses also indicative of fast rotation. The rotational variables have unusual abundances suggestive of analysis problems. Semi-regular variables tend to be lower metallicity ($\rm [Fe/H]{\sim}-0.5$) than most giant stars. We find that the APOGEE DR16 temperatures of oxygen-rich semi-regular variables are strongly correlated with the $W_{RP}-W_{JK}$ color index for $\rm T_{eff}\lesssim3800$ K. Using abundance measurements from APOGEE DR16, we find evidence for Mg and N enrichment in the semi-regular variables. We find that the Aluminum abundances of the semi-regular variables are strongly correlated with the pulsation period, where the variables with $\rm P\gtrsim 60$ days are significantly depleted in Al. 
\end{abstract}


%%%%%%%%%%%%%%%%%%%%%%%%%%%%%%%%%%%%%%%%%%%%%%%%%%

%%%%%%%%%%%%%%%%% BODY OF PAPER %%%%%%%%%%%%%%%%%%
\clearpage
\section{Introduction}

Variable stars are useful astrophysical tools that can be used to study the lives and deaths of stars. Pulsating variables, including Cepheids, RR Lyrae stars and Mira variables are used as distance indicators as they follow distinct period-luminosity relationships (e.g., \citealt{1908AnHar..60...87L,2006MNRAS.370.1979M,2018SSRv..214..113B,2008MNRAS.386..313W}, and references therein). Eclipsing binary stars allow for the derivation of dynamical information and fundamental stellar parameters, including the masses and radii of the stars \citep{2010A&ARv..18...67T}. The precise measurements afforded by  eclipsing binaries allow for tests of stellar theory across the Hertzsprung-Russell diagram. Variable stars are also used to study stellar populations and Galactic structure \citep{2018MNRAS.479..211M,2018IAUS..334...57M,2014IAUS..298...40F}.

Modern large scale sky surveys such as the All-Sky Automated Survey (ASAS; \citealt{2002AcA....52..397P}), the All-Sky Automated Survey for SuperNovae (ASAS-SN, \citealt{2014ApJ...788...48S, 2017PASP..129j4502K,Jayasinghe2018}), the Optical Gravitational Lensing Experiment (OGLE; \citealt{2003AcA....53..291U}), the Northern Sky Variability Survey (NSVS; \citealt{2004AJ....127.2436W}), MACHO \citep{1997ApJ...486..697A}, EROS \citep{2002A&A...389..149D}, the Catalina Real-Time Transient Survey (CRTS; \citealt{2014ApJS..213....9D}), the Asteroid Terrestrial-impact Last Alert System (ATLAS; \citealt{2018PASP..130f4505T,2018AJ....156..241H}), Gaia \citep{2018A&A...616A...1G,2018A&A...618A..30H,2019A&A...623A.110G}, and the Zwicky Transient Facility \citep{2019PASP..131a8002B,2020arXiv200508662C} have revolutionized the study of stellar variability. Amateur astronomers have also contributed to these discoveries over the years. As of May 2020, the International Variable Stars Index (VSX,\citealt{2006SASS...25...47W}) hosted by the American Association of Variable Star Observers (AAVSO) lists ${\sim}1.4\times 10^6$ variable stars.

In addition to these modern photometric surveys, large-scale wide-field spectroscopic surveys such as the Apache Point Observatory Galactic Evolution Experiement (APOGEE; \citealt{2006AJ....131.2332G,2017AJ....154...28B,2019PASP..131e5001W}), Large Sky Area Multi-Object Fibre Spectroscopic Telescope (LAMOST \citealt{2012RAA....12.1197C}), GALactic Archaeology with HERMES (GALAH; \citealt{2015MNRAS.449.2604D,2018MNRAS.478.4513B}) and the RAdial Velocity Experiment (RAVE;\citealt{2017ApJ...840...59C}) have been making medium/high resolution spectroscopic observations of millions of Galactic stars. Spectroscopic observations of Galactic stars are invaluable for deciphering the chemical evolution of our Galaxy (see for e.g.,\citealt{2019ApJ...874..102W,2019ApJ...886...84G}), and for evolved stars they also provide clues to understanding chemical enrichment caused by dredge up episodes (see for e.g., \citealt{2015A&A...583A..87S,2019ApJ...872..137S}. Variable stars that have both extensive time-series data and spectroscopic observations will allow for the study of stellar evolution to great detail.

ASAS-SN monitored the visible sky to a depth of $V\lesssim17$ mag with a cadence of 2-3 days using two units in Chile and Hawaii each with 4 telescopes from 2014-2018. Since then, ASAS-SN has expanded to 5 units with 20 telescopes and is currently monitoring the sky in the $g$-band to a depth of $g\lesssim18.5$ mag with a cadence of $\sim1$ day. The ASAS-SN telescopes are hosted by the Las Cumbres Observatory (LCO; \citealt{2013PASP..125.1031B}) in Hawaii, Chile, Texas and South Africa. The primary focus of ASAS-SN is the detection of bright supernovae and other transients (e.g., tidal disruption events, cataclysmic variables, AGN flares, stellar flares, etc.) with minimal bias (e.g., \citealt{2014MNRAS.445.3263H,2016MNRAS.455.2918H,2017MNRAS.471.4966H}), but its excellent baseline and all-sky coverage allows for the characterization of stellar variability across the whole sky. 
 
In Paper I \citep{Jayasinghe2018}, we discovered ${\sim}66,000$ new variables that were flagged during the search for supernovae, most of which are located in regions that were not well-sampled by previous surveys. In Paper II \citep{Jayasinghe2019a}, we homogeneously analyzed ${\sim} 412,000$ known variables from the VSX catalog, and developed a versatile random forest variability classifier utilizing the ASAS-SN V-band light curves and data from external catalogues. As data from The Transiting Exoplanet Survey Satellite (TESS; \citealt{2015JATIS...1a4003R}) became available, we have explored the synergy between the two surveys. In Paper III \citep{Jayasinghe2019b}, we characterized the variability of ${\sim}1.3$ million sources within 18 deg of the Southern Ecliptic Pole towards the TESS continuous viewing zone and identified ${\sim} 11,700$ variables, including ${\sim} 7,000$ new discoveries. We have also explored the synergy between ASAS-SN and large scale spectroscopic surveys using data from APOGEE \citep{2015AJ....150..148H} with the discovery of the first likely non-interacting binary composed of a black hole with a field red giant \citep{2019Sci...366..637T} and the identification of 1924 APOGEE stars as periodic variables in Paper IV \citep{2019MNRAS.487.5932P}. In Paper V, we systematically searched for variable sources with $V<17$ mag in the southern hemisphere and identified ${\sim}220,000$ variable sources, of which ${\sim}88,300$ were new discoveries \citep{Jayasinghe2019c}. In Paper VI, we derived period--luminosity relationships for $\delta$ Scuti stars \citep{Jayasinghe2020b}. We studied contact binaries in Paper VII \citep{Jayasinghe2020a}. In Paper VIII, we identified 11 new ``dipper'' stars in the Lupus star forming region \citep{2020arXiv200514201B}.

Here, we summarize the results of our $V$-band variability search based on ${\sim}61.5$ million ASAS-SN light curves of sources from the AAVSO Photometric All-Sky Survey (APASS; \citealt{2015AAS...22533616H}) DR9 catalog with $V<17$ mag and the ATLAS All-Sky Stellar Reference Catalog (\verb"refcat2", \citealt{2018ApJ...867..105T}) catalog with $g<17$ mag. In this work, we describe our $V$-band variability catalog of ${\sim}426,000$ variable sources, of which ${\sim}219,000$ are new discoveries. In Section $\S2$, we discuss the ASAS-SN observations, our variable star identification and classification procedure and summarize the final $V$-band catalog. Section $\S3$ discusses the cross-matches made to various spectroscopic catalogs and the general spectroscopic properties of the ASAS-SN variable stars. In Section $\S4$, we discuss the eclipsing binaries, rotational variables and semi-regular variables in greater depth and present a summary of our work in Section $\S5$.

\section{The ASAS-SN $V$-band Catalog of Variable Stars}

\begin{figure*}
	\includegraphics[width=\textwidth]{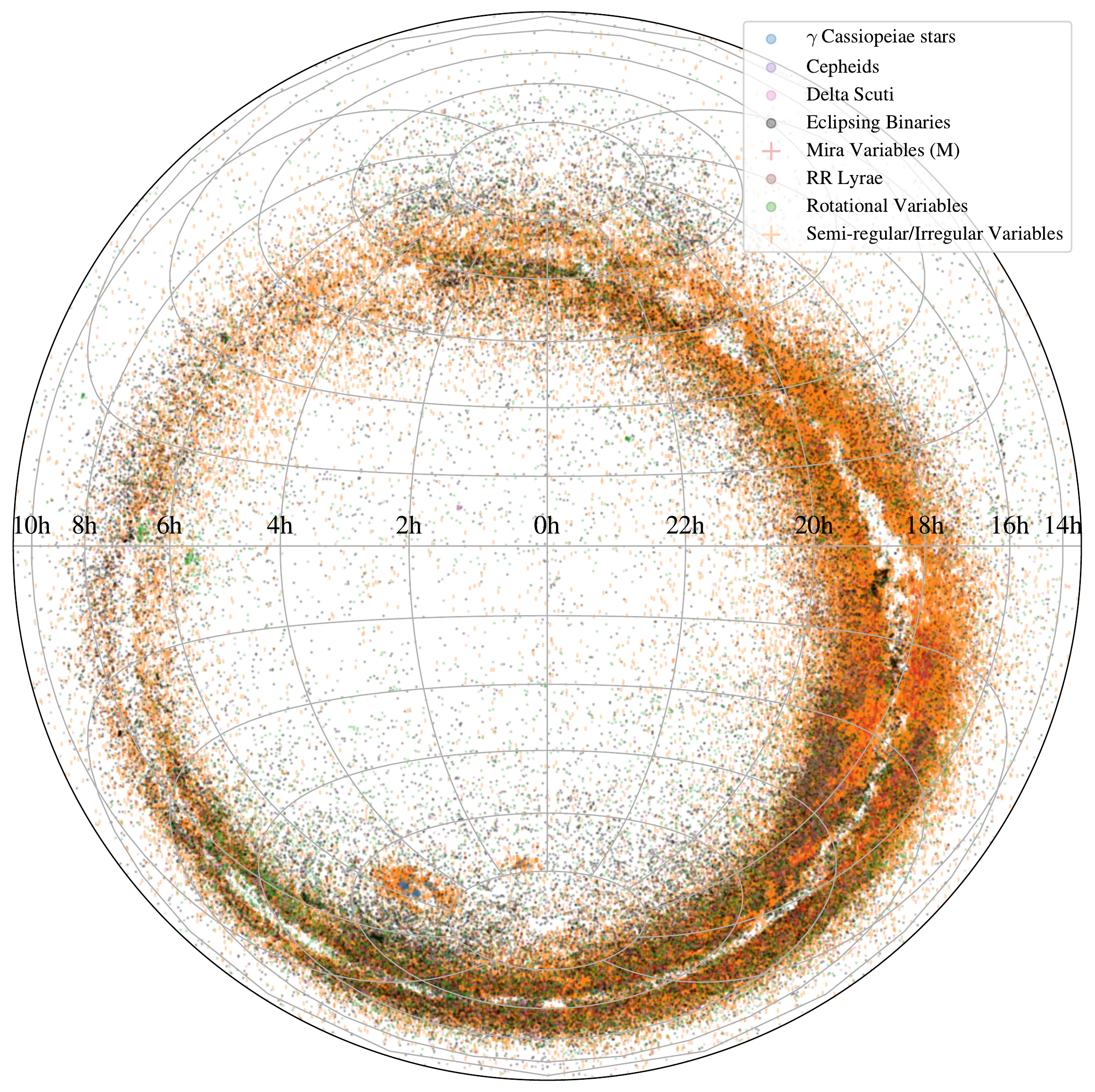}
    \caption{Projected distribution of the $\sim 219,200$ new ASAS-SN discoveries in Equatorial coordinates (Lambert projection). The points are colored by the variability type.}
    \label{fig:fig1}
\end{figure*}

\begin{figure*}
	\includegraphics[width=\textwidth]{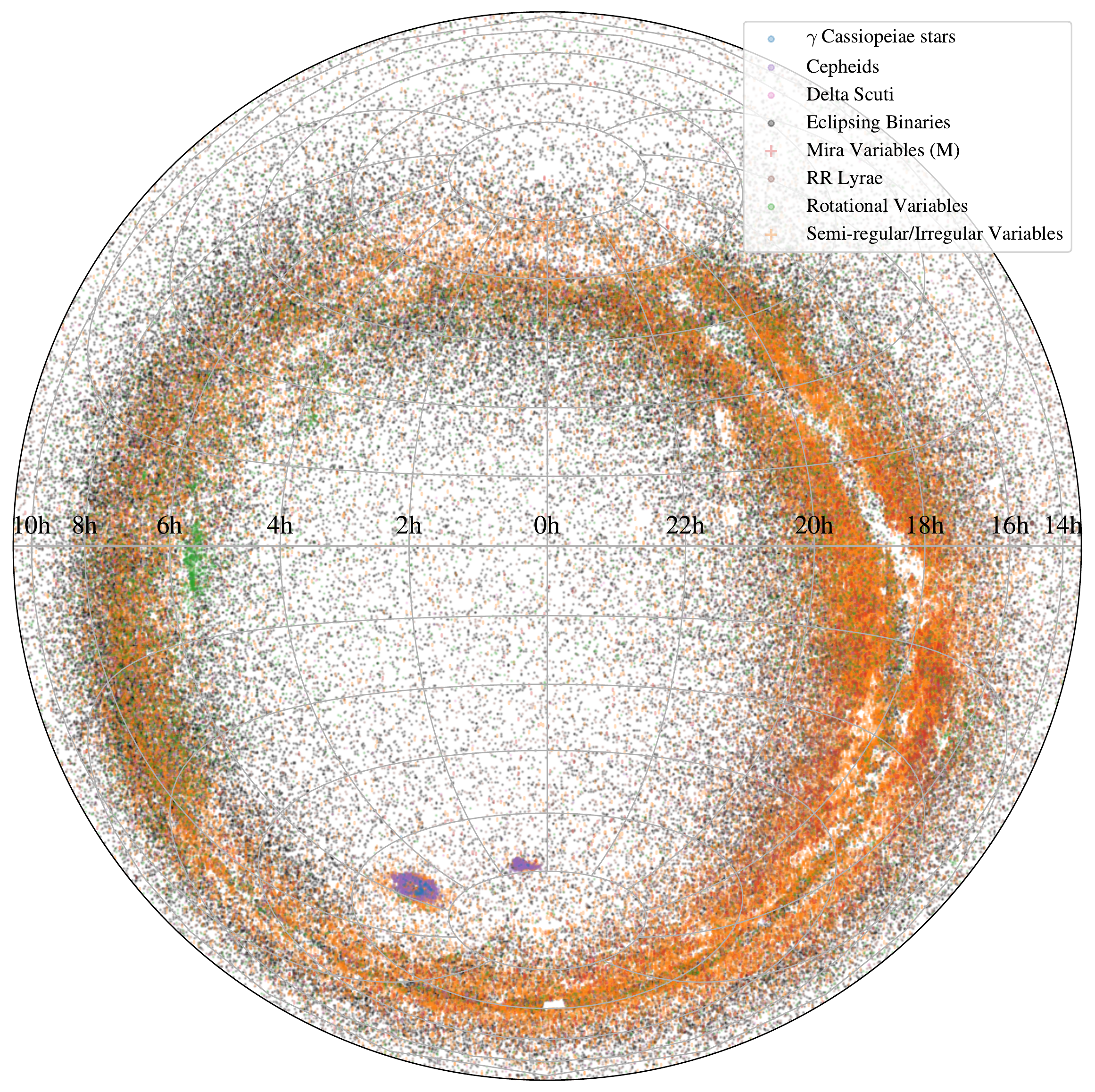}
    \caption{Projected distribution of the $\sim 207,200$ known variable stars recovered through ASAS-SN in Equatorial coordinates (Lambert projection). The points are colored by the variability type.}
    \label{fig:fig2}
\end{figure*}

The ASAS-SN V-band observations used in this work were made by the ``Brutus" (Haleakala, Hawaii) and ``Cassius" (CTIO, Chile) quadruple telescopes between 2013 and 2018. Each ASAS-SN V-band field is observed to a depth of $V\lesssim17$ mag. The field of view of an ASAS-SN camera is 4.5 deg$^2$, the pixel scale is 8\farcs0 and the FWHM is typically ${\sim}2$ pixels. ASAS-SN tends to saturate at ${\sim} 10-11$ mag, but we attempt to correct the light curves of saturated sources for bleed trails (see \citealt{2017PASP..129j4502K}). The V-band light curves were extracted as described in \citet{Jayasinghe2018} using image subtraction \citep{1998ApJ...503..325A,2000A&AS..144..363A} and aperture photometry on the subtracted images with a 2 pixel radius aperture. We corrected the zero point offsets between the different cameras as described in \citet{Jayasinghe2018}. The photometric errors were recalculated as described in \citet{Jayasinghe2019b}.

As we did in Paper V, we started with the APASS DR9 catalog \citep{2015AAS...22533616H} as our input source catalog for the northern hemisphere. We selected ${\sim}23.1$M APASS sources with $V<17$ mag in the northern hemisphere ($\delta>0$ deg). However, there are regions towards the Galactic plane that are missing in the APASS DR9 catalog \citep{2015AAS...22533616H,2019A&A...621A.144M}. To address the issue of incomplete sky coverage, we used the \verb"refcat2" catalog \citep{2018ApJ...867..105T} to produce light curves for the sources missing from APASS DR9. From the \verb"refcat2" catalog, we selected ${\sim}7.1$M sources with \verb"r1"$>30"$ and $G<17$ mag, where \verb"r1" is the radius at which the cumulative $G$ flux in the aperture exceeds the flux of the source being considered and is a measure of blending around a star. We use this cut in \verb"r1" to reduce the number of heavily blended sources. This does bias the selected \verb"refcat2" sources towards more isolated stars, and will reduce the completeness of the catalog, especially towards the Galactic plane. In total, we produced ${\sim}61.5$M $V$-band light curves. The V-band light curves of all ${\sim}61.5$M sources are available online at the ASAS-SN Photometry Database (\url{https://asas-sn.osu.edu/photometry}).

We applied the trained random forest classifier from \citet{Jayasinghe2019c} to identify candidate variables. From this sample, blended sources were identified and removed as described in \citet{Jayasinghe2019c}. Following these procedures, we used the variability classifier implemented in \citet{Jayasinghe2019a}, which consists of a random forest classifier plus several refinement steps, in order to classify the candidate variables. We applied additional quality checks to improve the purity of our catalog (summarized in Table 4 from \citealt{Jayasinghe2019c}). Peak-to-peak variability amplitudes were estimated by fitting a random forest regression model to the light curves \citep{Jayasinghe2019c}. We previously noted that light curves that are contaminated by systematics tend to be classified as irregular or generic variables. Thus, we visually reviewed all the sources that were classified as L, VAR, GCAS, or YSO to improve the purity of our catalog. Following this, we identified ${\sim}124,000$ and ${\sim}40,000$ variables among the northern APASS sources and the \verb"refcat2" sources, respectively. As described in \citet{Jayasinghe2019c}, we cross-matched these variables to the various catalogs of known variable stars. Following this, we identified ${\sim}30,900$ and ${\sim}28,200$ variables as new discoveries among the northern APASS sources and the \verb"refcat2" sources, respectively.

Here, we present the final catalog of variables that we identified in the ASAS-SN $V$-band data. In total, we have identified ${\sim}426,000$ variables in the $V$-band data, out of which ${\sim}219,000$ were new ASAS-SN discoveries. The complete catalog of ${\sim}426,000$ variables and their light curves are available at the ASAS-SN Variable Stars Database (\href{https://asas-sn.osu.edu/variables}{https://asas-sn.osu.edu/variables}) along with the $V$-band light curves for each source. Both this database and the VSX catalog have included these Northern variables and light curves since September 2019. The ASAS-SN variable stars have also been included in the VSX catalog\footnote{VSX: https://www.aavso.org/vsx/index.php}. Table \ref{tab:var} lists the number of sources of each variability type in the ASAS-SN $V$-band catalog of variable stars. 

The definition of an ASAS-SN discovery is that we could find no reference to the system as a variable star at the time we publicly released the new discoveries from papers I-IX. We directly matched to many large catalogs like the ASAS \citep{2002AcA....52..397P}, CRTS \citep{2014ApJS..213....9D} and WISE \citep{2018ApJS..237...28C} catalogs of variables and also to the AAVSO's VSX catalog \citep{2006SASS...25...47W}, which attempts to be a complete summary of variable stars.  Because the final accounting of ${\sim}220,000$ new variables comes from this series of papers that began in 2018, how they overlap with variable star catalogs published in the interim depends on which survey first released the variable. As a particular example, Paper I \citep{Jayasinghe2018} appeared before the Gaia DR2 variable star catalog \citep{2019A&A...623A.110G}, so a previously unknown variable from Paper I is a new ASAS-SN variable even if it is in the later Gaia DR2 catalog. In the later papers, the Gaia DR2 catalog was included as one of the catalogs to check before identifying a source as a newly discovered variable. So, for example, the roughly ${\sim}60,000$ new variables released as part of this paper contain no stars in the Gaia catalog. The same approach was taken for the ATLAS variable star catalog \citep{2018AJ....156..241H}.

With the completion of our V-band variability survey, ASAS-SN has significantly increased the numbers of semi-regular variables ($+224\%$), $\delta$ Scuti variables ($+101\%$), rotational variables ($+116\%$) and detached eclipsing binaries ($+90\%$). Combined, we have increased the census of semi-regular and irregular variables by $+235\%$ and $\delta$ Scuti variables (both the DSCT and HADS sub-types) by $+81\%$. The majority of the semi-regular variables, $\delta$ Scuti variables, and rotational variables have small variability amplitudes, which make their discovery non-trivial. Furthermore, semi-regular variables have periods between ${\sim}10-10^3$ days, and their detection requires long duration light curves with good cadence. The long baselines (${\sim}500-2000$ days, depending on the field) and the cadence (${\sim}2-3$ days) of the ASAS-SN $V$-band light curves enable the discovery of these long period variables with good efficiency. We illustrate the sky distribution of the new ASAS-SN variables in Figure \ref{fig:fig1}. Variables with large amplitudes and strong periodicity are relatively easily discovered and characterized by wide field photometric surveys and amateur astronomers, so the existing completeness of these variable types is very high. Nevertheless, we make considerable additions to the census of these variable types, with the smallest percentage increase coming from the first-overtone classical Cepheids (DCEPS) at $+9\%$. Overall, we have increased the numbers of variable stars across the entire sky with $V$-band magnitudes $11<V<17$ mag by $+106\%$. We also show the sky distribution of the known variables identified in our work in Figure \ref{fig:fig2}. Most (${\sim}74\%$) of our discoveries were in the Southern hemisphere. Given the relative scarcity of wide-field photometric surveys that monitor the Southern hemisphere, this is not surprising. 

\begin{figure*}
	\includegraphics[width=1\textwidth]{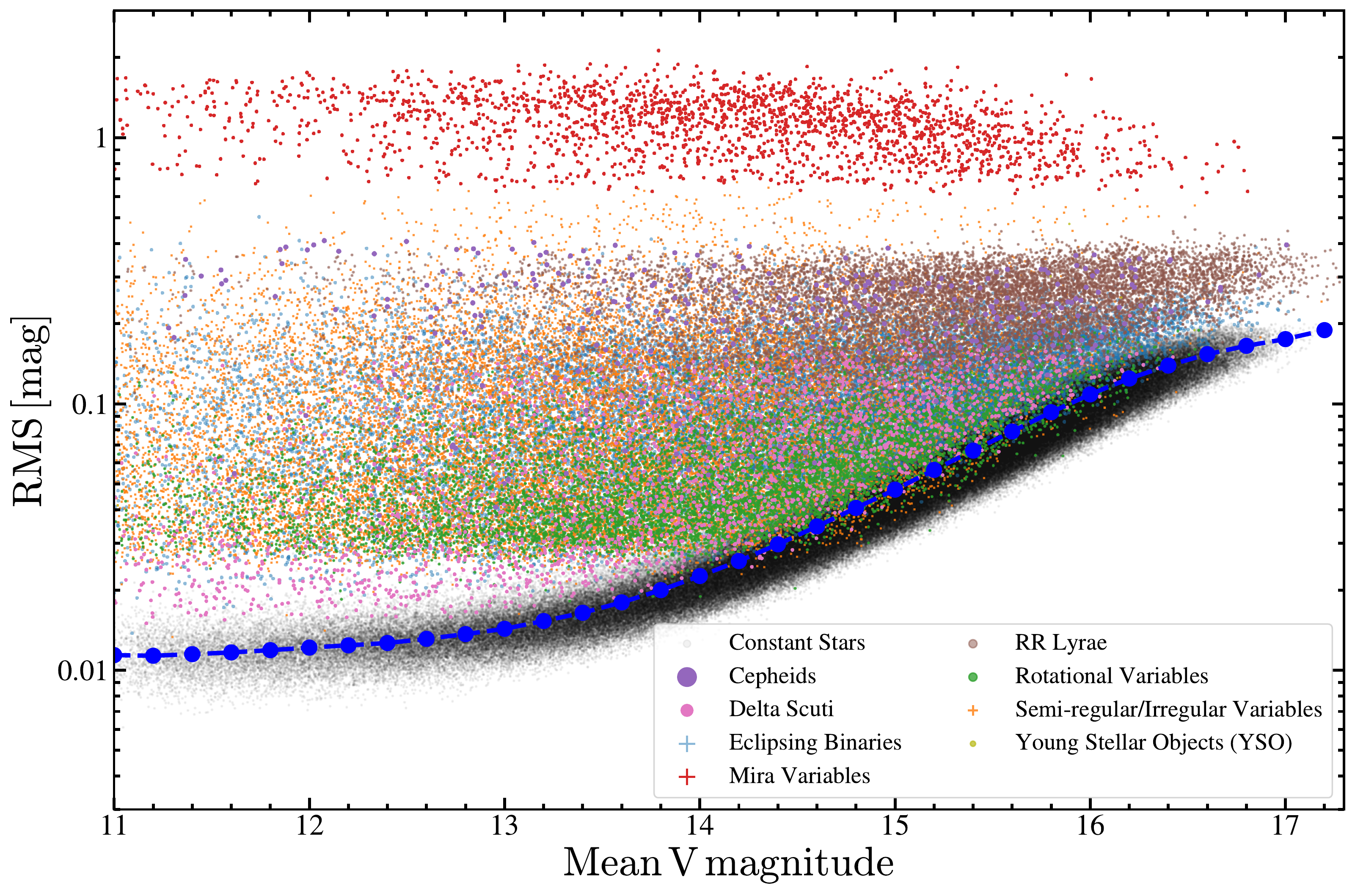}
    \caption{Root-mean-square (RMS) as a function of mean magnitude for the ASAS-SN V-band variable stars with $\rm Prob>0.95$. A sample of $300,000$ constant sources are shown in black. The median RMS of the constant stars is shown in blue for bins of 0.2 mag.}
    \label{fig:fig32}
\end{figure*}

Figure \ref{fig:fig32} shows the root-mean-square (RMS) variability of a sample of V-band variables as a function of their mean brightness along with $300,000$ constant sources. As expected, Mira variables have the largest RMS variability. At the bright end of the survey, ASAS-SN is able to detect variability at the level of ${\sim}1\%$, dropping to ${\sim}10-20\%$ at the faintest magnitudes.

\begin{figure}
	\includegraphics[width=0.5\textwidth]{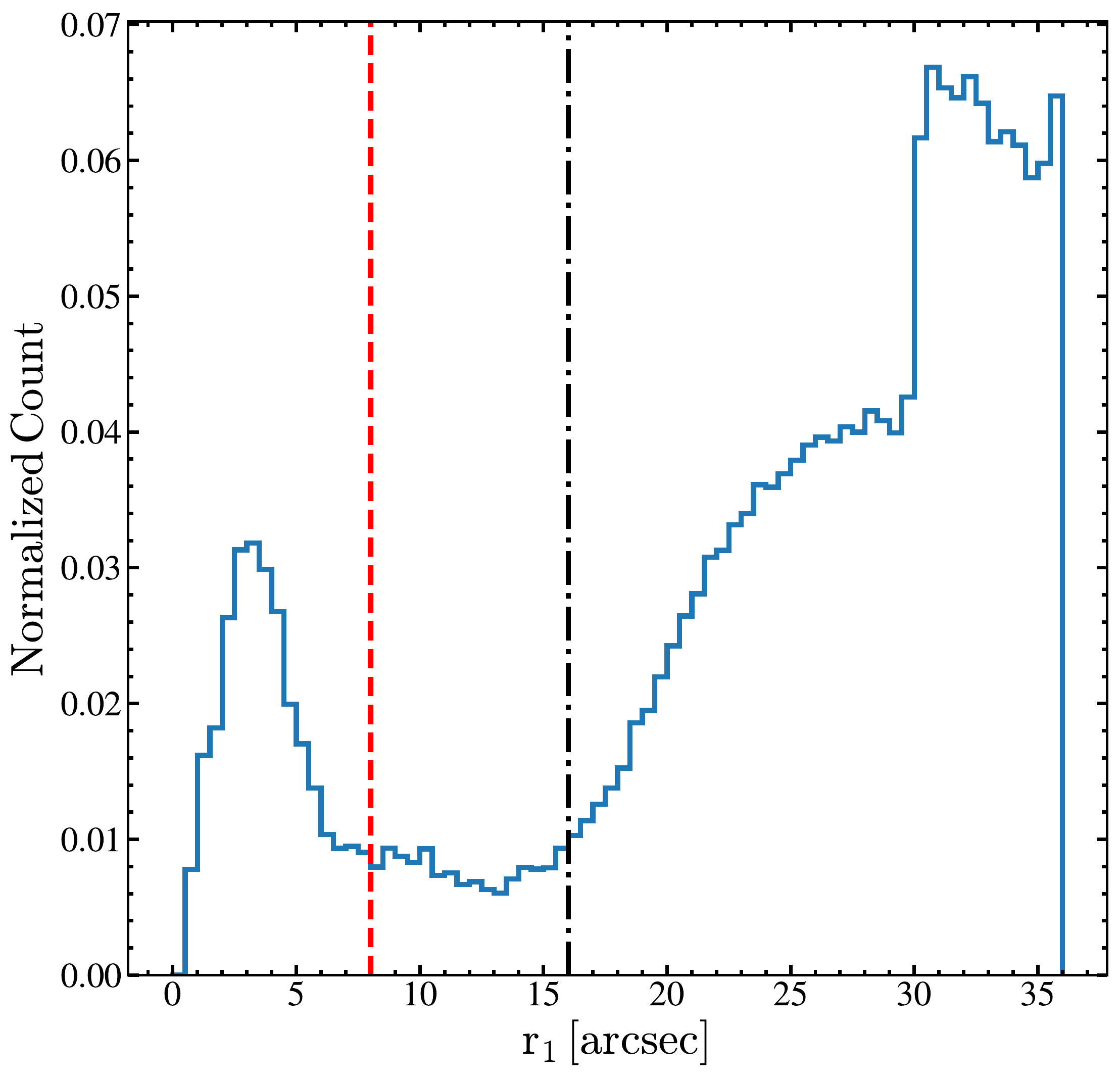}
    \caption{Distribution of the radius at which the cumulative \textit{Gaia} $G$ flux in the aperture exceeds the flux of the variable source ($\rm r1$ from \citet{2018ApJ...867..105T}) for the ASAS-SN V-band variable stars. The ASAS-SN pixel scale (${\sim}8$\farcs0) and FWHM (${\sim}16$\farcs0) are shown as red dashed and black dot-dashed lines respectively.}
    \label{fig:fig34}
\end{figure}

We can estimate the amount of flux blended with the variables using the proximity statistics from \verb"refcat2" \citep{2018ApJ...867..105T}. Figure \ref{fig:fig34} shows the distribution of the radius at which the cumulative \textit{Gaia} $G$-band flux in the aperture exceeds the flux of the variable source (\verb"r1"). A value of 99\farcs9 is assigned when the star is so isolated that the cumulative flux never exceeds the required threshold within the 36\farcs0 search radius employed by \citet{2018ApJ...867..105T}. A large fraction (${\sim}72\%$) of the V-band variables met this criterion. Another ${\sim}11\%$ of the variables had \verb"r1"$>30"$, and are effectively isolated. Thus, in the context of blending, most (${\sim}83\%$) of the variables can be considered to be relatively isolated. However, a peak corresponding to heavily blended sources is seen at ${\sim}3$\farcs0 in Figure \ref{fig:fig34}. Only ${\sim}4\%$ and ${\sim}6\%$ of the variables exceed their flux within an aperture corresponding to the ASAS-SN pixel scale (${\sim}8$\farcs0) and FWHM (${\sim}16$\farcs0) respectively. The main effect of blending is to reduce the apparent variability amplitude.

\begin{table*}
	\centering
	\caption{Variables by type}
	\label{tab:var}
\begin{tabular}{llrrr}
		\hline
		VSX Type & Description & Known & New & New/Known \\
		\hline
CWA   & W Virginis type variables with $P>8$ d & 243 & 108 & 0.44\\
CWB   & W Virginis type variables with $P<8$ d & 182 & 72 & 0.40\\
DCEP  & Fundamental mode Classical Cepheids& 1466 & 162 & 0.11\\
DCEPS & First overtone Cepheids & 529 & 49 & 0.09\\
DSCT  & $\delta$ Scuti variables & 2203 & 2227 & 1.01\\
EA    & Detached Algol-type binaries & 21399 & 19329 & 0.90\\
EB    & $\beta$ Lyrae-type binaries & 17078 & 6956 & 0.41\\
EW    & W Ursae Majoris type binaries & 56431 & 15415 & 0.27\\
HADS  & High amplitude $\delta$ Scuti variables & 2541 & 1626 & 0.64\\
M    & Mira Variables & 4612 & 1661 & 0.36\\
ROT   & Rotational variables & 15218 & 17624 & 1.16\\
RRAB  & RR Lyrae variables (Type ab) & 16683 & 5136 & 0.31\\
RRC   & First Overtone RR Lyrae variables &4380 & 1808 & 0.41\\
RRD   & Double Mode RR Lyrae variables & 299& 71 & 0.24\\
RVA   & RV Tauri variables (Subtype A) & 44 & 22 & 0.50\\
SR    & Semi-regular variables & 51270 & 115088 & 2.24\\
\hline
L     & Irregular variables & 6584 & 21110 & 3.21\\
GCAS  & $\gamma$ Cassiopeiae variables & 146 & 258 & 1.77\\
YSO   & Young stellar objects & 2173 & 2353 & 1.08\\
\hline
GCAS:  & Uncertain $\gamma$ Cassiopeiae variables & 37 & 71 & 1.92\\
VAR  & Generic variables & 2668 & 8193 & 3.07\\
\hline
Total  &  & 206218 & 219351 & 1.06\\
\hline
\end{tabular}
\end{table*}

During the process of variability classification, we cross-matched the variables with Gaia DR2 \citep{2018A&A...616A...1G} using a matching radius of 5\farcs0. The sources were assigned distance estimates from the Gaia DR2 probabilistic distance estimates \citep{2018AJ....156...58B} by cross-matching based on the Gaia DR2 \verb"source_id". We also cross-matched these sources to the 2MASS \citep{2006AJ....131.1163S}, AllWISE \citep{2013yCat.2328....0C,2010AJ....140.1868W} and GALEX \citep{2017ApJS..230...24B} catalogues using a matching radius of 10\farcs0. We used \verb"TOPCAT" \citep{2005ASPC..347...29T} for this process. For each source, we also calculate the total line of sight Galactic reddening $E(B-V)$ from the recalibrated `SFD' dust maps \citep{2011ApJ...737..103S,1998ApJ...500..525S}.  We calculated the absolute, reddening-free Wesenheit magnitudes \citep{1982ApJ...253..575M,2018A&A...616L..13L} \begin{equation}
    W_{RP}=M_{\rm G_{RP}}-1.3(G_{BP}-G_{RP}) \,, 
	\label{eq:wrp}
\end{equation} 
and
\begin{equation}
    W_{JK}=M_{\rm K_s}-0.686(J-K_s) \,,
	\label{eq:wk}
\end{equation}for each source, where the $G_{BP}$ and $G_{RP}$ magnitudes are from Gaia DR2 \citep{2018A&A...616A...1G} and the $J$ and $K_s$ magnitudes are from 2MASS \citep{2006AJ....131.1163S}. The Wesenheit magnitudes are important for refining variable type classifications (see \citealt{Jayasinghe2019a}).

\begin{figure*}
	\includegraphics[width=\textwidth]{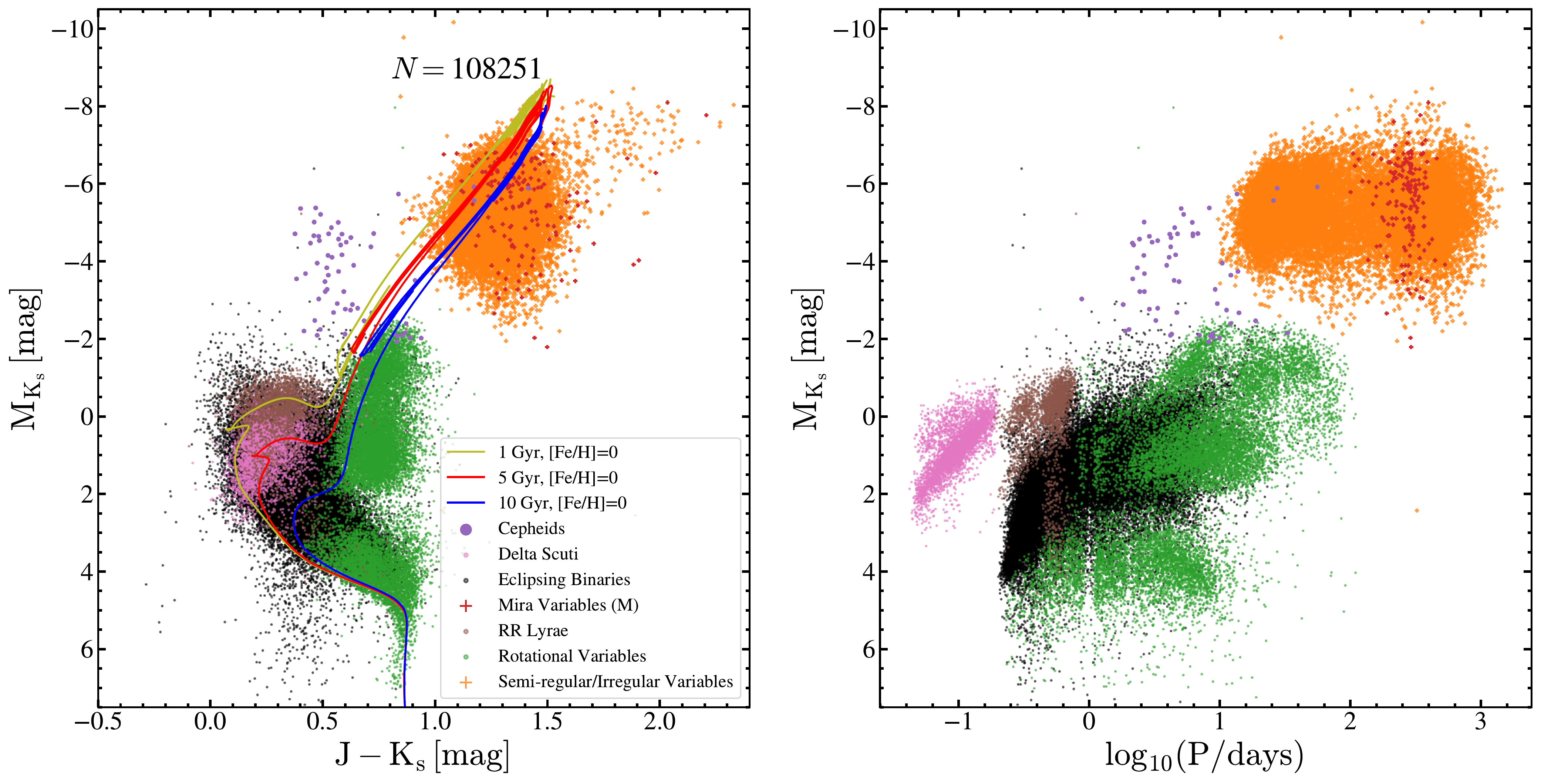}
    \caption{NIR color-magnitude (left) and period-luminosity (right) diagrams for the ASAS-SN V-band variable stars with $\rm Prob>0.95$, $A_V<2$ mag and parallaxes better than $20\%$. MIST isochrones \citep{2016ApJ...823..102C,2016ApJS..222....8D} for single stars with $\rm [Fe/H]=0$ at 1 Gyr, 5 Gyr and 10 Gyr are shown for comparison.}
    \label{fig:fig3}
\end{figure*}

The near-infrared (NIR) $M_{K_s}$ vs. $J-K_s$ color-magnitude diagram and the $\rm M_{Ks}$ period--luminosity relationship (PLR) diagram for all the variables with variable type classification probabilities $\rm Prob>0.95$, $A_V<2$ mag and \textit{Gaia} DR2 parallaxes better than $20\%$ are shown in Figure \ref{fig:fig3}. Generic and uncertain variable types are not shown. We have sorted the variables into groups to highlight the different classes of variable sources. Rotational variables in our catalog consist of spotted stars on the main-sequence (MS) as well as evolved stars on the red-giant branch (RGB). Semi-regular variables and Mira variables lie on the asymptotic giant branch (AGB). Several PLR sequences are seen in Figure \ref{fig:fig3}. We studied the PLR sequences for the $\delta$ Scuti variables and contact binaries in Papers VI and VII respectively. The slight deficits of variables at the aliases of a sidereal day (e.g., $P\approx1$ d, $P\approx2$ d, $P\approx30$ d, etc.) are due to the quality checks from \citet{Jayasinghe2019c}.

\begin{figure}
	\includegraphics[width=0.5\textwidth]{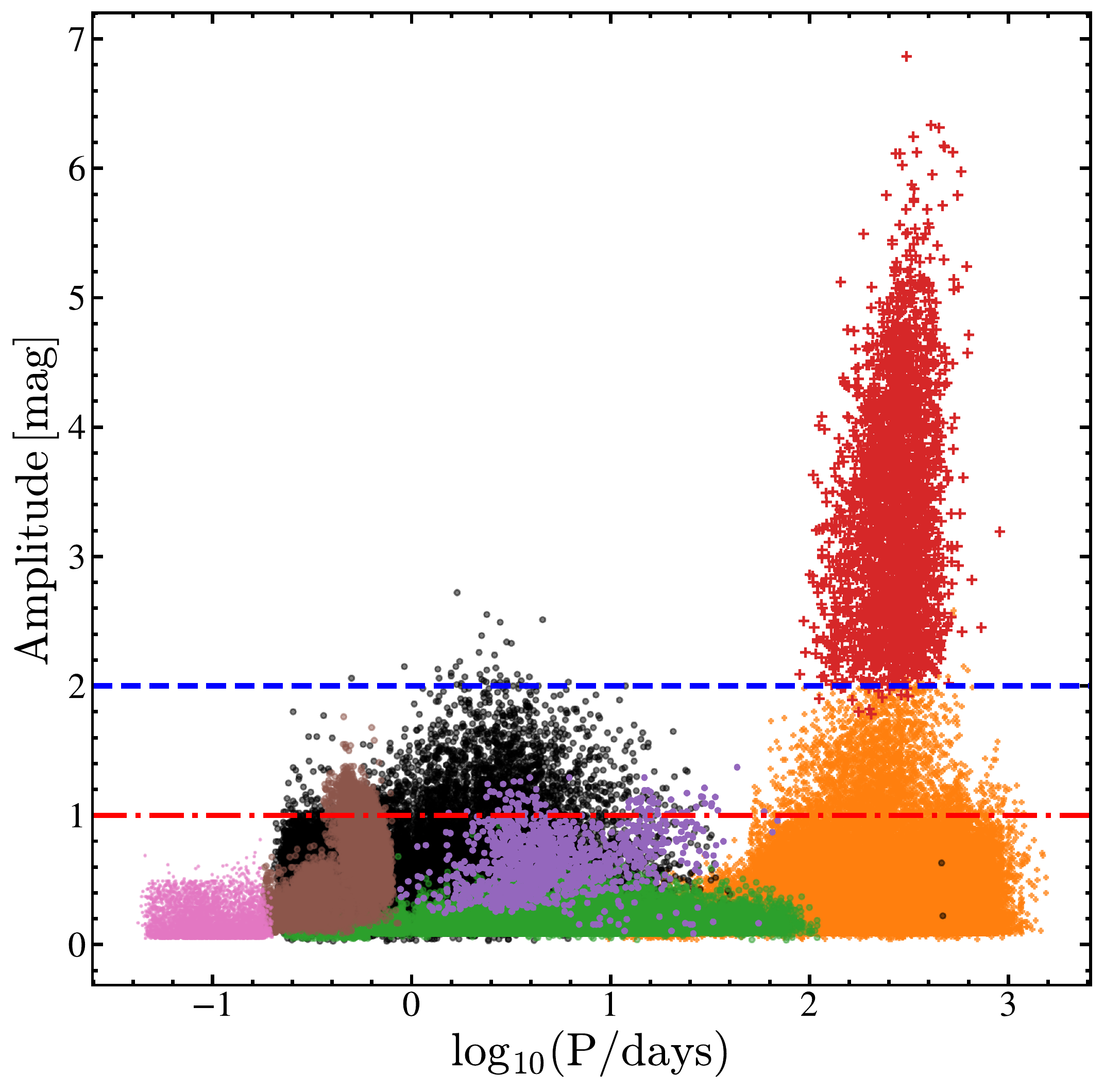}
    \caption{Period-amplitude distributions of the periodic ASAS-SN V-band variables. Reference amplitudes of 1 and 2 mag are shown in red and blue respectively.  }
    \label{fig:fig33}
\end{figure}

The period-amplitude distribution of the periodic variables is shown in Figure \ref{fig:fig33}. Most variables (${\sim}96\%$) have amplitudes $A<1$ mag. Of the variables with amplitudes $A>1$ mag, ${\sim}59\%$ are Mira and semi-regular variables. Eclipsing binaries (${\sim}22\%$) and RR Lyrae (${\sim}17\%$) are also prominent in this group. Variables with large amplitudes $A>2$ mag are rarer (${\sim}1\%$) and these high amplitude variables are almost exclusively Mira variables.

\section{ASAS-SN Variables in Wide-field spectroscopic surveys}

ASAS-SN significantly overlaps with modern major wide-field spectroscopic surveys owing to its all-sky coverage and the magnitude range of the survey. We cross-matched our catalog with the APOGEE DR16 catalog \citep{2015AJ....150..148H, 2016AJ....151..144G, 2019arXiv191202905A,2020arXiv200705537J}, the RAVE-on catalog \citep{2017ApJ...840...59C}, the LAMOST DR5 v4 catalog \citep{2012RAA....12.1197C} and the GALAH DR2 catalog \citep{2015MNRAS.449.2604D,2018MNRAS.478.4513B} using a matching radius of 5\farcs0. LAMOST only reports $\rm T_{eff}$, $\rm \log (g)$, and $\rm [Fe/H]$ for A, F, G and K stars. We identified 39811 (39036) total (unique) matches to the catalogs from the LAMOST (17381), GALAH (3067), RAVE (15050) and APOGEE (4313) spectroscopic surveys. 

These spectroscopic surveys differ in their targeting strategy, spectral resolution and data reduction. APOGEE is a NIR survey with a spectral resolving power of $R{\sim}22,500$. LAMOST, RAVE and GALAH are optical surveys with $R{\sim}1,800$, $R{\sim}7,000$ and $R{\sim}28,000$ respectively. The different pipelines that are used in the data reduction process can result in survey specific offsets in the spectroscopic parameters for similar stars. However, efforts have been made to compare the various spectroscopic parameters amongst these surveys, and the parameters are generally similar for most stars \citep{2017ApJ...840...59C,2018MNRAS.478.4513B}. The targeting strategies are also different, with APOGEE focusing more on observing red stars than the other surveys. Additionally, variability might impact the derivation of spectroscopic parameters, particularly when sources are in double-lined spectroscopic binaries (eclipsing binaries). Thus, we cannot rely on a single survey to study all the different classes of variable stars. 

\begin{figure*}
	\includegraphics[width=\textwidth]{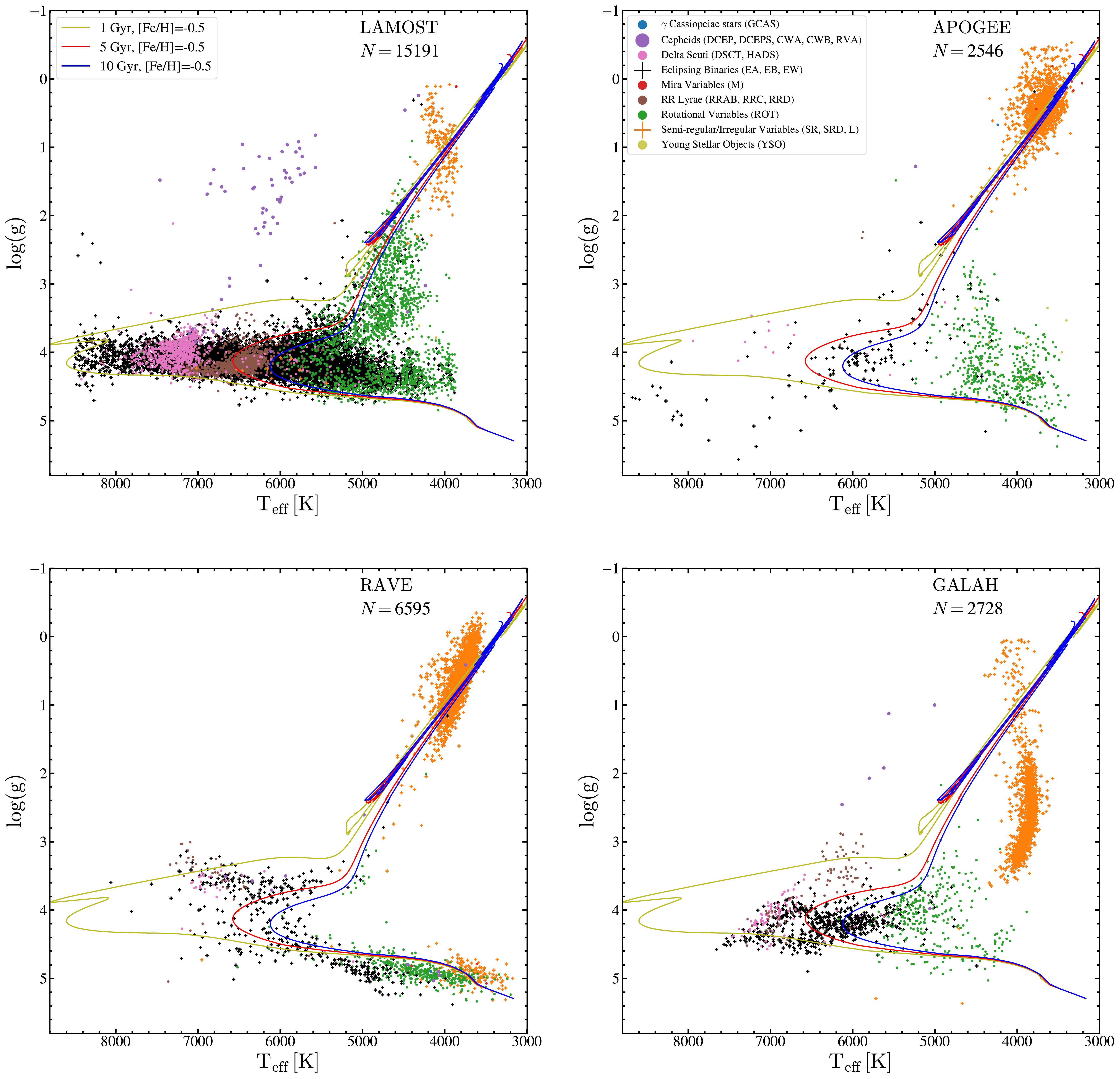}
    \caption{Distributions of the ASAS-SN V-band variables in $\rm \log (g)$ and $\rm T_{eff}$ across the LAMOST, APOGEE, RAVE and GALAH datasets. The points are colored by the variable type. Spectroscopic data quality cuts are not included. MIST isochrones \citep{2016ApJ...823..102C,2016ApJS..222....8D} for single stars with $\rm [Fe/H]=-0.50$ at 1 Gyr, 5 Gyr and 10 Gyr are shown for comparison.}
    \label{fig:fig4}
\end{figure*}

We illustrate the distribution of the ASAS-SN variable stars with classification probabilities $\rm Prob>0.95$ in $\rm T_{eff}$ and $\rm \log (g)$ (Kiel Diagram) across these surveys in Figure \ref{fig:fig4}. We will only consider sources with $\rm Prob>0.95$ in all of our examinations of spectroscopic properties. We have generally not implemented any cuts on the various flags that are available across these data sets. The LAMOST survey provides the only data set that samples all the major variability classes in our catalog. Most of the cross-matches to APOGEE come from the semi-regular variables and rotational variables. If we implement the data quality cut \verb"ASPCAPFLAG=0", ${\sim}65\%$ of the APOGEE sources remain. We note that some fraction of the SR variables with temperatures $\rm T_{eff}<3800$ K in the RAVE data set have values of $\rm \log (g)$ that are inconsistent with being evolved stars on the AGB. If we implement the quality cut \verb"QC=0" from \citet{2017ApJ...840...59C}, ${\sim}95\%$ of the RAVE sources remain, but this issue with the location of the giants persists. It is also clear that the vast majority of the semi-regular variables in GALAH have incorrect spectroscopic parameters as they populate a non-physical locus in the Kiel diagram. \citet{2018MNRAS.478.4513B} noted this issue with their pipeline for cool giants with $\rm T_{eff}<4500 \,K$. Implementing the data quality cut \verb"FLAG_CANNON=0" suggested by \citet{2018MNRAS.478.4513B}, eliminates all but ${\sim}5\%$ of the GALAH sources, which suggests that the GALAH data set is sub-optimal for our purpose of studying variable stars. We find that the data from the LAMOST and RAVE surveys are best suited to characterizing pulsators, eclipsing binaries and rotational variables. APOGEE data are excellent for the characterization of the cooler semi-regular and irregular variables.

\begin{figure*}
	\includegraphics[width=\textwidth]{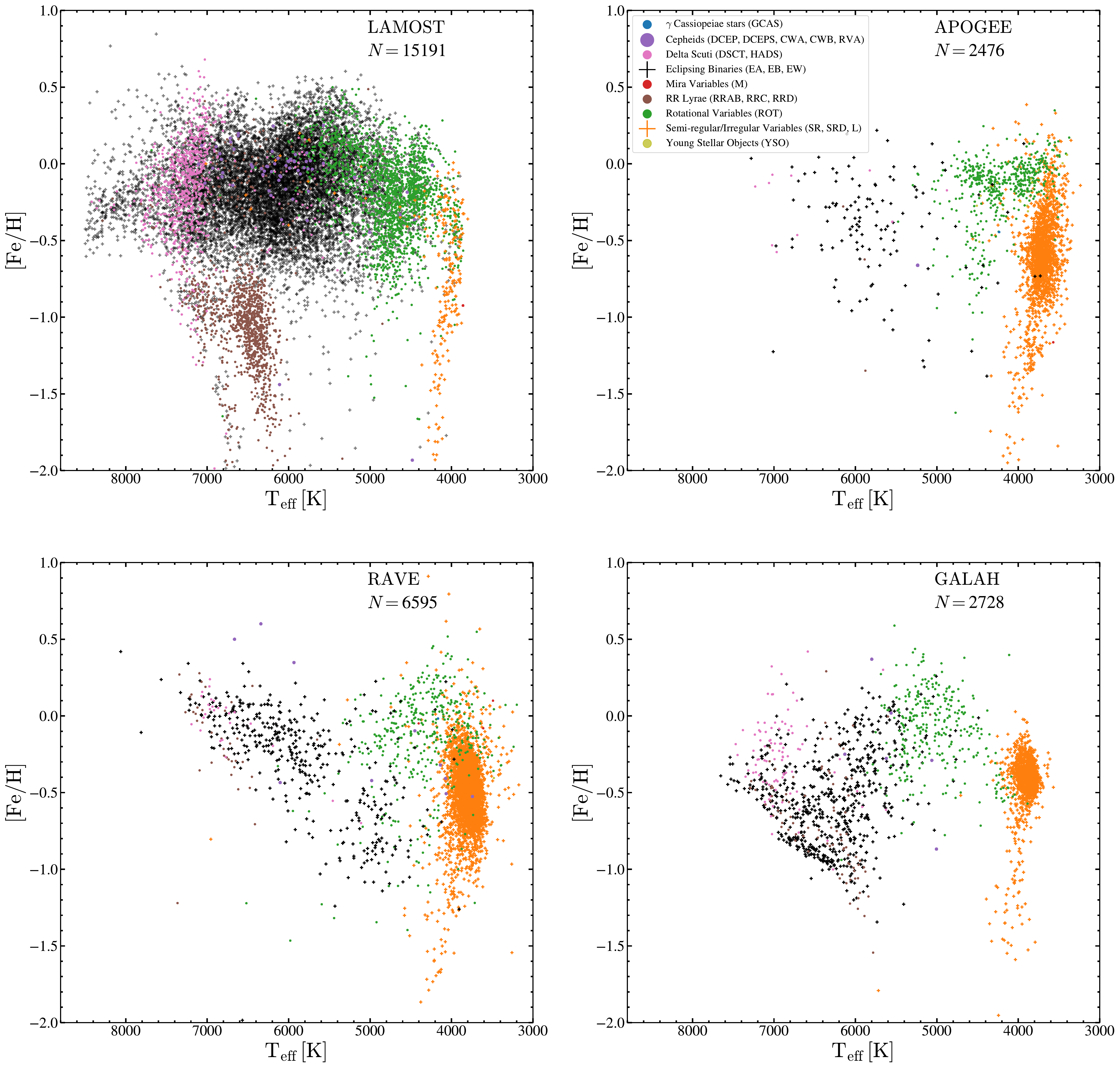}
    \caption{Distributions of the ASAS-SN V-band variables in $\rm T_{eff}$ and $\rm [Fe/H]$ across the LAMOST, APOGEE, RAVE and GALAH datasets. The points are colored by the variable type.}
    \label{fig:fig5}
\end{figure*}

The distribution of the variables in $\rm T_{eff}$ and $\rm [Fe/H]$ is shown in Figure \ref{fig:fig5}. We note that the eclipsing binaries in the GALAH sample have metallicities that are largely inconsistent with the eclipsing binaries in the other catalogs. However, only a tiny fraction of this sample passes the quality cut \verb"FLAG_CANNON=0". In order to improve the accuracy of our work, we will only consider GALAH sources with \verb"FLAG_CANNON=0". We also restrict the RAVE sample of semi-regular variables to those stars with $\rm \log (g)<2$.

\begin{figure*}
	\includegraphics[width=0.8\textwidth]{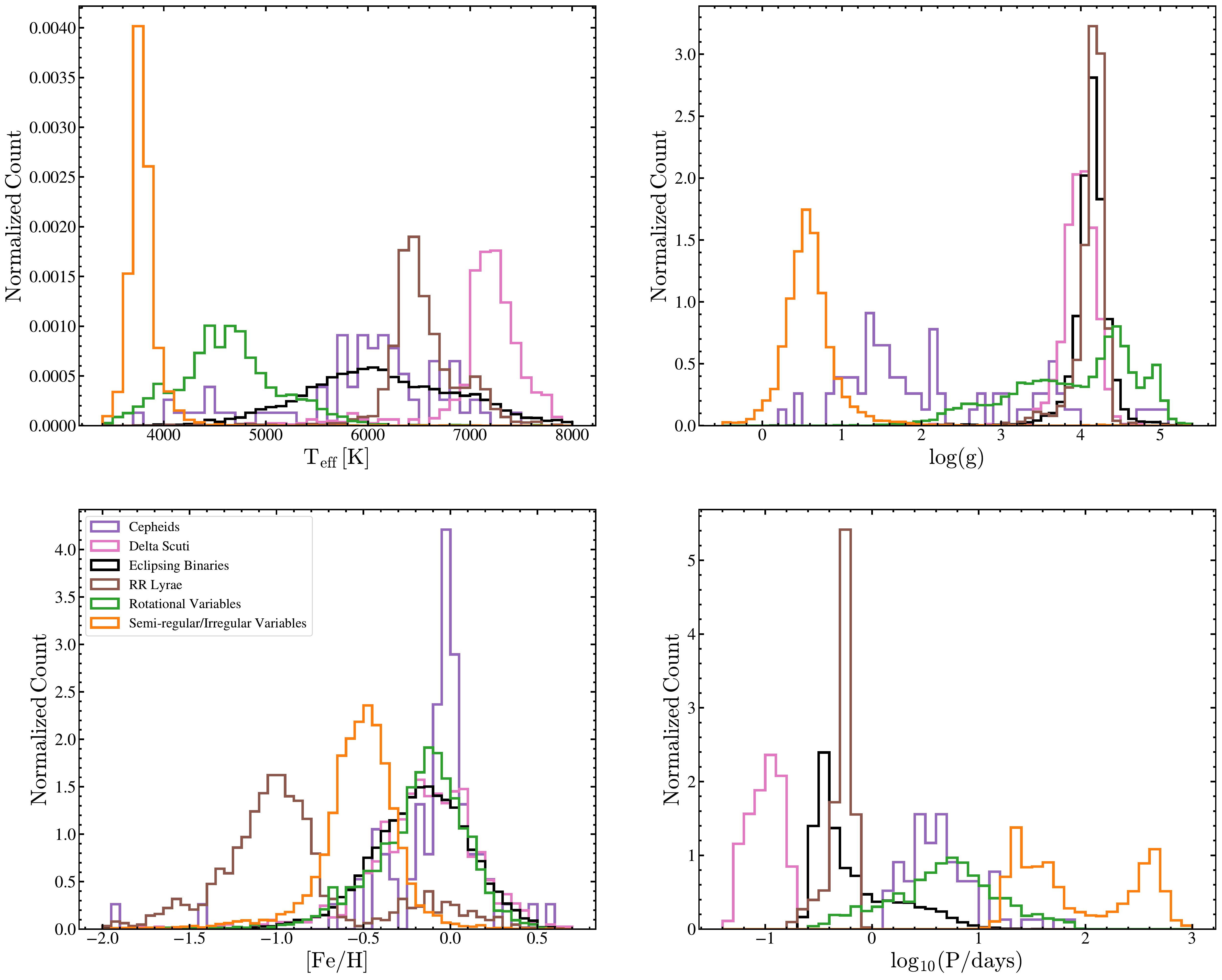}
    \caption{Distributions of the ASAS-SN V-band variables in $\rm T_{eff}$, $\rm \log (g)$, $\rm [Fe/H]$ and $\rm \log_{10}(P/days)$. The histograms are colored by the variable type.}
    \label{fig:fig6}
\end{figure*}

The combined distributions of the variables in $\rm T_{eff}$, $\rm \log (g)$, $\rm [Fe/H]$ and $\rm \log_{10}(P/days)$ are shown in Figure \ref{fig:fig6}. The median and standard deviation of the spectroscopic parameters for variable types with sample sizes $N>10$ are also summarized in Table \ref{tab:spec}. On average, eclipsing binaries have sub-solar metallicities ($\rm [Fe/H]{\sim}-0.2$). Cepheid variables have metallicities consistent with Solar metallicity. $\delta$ Scuti variables and rotational variables have metallicites that are slightly sub-Solar ($\rm [Fe/H]{\sim}-0.1$). Semi-regular variables are strongly peaked at $\rm [Fe/H]{\sim}-0.5$, with very few having Solar or super-Solar metallicities. The population II RRAB stars have very low metallicities with $\rm [Fe/H]{\sim}-1$. The average metallicity of the overtone RR Lyrae (RRC) variables in our sample is $\rm [Fe/H]{\sim}-0.3$ and has a large dispersion of $\sigma{\sim}0.8$ dex. \citet{1991ApJ...378..119W} found that, on average, both RRAB and RRC variables in Baade's window had $\rm [Fe/H]{\sim}-1$. This suggests that some fraction of the sources classified as RRC variables are in fact EW-type eclipsing binaries with higher metallicities ($\rm [Fe/H]{\sim}-0.2$). Without spectroscopic information, there can be non-negligible confusion between these two variable groups during the classification process due to their very similar and symmetrical light curve shapes.

The temperatures and surface gravities of the semi-regular/irregular variables are consistent with these stars being highly evolved AGB stars. Classical pulsators, such as the RR Lyraes, Cepheids and $\delta$ Scuti variables, have temperatures that fall within the instability strip for pulsations. Overtone Cepheids and RR Lyrae are hotter than the fundamental mode pulsators at fixed temperature. The eclipsing binaries in this sample mostly have surface gravities consistent with main sequence (MS) or slightly evolved stars. Figure \ref{fig:fig5} shows that the eclipsing binaries span a large range in effective temperature with $\rm 4000\,K<Teff<8000\, K$ (A-K spectral types). On average, $\beta$ Lyrae-type semi-detached binaries (EB) have hotter effective temperatures than both contact binaries and detached eclipsing binaries. The surface gravities of the rotational variables peak near the MS, but the dispersion of $\sigma{\sim}0.8$ dex is large because it is a much more diverse population of sources, including spotted stars on the RGB. We will further investigate the populations of eclipsing binaries and rotational variables in $\S4$.

\begin{figure*}
	\includegraphics[width=0.5\textwidth]{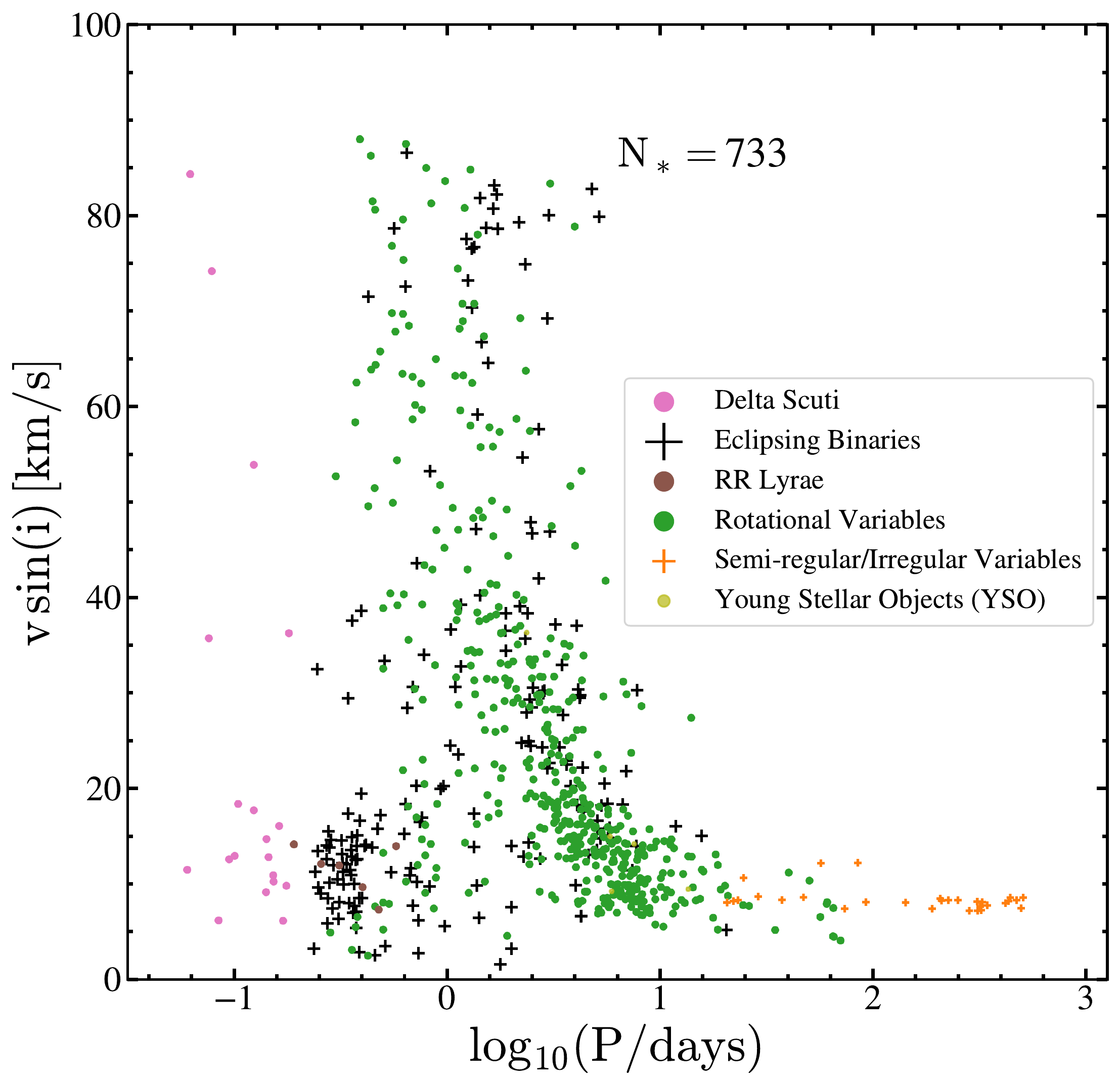}
    \caption{Distributions of the ASAS-SN V-band variables in $v\sin(i)$ vs. $\rm \log_{10}(P/days)$. The points are colored by the variable type.}
    \label{fig:fig7}
\end{figure*}

Figure \ref{fig:fig7} shows the correlation between APOGEE and GALAH estimates of $v\sin(i)$ and the ASAS-SN $\rm \log_{10}(P/days)$ for the variables in the spectroscopic sample. As expected for the rotational variables, $v\sin(i)$ decreases with the period. For the 32 semi-regular variables with $v\sin(i)$ measurements from GALAH (APOGEE does not report $v\sin(i)$ for very evolved stars), the median was $v\sin(i){\sim}8\, \rm km/s$. $\delta$ Scuti variables have a broad distribution in $v\sin(i)$, consistent with previous measurements \citep{1997A&AS..122..131S}. Most short period ($\rm P<1$ d) eclipsing binaries have $v\sin(i){\lesssim}20\, \rm km/s$.

\begin{table}
	\centering
	\caption{Distribution of $\rm [Fe/H]$, $\rm T_{eff}$ and $\rm \log (g)$ for variable types with at least 10 members with classifications. The variable types are defined in Table \ref{tab:var}. The median, and standard deviation for each spectroscopic parameter is shown.}
	\label{tab:spec}
\begin{tabular}{lrrrr}
		\hline
		VSX Type & $N$ & $\rm T_{eff}$ & $\rm \log (g)$ & $\rm [Fe/H]$\\
		\hline
DCEP & 40 & 6094$\pm$453 & 1.5$\pm$0.8 & $-0.02\pm$0.28\\
DCEPS & 17 & 6618$\pm$428 & 2.9$\pm$0.8 & 0.01$\pm$0.16\\
		\hline
DSCT & 537 & 7194$\pm$408 & 4.0$\pm$0.2 & $-0.09\pm$0.27\\
HADS & 267 & 7256$\pm$538 & 4.1$\pm$0.2 & $-0.24\pm$0.34\\
		\hline
EA & 3115 & 6283$\pm$851 & 4.1$\pm$0.3 & $-0.13\pm$0.27\\
EB & 1948 & 6725$\pm$771 & 4.1$\pm$0.3 & $-0.16\pm$0.27\\
EW & 6388 & 5905$\pm$656 & 4.2$\pm$0.2 & $-0.17\pm$0.32\\
		\hline
SR & 7232 & 3777$\pm$158 & 0.6$\pm$0.3 & $-0.51\pm$0.24\\
L & 371 & 3773$\pm$212 & 0.5$\pm$0.4 & $-0.47\pm$0.24\\
		\hline
ROT & 2748 & 4616$\pm$503 & 4.1$\pm$0.8 & $-0.14\pm$0.28\\
YSO & 27 & 4188$\pm$497 & 4.0$\pm$0.9 & $-0.21\pm$0.25\\
		\hline
RRAB & 1086 & 6477$\pm$297 & 4.2$\pm$0.2 & $-1.01\pm$0.37\\
RRC & 161 & 6960$\pm$554 & 4.2$\pm$0.3 & $-0.33\pm$0.60\\
		\hline
VAR & 17 & 5811$\pm$1262 & 4.1$\pm$1.8 & $-0.90\pm$0.63\\
		\hline
\end{tabular}
\end{table}

\clearpage
\clearpage

\section{Discussion}

Here we discuss eclipsing binaries, rotational variables and semi-regular variables using both the photometric information from ASAS-SN and other surveys (including \textit{Gaia}, \textit{WISE}, 2MASS and \textit{GALEX}) and spectroscopic information from the cross-matching in $\S3$ in more detail. In Section $\S4.1$, we study the temperature and metallicity dependences of the periods of eclipsing binaries. In Section $\S4.2$, we examine our catalog of rotational variables and particularly the rapidly rotating evolved stars. In Section $\S4.3$, we discuss the semi-regular variables and their chemical properties using the measurements from APOGEE DR16.

\subsection{Eclipsing Binaries}
Eclipsing binaries are useful astrophysical tools that can be used to measure the masses and radii of stars across the Hertzsprung-Russell diagram (see \citealt{2010A&ARv..18...67T}, and references therein). We classified ${\sim}136,000$ eclipsing binaries in our catalog into the VSX types: EW (W UMa), EB ($\beta$-Lyrae) and EA (Algol). These classifications were made using a random forest classifier with features derived from light curve characteristics, including Fourier parameters (see \citealt{Jayasinghe2019a}). EW binaries have light curves with similar primary/secondary eclipse depths whereas EB binaries tend to have eclipses with significantly different depths. Both the EW and EB binaries transition smoothly from the eclipse to the out-of-eclipse state. EA (Algol) binaries are systems where the exact onset and end of the eclipses are easily defined. EA binaries may or may not have a secondary minimum.

\begin{figure*}
	\includegraphics[width=\textwidth]{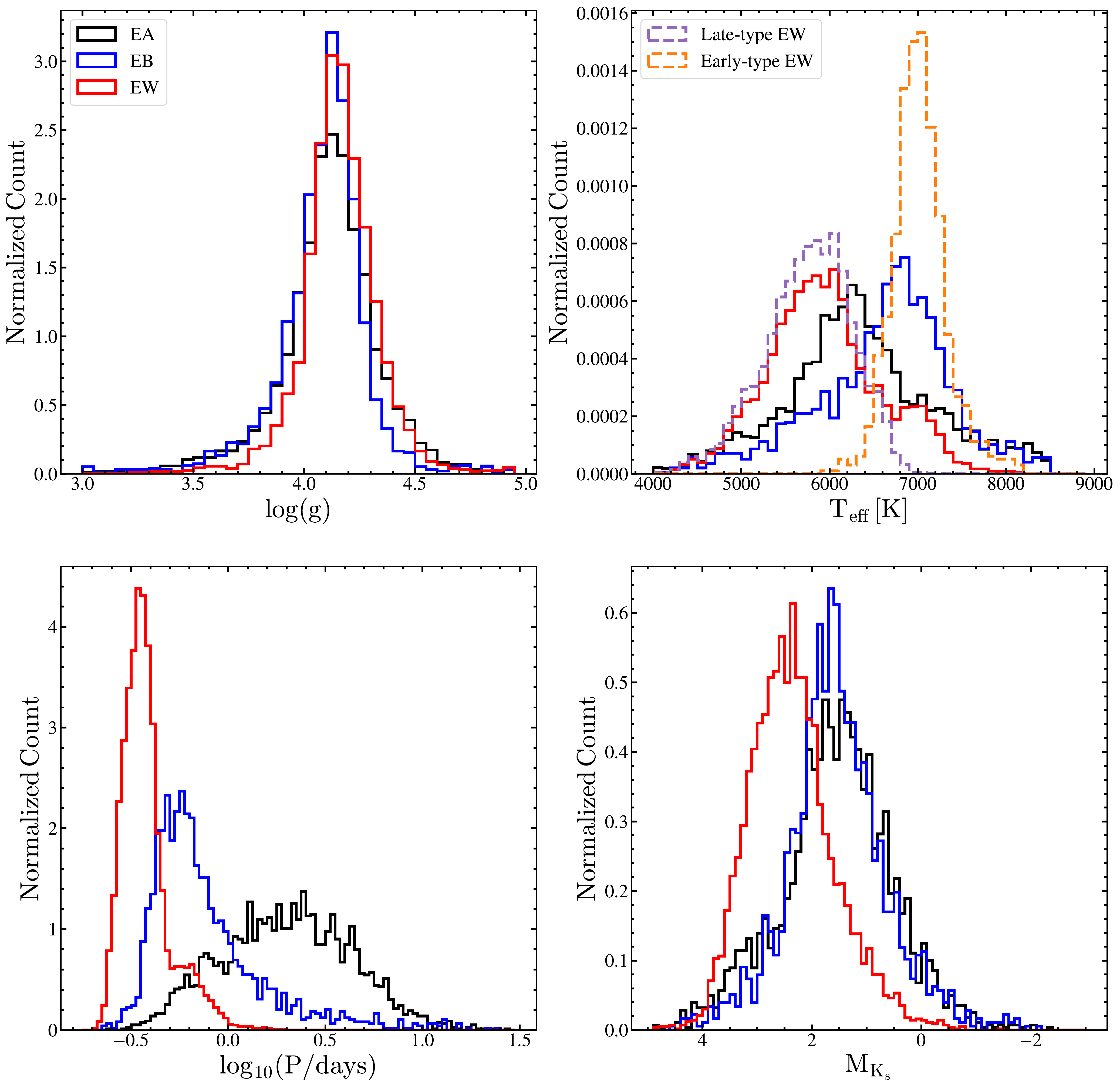}
    \caption{Distributions of the EW (red), EB (blue) and EA (black) binaries in $\rm \log (g)$, $\rm T_{eff}$, $\rm \log_{10}(P/days)$ and $\rm M_{Ks}$. The distributions in $\rm T_{eff}$ for the early (orange) and late-type (purple) EW binaries from \citet{Jayasinghe2020a} are shown as the dashed histograms. }
    \label{fig:fig8}
\end{figure*}

Figure \ref{fig:fig8} shows the distributions of the eclipsing binaries in $\rm \log (g)$, $\rm T_{eff}$, $\rm \log_{10}(P/days)$ and $\rm M_{Ks}$. The surface gravity distributions of the three sub-types are very similar. The biggest differences between the different sub-types are in their effective temperatures. On average, EW type contact binaries are significantly cooler ($\rm T_{eff}{\sim}5900$ K) than both the EB ($\rm T_{eff}{\sim}6700$ K) and the EA systems ($\rm T_{eff}{\sim}6300$ K). EW binaries peak at $\rm \log_{10}(P/days){\sim}-0.5$ and drop sharply at both longer and shorter periods. There are very few EW systems with $\rm \log_{10}(P/days)>0$. EB systems peak at $\rm \log_{10}(P/days){\sim}-0.8$ and span a larger range in period. The detached EA systems peak at $\rm \log_{10}(P/days){\sim}0.4$ and are more evenly distributed in their orbital periods than the EB and EW systems. EW systems are fainter, with a peak at $\rm M_{Ks}{\sim}2.4$ mag, whereas the EB and EA systems peak at a similar $\rm M_{Ks}{\sim}1.6$ mag.

In Paper VII \citep{Jayasinghe2020a}, we analyzed a sample of ${\sim}71,000$ EW systems, and noted a clear dichotomy between the early and late-type EW systems. We found that the period distribution had a clear minimum at $\rm \log (\rm P/d) =-0.30$ which also corresponded to a break in the slope of the period-luminosity relation. The distinction between the populations was even clearer in the space of period and effective temperature, with a gap along the line $\rm T_{eff}=6710K-1760K\,\log(P/0.5\,d)$. The median temperature of the early-type contact binaries ($\rm T_{eff}{\sim}6900$ K) was significantly hotter than the late-type contact binaries ($\rm T_{eff}{\sim}5800$ K). We further noted that the Kraft break \citep{1967ApJ...150..551K} appeared to determine the observed dichotomy of the contact binaries. Early-type systems form due to stellar evolution and the subsequent expansion of a more massive component that is above the Kraft break (${\sim}1.3 M_{\odot}$). In contrast, the less massive late-type systems can come into contact due to efficient angular momentum loss during the detached phase \citep{2014MNRAS.437..185Y}. 

\begin{figure*}
	\includegraphics[width=\textwidth]{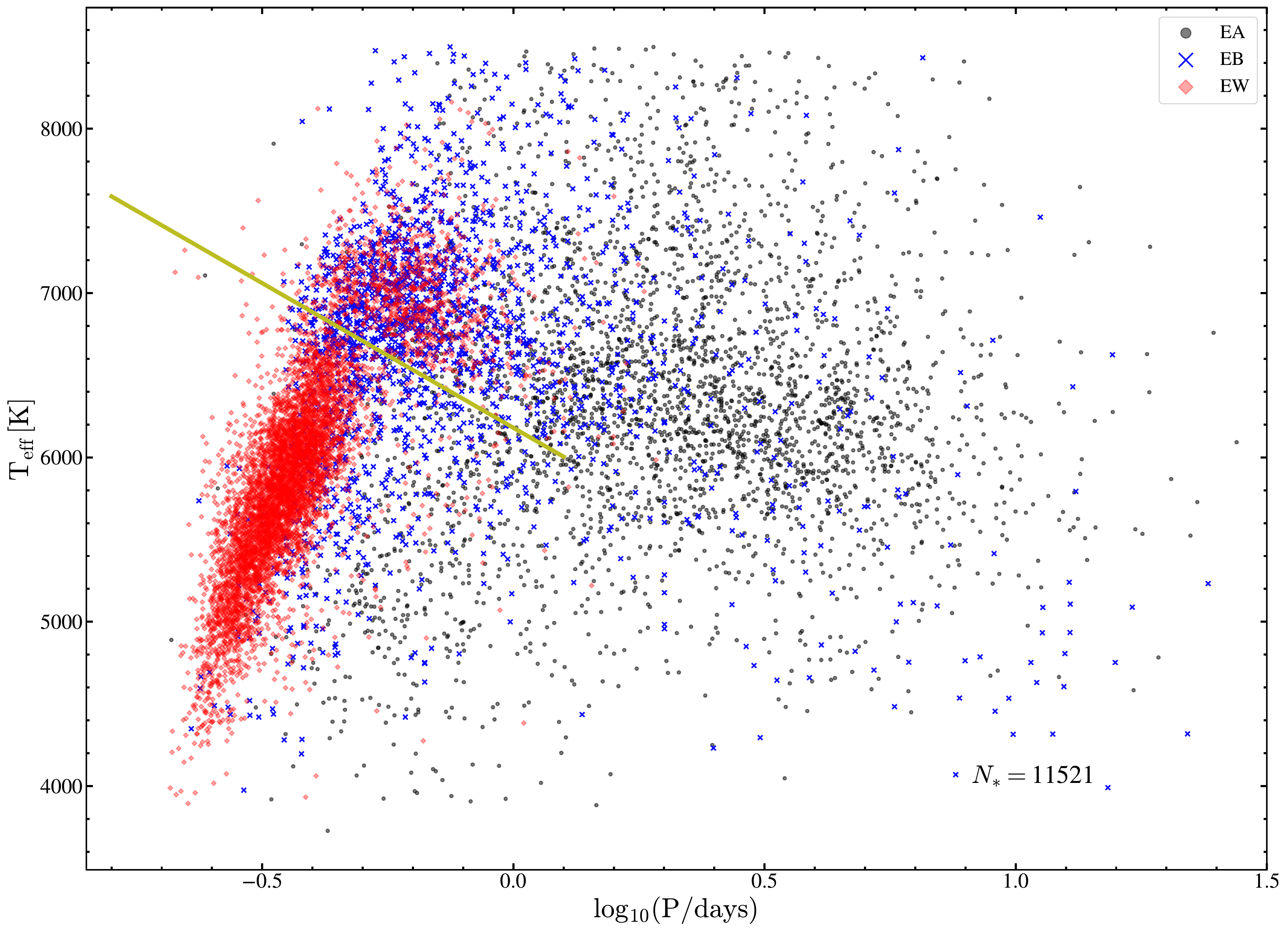}
    \caption{$\rm T_{eff}$ vs. $\rm \log (\rm P/d)$ for EW (red), EB (blue) and EA (black) binaries. The yellow line in $\rm T_{eff}$ and $\rm \log (\rm P/d)$ that separates early and late-type EW binaries is derived in \citet{Jayasinghe2020a}.}
    \label{fig:fig9}
\end{figure*}

Figure \ref{fig:fig9} shows the different sub-types of eclipsing binaries in the space of $\rm \log (\rm P/d)$ and $\rm T_{eff}$. There is significant overlap in period-temperature space between the EW and EB binaries, with most EB binaries falling above the cut defined in \citet{Jayasinghe2020a} for the early-type EW systems. The overlap between the EW and EB binaries is most significant in the period---temperature space above this cut. The EA binaries are distributed randomly in this space. The distribution of EB binaries peaks at a similar temperature to the early-type EW binaries ($\rm T_{eff}{\sim}6900$ K, see Figure \ref{fig:fig8}), which suggests that the early-type EW are drawn from the population of EB binaries. However, these two populations differ in the ratio between the primary and secondary light curve minima. The degree of thermal equilibrium between the two stars dictates the difference in eclipse depths \citep{2006MNRAS.368.1311P}, and the very similar eclipses seen in EW binaries suggest that these stars are in better thermal contact than the EB binaries that have different eclipse depths. 

In our previous work, we hypothesized that the early-type EW systems have a massive component above the Kraft break (${\gtrsim}1.3 M_{\odot}$) that is in thermal contact with the secondary. The EB systems are likely to have similar massive components, but the two components in these systems diverge significantly from thermal contact, unlike those in the late-type EW systems. Models of thermal relaxation oscillations (TRO, see, for e.g., \citealt{1976ApJ...205..208L,1976ApJ...205..217F,2005ApJ...629.1055Y}) predict a population of EB eclipsing binaries with unequal minima that overlap in period and temperature with the EW systems \citep{2003ASPC..293...76W}. In the TRO cycle, an eclipsing binary oscillates between the contact and semi-detached states. The semi-detached phase spans a period and temperature range that is comparable to the contact phase, and the two components develop different effective temperatures during the EB stage of the cycle \citep{2003ASPC..293...76W}. The overlap between the early-type EW and the EB systems appear to be consistent with the predictions of the TRO models for contact binaries.

\begin{figure*}
	\includegraphics[width=\textwidth]{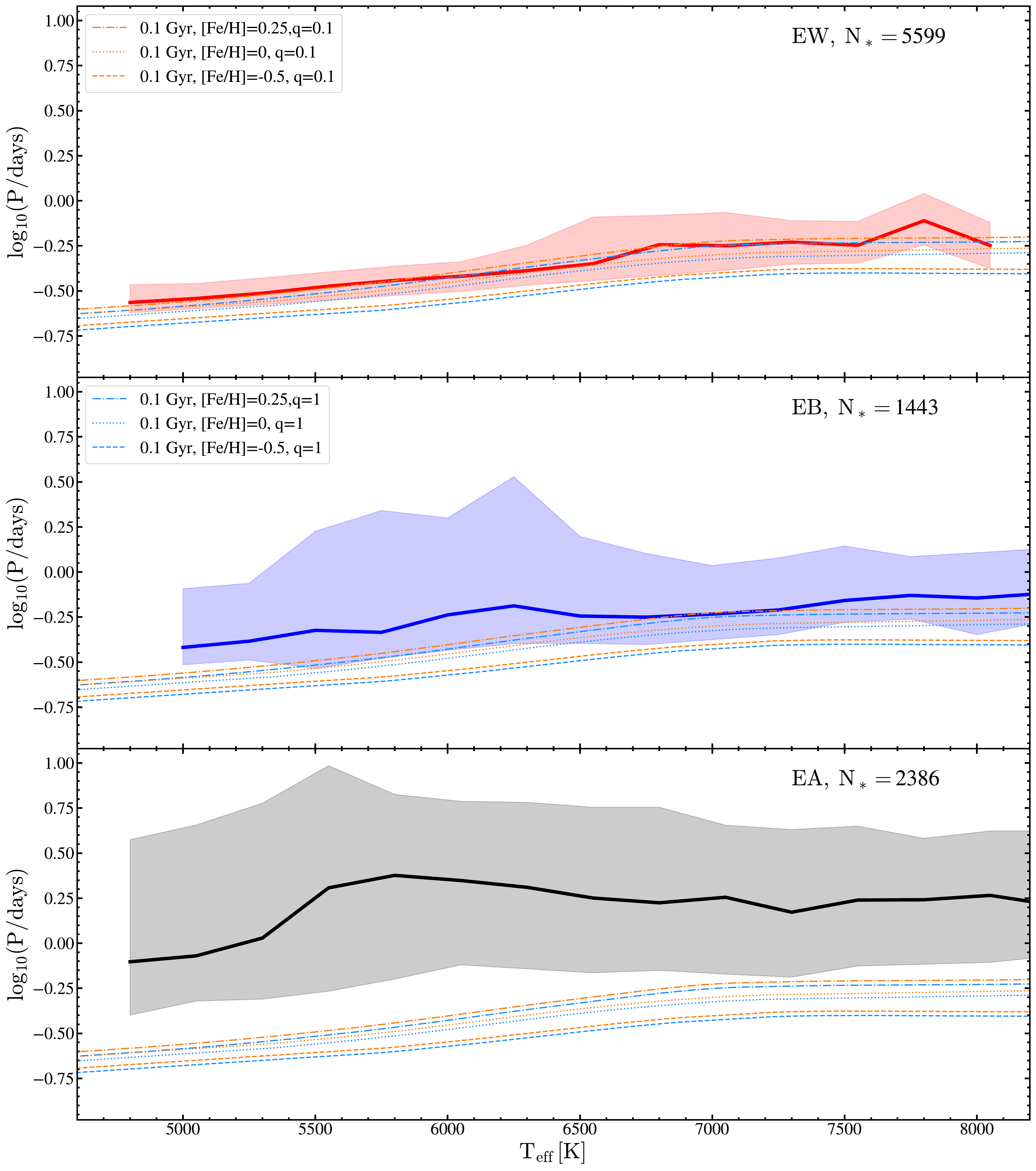}
    \caption{The median $\rm \log (\rm P/d)$ as a function of $\rm T_{eff}$ distributions for EW (top), EB (middle) and EA (bottom) binaries. The shaded regions correspond to the $5\%$ to $95\%$ ranges of the periods. The predicted period-temperature relationships for MS Roche contact binaries with mass ratios $q=1$ (light blue) and $q=0.1$ (orange) is derived using the MIST isochrones \citep{2016ApJ...823..102C,2016ApJS..222....8D} and are shown for metallicities of $\rm [Fe/H]=-0.50$ (dashed), $\rm [Fe/H]=0$ (dotted) and $\rm [Fe/H]=0.25$ (dot dashed) for a $10^8$ yr old stellar population.}
    \label{fig:fig10}
\end{figure*}

The orbital period at which a star overflows its Roche Lobe in a binary consisting of two equal mass stars with a mass ratio of $q=1$ is approximately \citep{1983ApJ...268..368E} \begin{equation}
     { P \over \rm days} \simeq 0.351 \left( { R_* \over R_\odot } \right)^{{-3}/2} \left( { M_* \over M_\odot } \right)^{-1/2}.
         \label{roche}
\end{equation} We used the MIST single star isochrones \citep{2016ApJ...823..102C,2016ApJS..222....8D} to predict the period-temperature relationship at metallicities of $\rm [Fe/H]=-0.5, \,0$ and $\,0.25$ for a MS population with an age of $10^8$ yr. The minimum period for one of the stars to be in Roche contact is weakly dependent on the mass-ratio $q$ (see \citealt{1983ApJ...268..368E}). We illustrate this for a mass ratio of $q=0.1$. The predicted minimum period-temperature relationship depends more strongly on the metallicity--- binaries at lower metallicities can have a shorter minimum orbital period at fixed temperature. This relationship is steep for binaries with $\rm T_{eff}\lesssim7000\, K$ and flattens at higher temperatures. Figure \ref{fig:fig10} shows the median and $5\%$ to $95\%$ ranges for the periods of eclipsing binaries as a function of temperature. We selected sources with $\log(g)>4$ to restrict the sample to MS binaries. As one would expect, contact EW binaries at fixed $\rm T_{eff}$ have shorter periods than the EB and EA binaries. We see that the lower edge of the EW distribution hugs the Roche limit as expected.

\begin{figure*}
	\includegraphics[width=\textwidth]{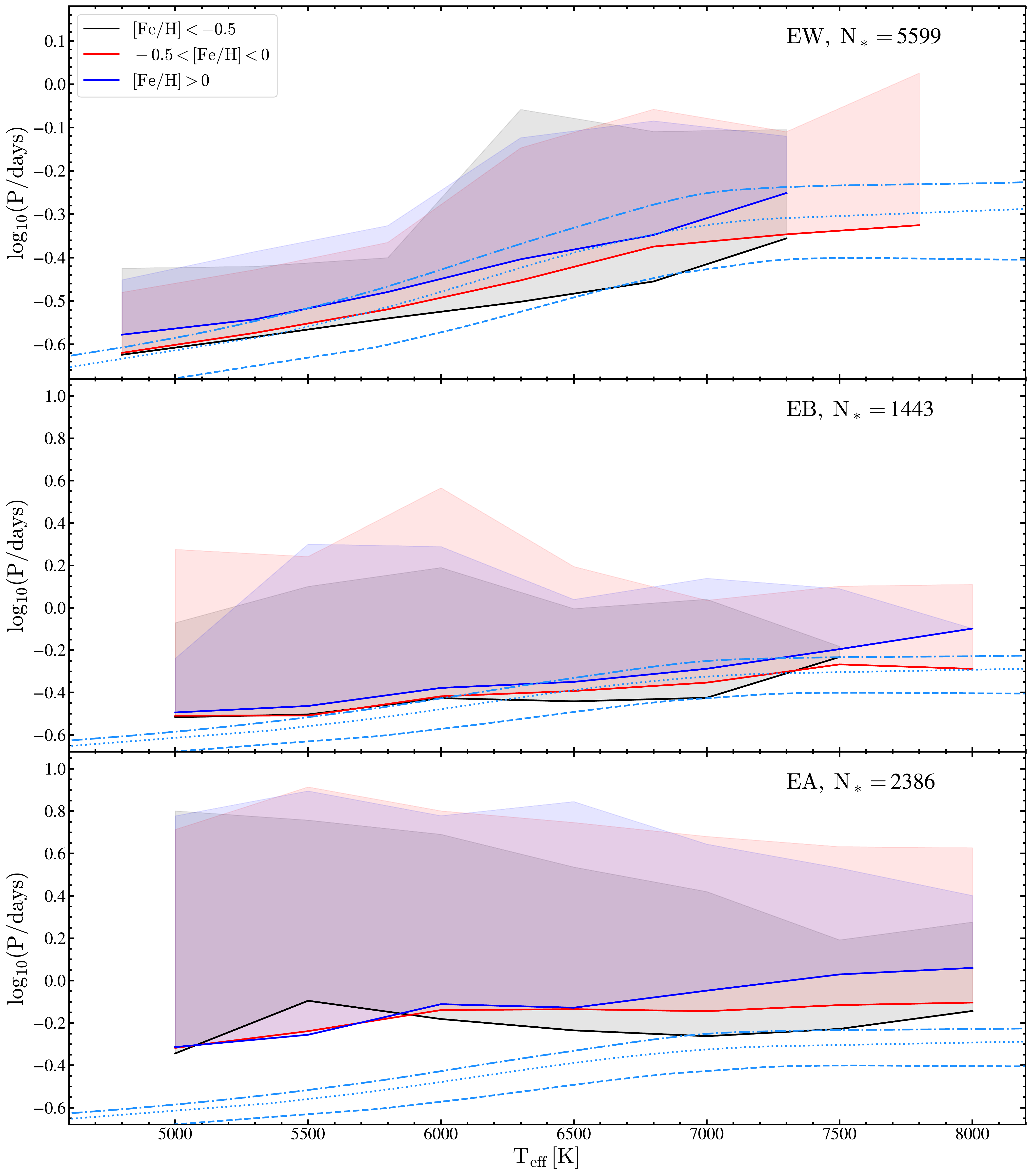}
    \caption{The distribution of the $5^{\rm th}$ percentile in $\rm \log (\rm P/d)$ vs $\rm T_{eff}$ for EW (top), EB (middle) and EA (bottom) binaries in metallicity bins of $\rm [Fe/H]<-0.5$ (black), $\rm-0.5<[Fe/H]<0$ (red) and $\rm [Fe/H]>0$ (blue). The shaded regions correspond to the $5\%$ to $95\%$ ranges of the periods. The predicted period-temperature relationships (light blue) for equal mass MS binaries overflowing their Roche Lobes is derived using the MIST isochrones \citep{2016ApJ...823..102C,2016ApJS..222....8D} and are shown for metallicities of $\rm [Fe/H]=-0.50$ (dashed), $\rm [Fe/H]=0$ (dotted) and $\rm [Fe/H]=0.25$ (dot dashed) for a $10^8$ yr old stellar population.}
    \label{fig:fig11}
\end{figure*}

In the MIST model shown in Figure \ref{fig:fig10}, we see that the low metallicity binaries can be more compact before reaching the Roche limit. In Figure \ref{fig:fig11}, we divide each class of eclipsing binaries into metallicity bins of $\rm [Fe/H]<-0.5$, $\rm-0.5<[Fe/H]<0$ and $\rm [Fe/H]>0$, and show the period ranges as a function of temperature for each bin. Most (${\sim}63\%$) eclipsing binaries have metallicities in the range $\rm-0.5<[Fe/H]<0$. Similarly, \citet{2019MNRAS.487.5932P} found that the ASAS-SN eclipsing binaries had a median $\rm[Fe/H]$ 0.2 dex lower than the entire APOGEE DR14 sample. In general, we see that the lower metallicity binaries have shorter periods than the binaries that are metal-rich at fixed temperature.

\clearpage
\clearpage
\subsection{Rotational Variables}

The ${\sim}33,000$ Rotational (ROT) variables in our V-band catalog are drawn from a variety of rotational variable types, including $\alpha^2$ Canum Venaticorum variables (ACV), RS Canum Venaticorum-type (RS) binary systems, BY Draconis-type variables (BY), FK Comae Berenices-type variables (FKCOM), rotating ellipsoidal variables (ELL) and spotted T Tauri stars showing periodic variability (TTS/ROT). Rotational variables are distributed across the Hertzsprung-Russell diagram, but the detectability of a rotational signal largely depends on their evolutionary state. \citet{2017A&A...605A.111C} studied ${\sim}17,400$ \textit{Kepler} red giants and noted that ${\sim}2\%$ of these sources had a detectable rotational signal in their light curves. They also studied ${\sim}600$ red clump stars in their sample and found that ${\sim}15\%$ of these sources had a detectable rotational signal. In contrast, \citet{2014ApJS..211...24M} detected periodic rotational signals in the \textit{Kepler} light curves of ${\sim}25.6\%$ of 133,030 main-sequence stars. Rotational variables found by ASAS-SN must be higher amplitude than the typical examples found by \textit{Kepler}, favoring spotted stars with large spot covering fractions and/or asymmetric spot coverage \citep{2019ApJ...879..114I}. 

\begin{figure*}
	\includegraphics[width=0.8\textwidth]{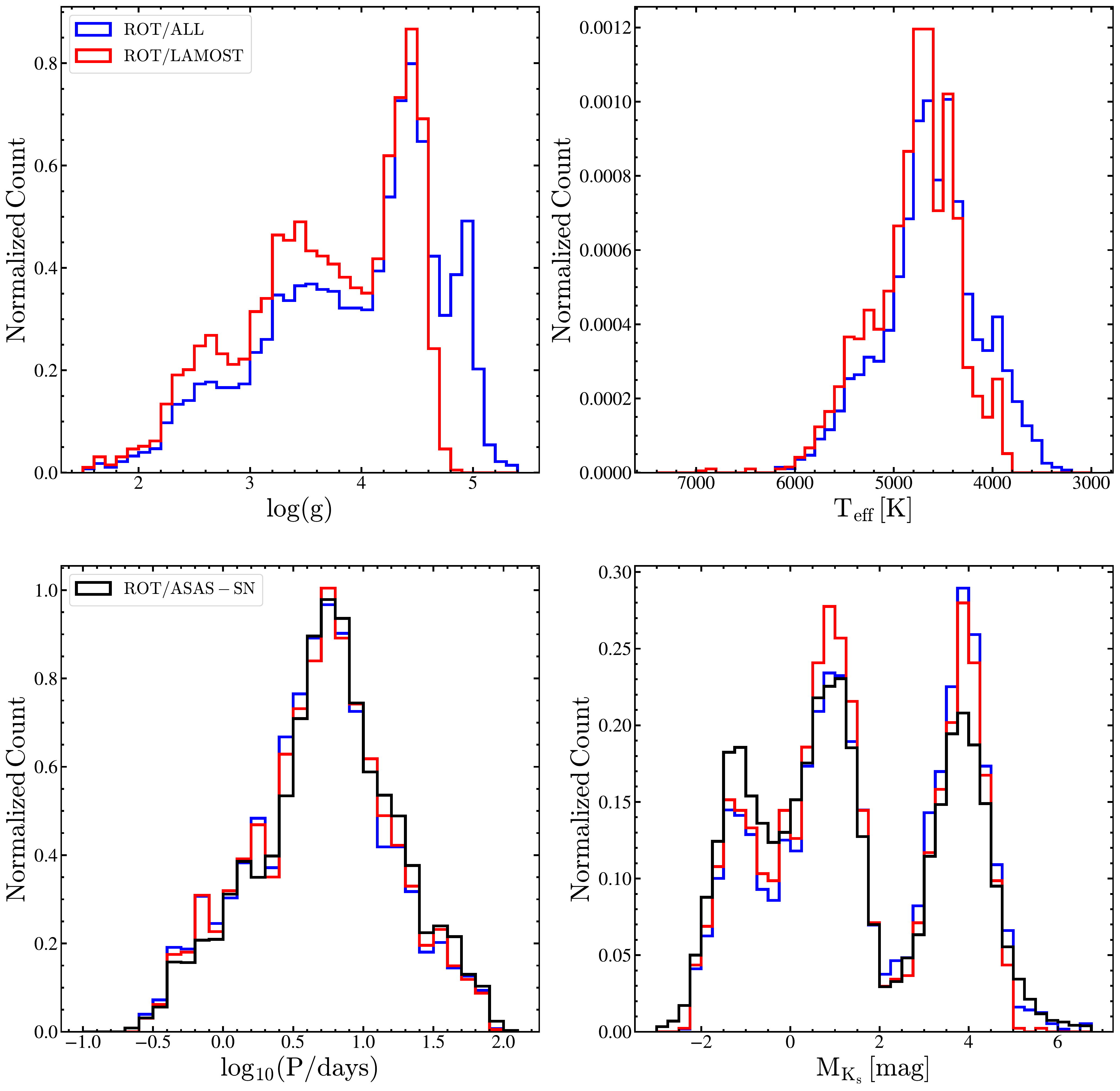}
    \caption{Distributions of the rotational variables for the complete spectroscopic sample (blue) and the LAMOST sample (red) in $\rm \log (g)$, $\rm T_{eff}$, $\rm \log_{10}(P/days)$ and $\rm M_{Ks}$. The $\rm \log_{10}(P/days)$ and $\rm M_{Ks}$ distributions of the rotational variables in the entire ASAS-SN sample are shown in black.}
    \label{fig:fig12}
\end{figure*}

Figure \ref{fig:fig12} shows the distributions of the ROT variables in $\rm \log (g)$, $\rm T_{eff}$, $\rm \log_{10}(P/days)$ and $\rm M_{Ks}$. The rotational variables in LAMOST are shown separately, as most spectroscopic cross-matches to the rotational variables come from this catalog. The LAMOST catalog is incomplete for the M-dwarf rotational variables with $\rm \log (g)>4.8$, $\rm T_{eff}<3800$ K. As expected, there are three distinct classes of rotational variables in the $\rm \log (g)$ and $\rm M_{Ks}$ distributions. These correspond to sources on the MS/pre-MS ($\rm \log (g){\sim}4.5$, $\rm M_{Ks}{\sim}4$), sub-giants/giants toward the base of the RGB ($\rm \log (g){\sim}3.5$, $\rm M_{Ks}{\sim}1$) and the red clump ($\rm \log (g){\sim}2.6$, $\rm M_{Ks}{\sim}-1.5$). These three classes of rotational variables have been characterized in previous surveys (see for e.g., \citealt{2013ApJ...775L..11M,2014ApJS..211...24M,2017A&A...605A.111C}). The color-magnitude diagram shown in Figure \ref{fig:fig3} suggests that sources with $\rm M_{Ks}{\gtrsim}2$ mag are likely to be sub-giants or giants at the base of the RGB. Most of these stars at the base of the RGB will have a sub-giant luminosity class. Red clump (RC) stars have an absolute magnitude of $\rm M_{Ks}=-1.61\pm0.01$ mag \citep{2017MNRAS.471..722H}, which is consistent with the peak in the $\rm M_{Ks}$ distribution.

\begin{figure*}
	\includegraphics[width=\textwidth]{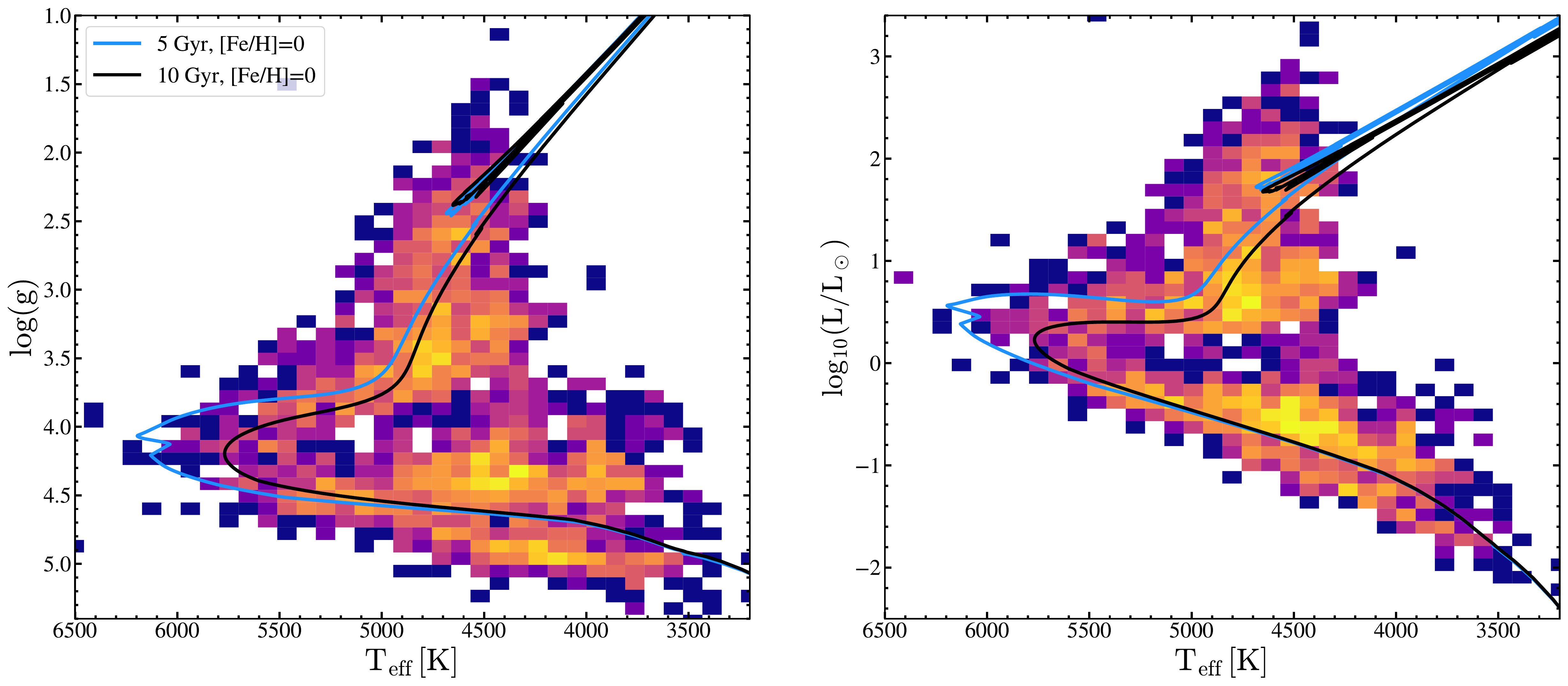}
    \caption{2-D histograms of the rotation variables in $\rm \log(g)$ (left) and luminosity (right) as a function of $\rm T_{eff}$. The bins are colored by the number density. Solar metallicity MIST isochrones \citep{2016ApJ...823..102C,2016ApJS..222....8D} for single stars at 5 Gyr and 10 Gyr are shown for comparison. }
    \label{fig:fig13}
\end{figure*}

Figure \ref{fig:fig13} shows the distribution of rotational variables in $\rm \log(g)$ and luminosity as a function of $\rm T_{eff}$. We use the calibrations from \citet{2010A&ARv..18...67T} to derive the radii of the rotational variables using $\rm T_{eff}$, $\rm \log (g)$ and $\rm [Fe/H]$. The \citet{2010A&ARv..18...67T} scalings agree with asteroseismic estimates \citep{2018ApJS..236...42Y}.  The luminosities are then derived using the radius and $\rm T_{eff}$. We find that the rotational variables are largely distributed along the MS and the RGB, with most of the evolved rotational variables located towards the base of the RGB at $\rm \log(g){\sim} 3.5$. Rotational variables are also located on the sub-giant branch (SGB). Sources on the red clump are also seen in both of these spaces. There are also many sources above the MS, which are likely spotted stars in binary systems or pre-MS spotted T Tauri stars (TTS/ROT). Of the rotational variables with spectroscopic information, ${\sim}53\%$ (${\sim}47\%$) had $\log(g)>4$ ($\log(g)<4$).

\begin{figure*}
	\includegraphics[width=\textwidth]{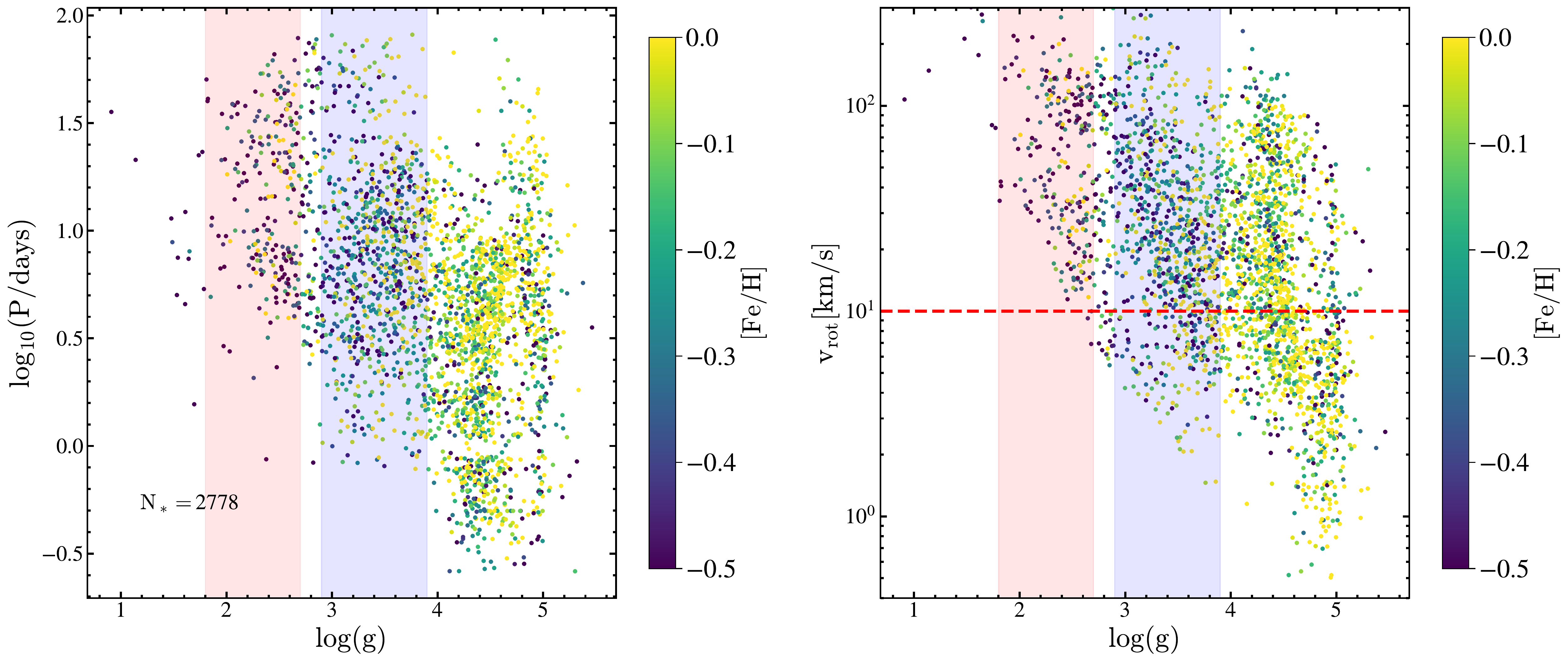}
    \caption{$\rm \log_{10}(\rm P/days)$ (left) and $v_{\rm rot}$ (right) as a function of $\rm \log(g)$  for the rotational variables. The points are colored by metallicity. Regions corresponding to the red clump ($\rm 1.8 \lesssim \log (g){\lesssim}2.8$) and the RGB ($\rm 2.9 \lesssim \log (g){\lesssim}3.9$) are shaded in red and blue, respectively. The red dashed line corresponds to the division between fast ($v_{\rm rot}>10$ km/s) and slow rotating giants.}
    \label{fig:fig14}
\end{figure*}

The distribution of the rotational variables in $\rm \log_{10}(P/days)$ or $v_{\rm rot}$ and $\rm \log (g)$ are shown in Figure \ref{fig:fig14}. We estimate the rotational velocities as \begin{equation}
      v_{\rm rot} = {2\pi R \over \rm P_{phot}},
\end{equation} where  $R$ is estimated using the \citet{2010A&ARv..18...67T} calibrations. Regions corresponding to the red clump ($\rm 1.8 {\lesssim} \log (g){\lesssim}2.8$) and the SGB/RGB ($\rm 2.9 {\lesssim} \log (g){\lesssim}3.9$) are shaded in red and blue, respectively. The stars in the SGB/RGB (RC) region have a median radius and luminosity of $R{\sim}3.4 \,R_\odot$ ($R{\sim}13.5 \,R_\odot$) and $\rm L{\sim}5.3 \,L_\odot$ ($\rm L{\sim}73.3 \,L_\odot$) respectively. SGB/RGB stars have shorter rotational periods than sources in the red clump, with ${\sim}62\%$ (${\sim}42\%$) of the RGB (RC) rotators having periods $\rm P_{rot}<10$ d. While most of the spectroscopic cross-matches to the rotational variables come from the LAMOST catalog, LAMOST has not released $v\sin(i)$ measurements. However, 78 evolved stars on the SGB/RGB had spectroscopic measurements of $v\sin(i)$ from APOGEE (${\sim}86\%$), or GALAH (${\sim}14\%$) (see Figure \ref{fig:fig7}). The majority (${\sim}81\%$) of these stars were fast rotators with $v\sin(i)>10 \rm \, km/s$. We find that ${\sim}80\%$ of the rotational variables on the SGB/RGB have $v_{\rm rot}>10 \rm \, km/s$, consistent with the rapid rotators from the sample with spectroscopic $v\sin(i)$. In contrast, ${\sim}98\%$ of the rotational variables on the red clump have $v_{\rm rot}>10 \rm \, km/s$. 

If we impose a more conservative cut on the red clump sample to minimize contamination from RGB stars by restricting our sample to stars with $\rm 2.3 \lesssim \log (g){\lesssim}2.6$, we find that nearly all of these stars have $v_{\rm rot}>10 \rm \, km/s$, implying that rapid rotation is more common for red clump stars than for stars on the RGB. To investigate the properties of intrinsically brighter RGB stars, we impose a cut on the SGB/RGB sample to only select RGB stars with $\rm 2.8 \lesssim \log (g){\lesssim}3.2$. These stars have a median radius and luminosity of $R{\sim}6.3 \,R_\odot$ and $\rm L{\sim}17.6 \,L_\odot$ respectively. We find that ${\sim}84\%$ of these giants have $v_{\rm rot}>10 \rm \, km/s$, which is comparable to the broader sample of rotational variables on the SGB/RGB.

Of the rapidly rotating red clump stars, ${\sim}30\%$ are metal-poor with $\rm [Fe/H]<-0.5$, whereas only ${\sim}8\%$ are metal-rich with $\rm [Fe/H]>0$. The multiplicity fraction is anti-correlated with metallicity (see for e.g., \citealt{2019ApJ...875...61M}), and the prevalence of metal-poor rapid rotators maybe a consequence of higher binary fractions at lower metallicity. Similarly, \citet{2015ApJ...807...82T} found a large fraction of rapidly-rotating stars on the red clump. This enhancement probably appears because stars are more likely to interact as they expand going up the giant branch, and stop interacting as they descend the giant branch to the red clump. The clump stars then retain the ``memory'' of the interactions allowed by the expansion. The multiplicity fraction at the red clump is also lower than that of the RGB, and is comparable to the multiplicity fraction at the tip of the RGB \citep{2018ApJ...854..147B,2020arXiv200200014P}. The reduced multiplicity fraction could be a sign of companion engulfment and spin up during a common envelope phase \citep{2013A&ARv..21...59I,2020arXiv200200014P}.

\begin{figure*}
	\includegraphics[width=\textwidth]{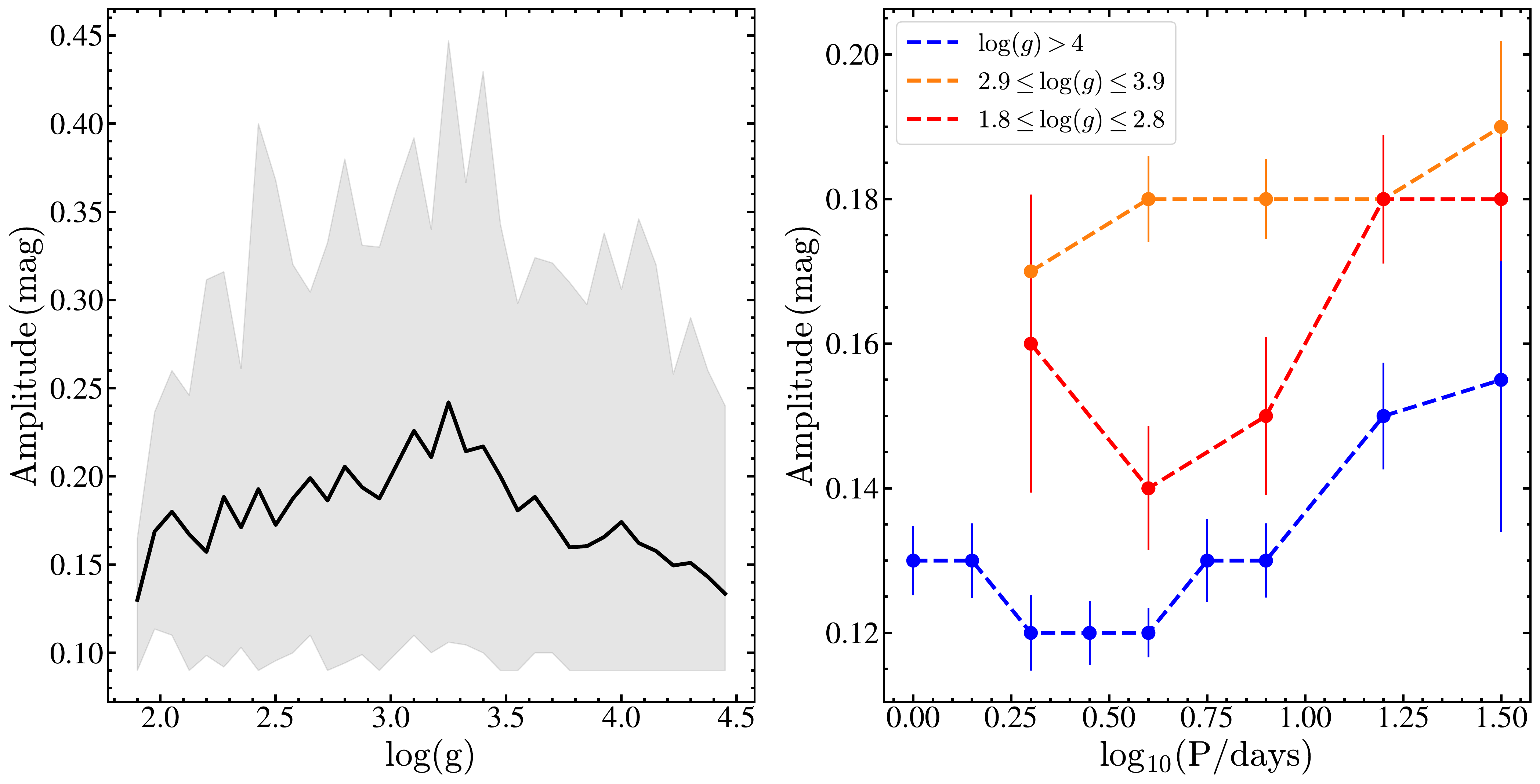}
    \caption{The $V$-band amplitude of the rotational variables in the spectroscopic sample as a function of $\rm \log (g)$ (left) and period (right). The shaded region corresponds to the $5\%$ to $95\%$ range of the amplitudes. The amplitude distributions with period are grouped by $\rm \log (g)$: $\log(g)>4$ (blue), $2.9 \leq \log(g) \leq 3.9$ (orange) and $1.8 \leq \log(g) \leq 2.8$ (red).}
    \label{fig:fig31}
\end{figure*}

\citet{2019ApJ...879..114I} studied 12,660 spotted stars in the Galactic bulge using OGLE data and noted the presence of two distinct groups consisting of rapidly rotating ($\rm P_{rot}<2$ d) stars with low amplitude variability ($A<0.2$ mag in the \textit{I}-band), and slowly rotating stars with large amplitudes (up to $A<0.8$ mag in the \textit{I}-band). The sample with large amplitudes mostly consisted of giants. We investigate the $V$-band variability amplitudes of the ASAS-SN rotational variables with spectroscopic information as a function of $\rm \log (g)$ and period in Figure \ref{fig:fig31}. The variability amplitudes of the rotational variables are dependent on their evolutionary state, with the main sequence stars having the lowest amplitudes. The variability amplitudes peak at $\rm \log (g){\sim}3.2$, and drop off again at lower $\rm \log (g)$. At any given period, the evolved rotational variables have larger amplitudes than stars on the MS. This is consistent with the results from \citet{2019ApJ...879..114I}. The amplitudes of the MS and evolved rotational variables increase at longer periods. \citet{2019ApJ...879..114I} noted similar trends in the OGLE sample. Most high amplitude MS rotators in the OGLE sample had long periods, and the spotted giants with long rotational periods showed larger brightness variations, consistent with the trends seen in Figure \ref{fig:fig31}. The large brightness variations in these stars can be attributed to large spot covering fractions and/or longitudinal asymmetries in the coverage of spots \citep{2019ApJ...879..114I}.

Of the MS rotational variables with $\log(g)>4$, ${\sim}54\%$ have $v_{\rm rot}>10 \rm \, km/s$, compared to ${\sim}83\%$ for evolved stars with $\log(g)<4$. The MS rotators and evolved stars have median periods of $\rm P_{rot}{\sim}3.5$ d and $\rm P_{rot}{\sim}8$ d respectively. A substantial fraction (${\sim}18\%$) of the MS stars have very short periods with $\rm P_{rot}<1$ d (Figure \ref{fig:fig14}). The MS rotators are also more metal-rich ($\rm [Fe/H]{\sim}-0.08$) than the evolved stars ($\rm [Fe/H]{\sim}-0.23$). The median effective temperature for these MS rotators ($\rm T_{eff} {\sim}4460$ K) corresponds to stars with late K spectral types. This is not surprising, as it has been shown that cool MS stars typically have large variability amplitudes and are more likely to show rotational signals than hotter stars \citep{2014ApJS..211...24M}, making their detection by ground-based surveys easier. In particular, \citet{2014ApJS..211...24M} found that ${\sim}69\%$ of the MS stars with $\rm 4000 \leq T_{eff}/K \leq 4500$ K showed periodic rotational variability in their \textit{Kepler} light curves.

\begin{figure*}
	\includegraphics[width=\textwidth]{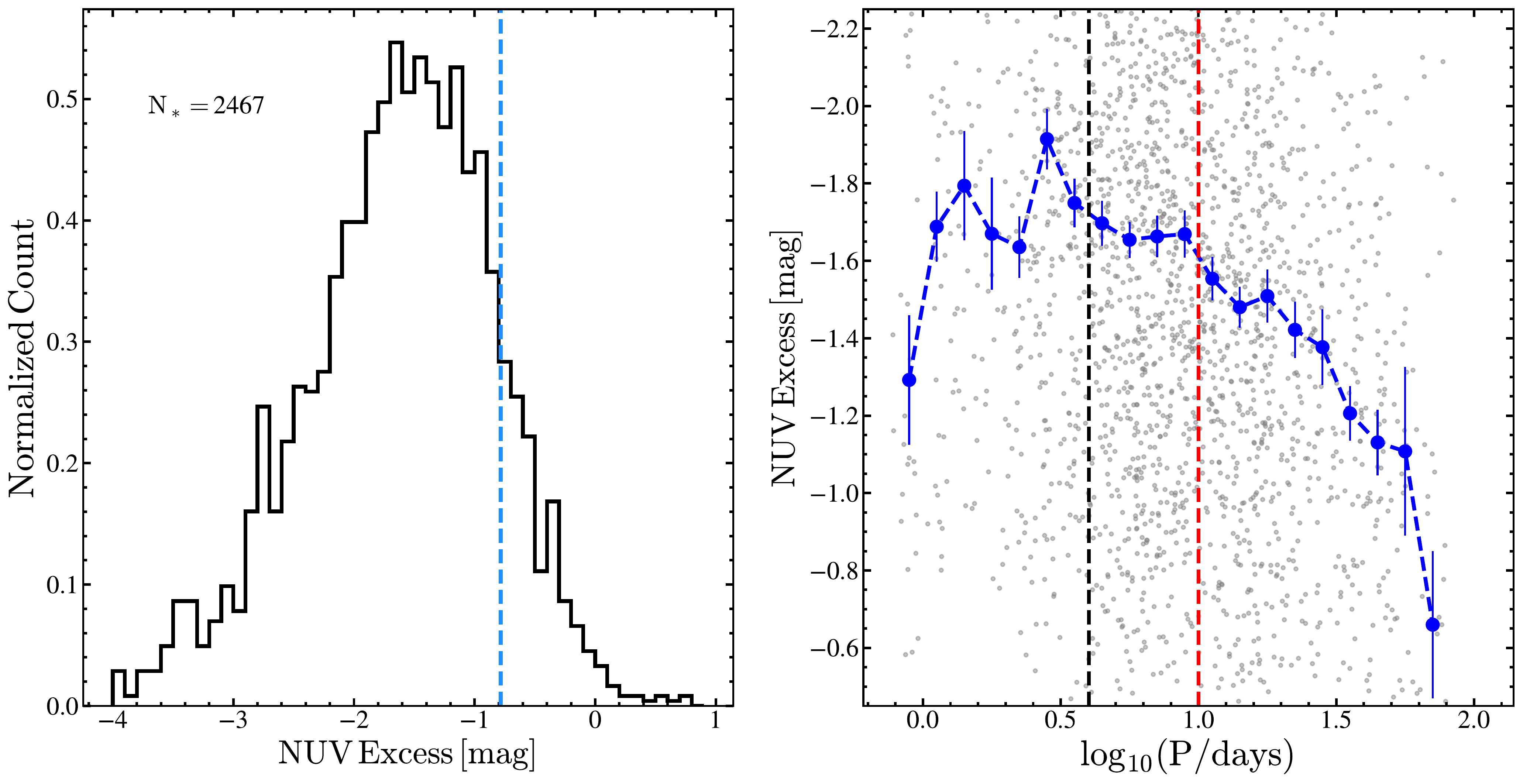}
    \caption{Distributions of the NUV excess (left) and its dependence on period (right) for the rotational variables with $\rm M_{Ks}<2$ mag. The blue dashed line is the expected NUV excess ($-0.78$ mag) for rotating giant stars with $v\sin(i)=10 \rm \, km/s$ from \citet{2020arXiv200500577D}. The red dashed line represents the break in the period-NUV excess relationship at $\rm P_{rot}=10$ days from \citet{2020arXiv200500577D}. The period at which rotating giants reach super-saturation is shown as a black dashed line at $\rm P_{rot}\approx 4$ days.}
    \label{fig:fig15}
\end{figure*}

\citet{2020arXiv200500577D} identified an empirical relationship between the near-UV (NUV) excess estimated by combining $GALEX$+2MASS data and $v\sin(i)$ for evolved stars in the APOGEE survey. Figure \ref{fig:fig15} shows the distribution of ${\sim}2,500$ evolved rotational variables with $\rm M_{Ks}<2$ mag (see Figure \ref{fig:fig3}) for which we could calculate the \citet{2020arXiv200500577D} NUV excess. Of the sources with \textit{GALEX} detections, ${\sim}58\%$ have periods $\rm P_{rot}<10$ d. \citet{2020arXiv200500577D} derived an empirical rotation-activity relation for giants of \begin{equation}
      {\rm NUV \, excess} = (-1.43\pm0.12) \log(v\sin(i)+ (0.647\pm0.131). 
\end{equation} So, giants with $v\sin(i)>10 \rm \, km/s$ will have an NUV excess $<-0.78$ mag. We find that ${\sim}87\%$ of the rotating giants have NUV excesses $<-0.78$ mag, which is consistent with our other estimates of the fraction of sources with rapid rotation ($v\sin(i)>10 \rm \, km/s$). If we restrict the sources to $\rm M_{Ks}<0$ mag, the fraction of sources with NUV excesses $<-0.78$ mag drops to ${\sim}80\%$. 

\citet{2020arXiv200500577D} argued that the dependence of the NUV excess on period could be divided into a saturated and a linear regime in activity with a break at $\rm P_{rot}=10$ d. Our sources follow a similar trend, with the median NUV excess decreasing with increasing period beyond $\rm P_{rot}=10$ d (Figure \ref{fig:fig15}). The median NUV excess is roughly flat between $\rm 4<P_{rot}/d<10$, as also found by \citet{2020arXiv200500577D}. In the super-saturated regime ($\rm P_{rot}\approx 4$ days for typical giants), active giant stars are expected to have decreased activity compared to giants in the saturated regime. We see some evidence for this super-saturated regime in Figure \ref{fig:fig15}.

\begin{figure*}
	\includegraphics[width=0.5\textwidth]{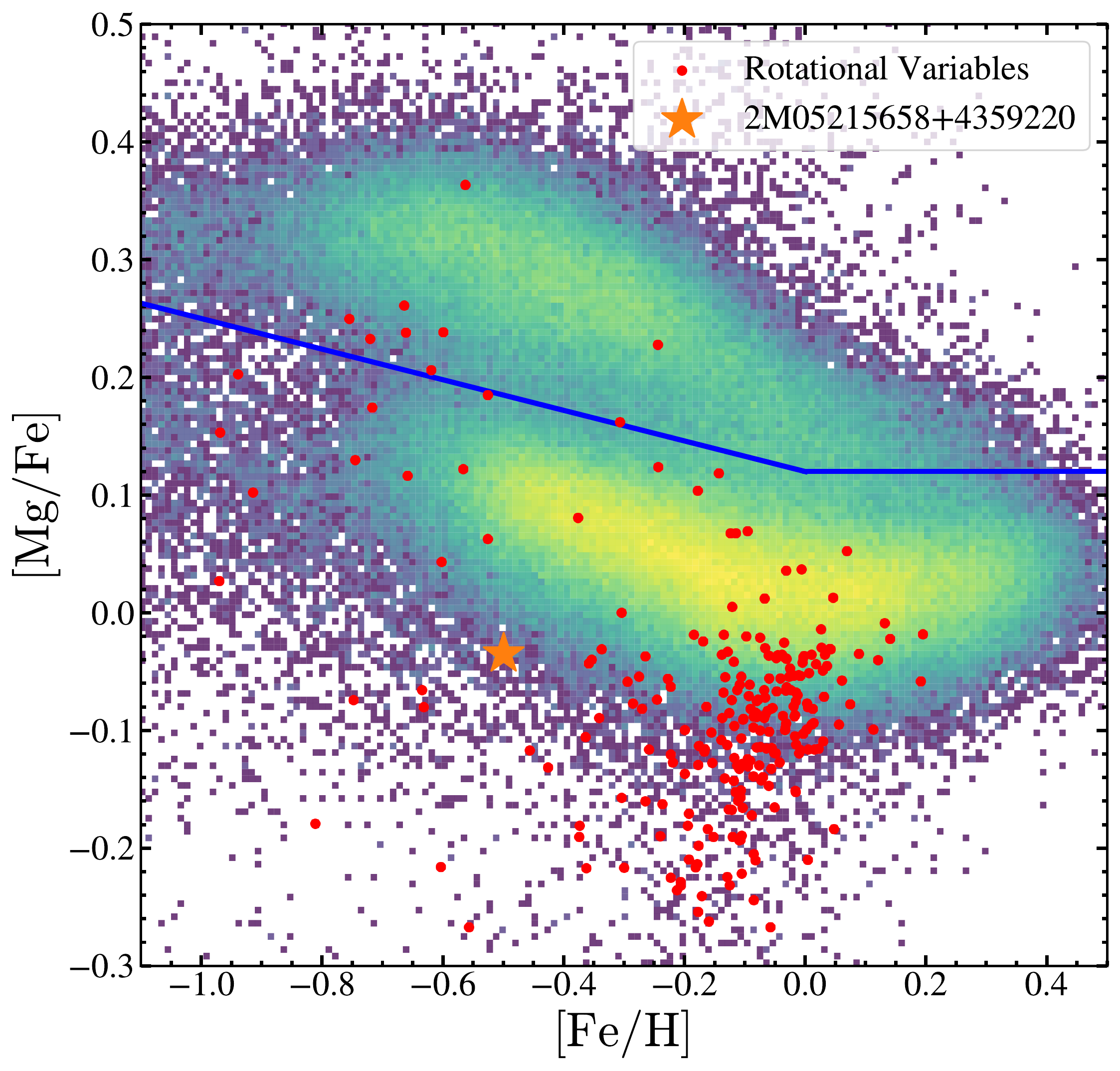}
    \caption{The distribution of all the rotational variables in APOGEE DR16 (red) and a reference sample of giants from APOGEE DR16 (2-d histogram) in the $\rm [Mg/Fe]$ vs. $\rm [Fe/H]$ plane. The blue line shows the division between the low-$\alpha$ and high-$\alpha$ populations from \citet{2019ApJ...874..102W}. The candidate giant star-black hole binary (2M05215658+4359220) from \citet{2019Sci...366..637T} is shown as a orange star.}
    \label{fig:fig16}
\end{figure*}

Figure \ref{fig:fig16} shows the distribution of all the 248 rotational variables in APOGEE with \verb"ASPCAP_FLAG=0" in the $\rm [Mg/Fe]$---$\rm [Fe/H]$ plane as compared to the ${\sim}248,000$ APOGEE DR16 giants with $\log(g)<3.8$ and \verb"ASPCAP_FLAG=0". We also show the division from \citet{2019ApJ...874..102W} between the ``high-$\alpha$'' and ``low-$\alpha$'' sequences. The high-$\alpha$ population is older, kinematically hotter and is located in the thick disk, whereas the low-$\alpha$ population is located in the thin disk \citep{2000AJ....120.2513P,1998A&A...338..161F,2019ApJ...874..102W}. The plateau in the high-$\alpha$ sequence at $\rm [Mg/Fe]{\sim}0.4$ is the average yield of core collapse supernovae and the drop to lower values of $\rm [Mg/Fe]$ is due to later Fe production in type Ia supernovae \citep{2019ApJ...874..102W}. The rotational variables are low-$\alpha$ stars strongly clustered towards $\rm [Mg/Fe]{\sim}-0.1$, and $\rm [Fe/H]{\sim}-0.1$. Their distribution in Figure \ref{fig:fig16} is very different from the typical giant stars or the SR variables and AGB stars we discuss in $\S 4.3.3$. The rotational variables occupy a sparsely populated area in the $\rm [Mg/Fe]$---$\rm [Fe/H]$ plane and this odd clustering of the rotational variables is very likely due to systematics in the ASPCAP abundance pipeline when dealing with rapidly rotating stars. We also highlight 2M05215658+4359220, the candidate giant star-black hole binary identified by \citet{2019Sci...366..637T} in Figure \ref{fig:fig16}. 2M05215658+4359220 is a rapidly rotating giant with $v\sin(i)\approx14.1 \pm 0.6 \rm \, km/s$ and is a clear outlier in the $\rm [Mg/Fe]$---$\rm [Fe/H]$ plane, even compared to the ROT variables. The ASPCAP pipeline includes the flag \verb"ROTATION_WARN" to warn of the presence of broadened lines in the APOGEE spectrum, suggesting that the abundances derived for this object could be suspect. The $\rm [C/N]$ abundance of this giant is unusual for a giant with $\rm M_{giant}\approx3.2\pm1.0 M_\odot$ and is more typical of a lower mass giant with $\rm M_{giant}{\sim}1 M_\odot$ \citep{2019Sci...366..637T}. \citet{2020arXiv200507653T} cautions against the use of the APOGEE $\rm [C/N]$ abundance in claiming a lower mass for the giant based on the systematic uncertainties in the determination of abundances for rapidly rotating giants. The distribution of the ASAS-SN rotational variables in Figure \ref{fig:fig16} supports this argument, as their abundances are unusual and suggestive of systematic issues in the ASPCAP pipeline when dealing with rapidly rotating stars. Indeed, if we examine the individual element abundances ($\rm [Fe/Mg]$, $\rm [C/Mg]$, $\rm [Mn/Mg]$ and $\rm [Ni/Mg]$) of the rotational variables, they are almost all peculiar, and those of 2M05215658+4359220 are more peculiar than most.
 
\clearpage
\clearpage
\subsection{Long Period Variables on the Asymptotic Giant Branch}

Low and intermediate mass stars end their lives after reaching the asymptotic giant branch (AGB). During the AGB phase, stellar evolution is characterized by hydrogen and helium shell burning on top of a degenerate carbon/oxygen core (see for e.g., \citealt{2005ARA&A..43..435H}). AGB stars are cool, luminous objects that chemically enrich the interstellar medium through nucleosynthetic processes and strong mass loss \citep{2005ARA&A..43..435H,2018A&A...616L..13L}. In particular, the thermally pulsing AGB (TP-AGB) stars are characterized by long-period pulsational variability and heavy mass loss rates \citep{2007A&A...469..239M}.

The long-period variables on the TP-AGB are broadly classified into the semi-regular (SR) variables and the Mira variables. Mira variables are luminous AGB stars \citep{1983ARA&A..21..271I} and have large variability amplitudes ($A>2.5$ mag in the V-band is the AAVSO definition). Previous surveys have argued that TP-AGB stars start off as pulsating semi-regular variables with small amplitudes and later evolve into the high amplitude Mira variables as they reach the tip of the AGB \citep{1999IAUS..191..151W,2007A&A...469..239M,2009AcA....59..239S}. Mira variables follow a period-luminosity relationship \citep{1981Natur.291..303G,2008MNRAS.386..313W}, although some Mira variables can undergo changes in their period over time \citep{1999PASP..111...94P}. Compared to the Mira variables, SR variables are less strictly periodic with irregularities in their light curves and multiple periods that can be used to study the dynamics of stellar interiors \citep{1999A&A...346..542K}. Microlensing surveys of the Magellanic clouds have shown that semi-regular variables fall along five distinct period-luminosity sequences (A-E; \citealt{1999IAUS..191..151W,2000PASA...17...18W}). More recent work has shown that the five sequences from \citet{1999IAUS..191..151W} are a result of an overlap of 14 or more period-luminosity sequences \citep{2007AcA....57..201S,2009AcA....59..239S,2017ApJ...847..139T}. In the ASAS-SN pipeline, Mira and semi-regular variables are classified on the basis of their periods, colors (optical and near-infrared) and absolute magnitudes (see \citealt{Jayasinghe2019a}). Semi-regular variables are separated from the Mira variables using their variability amplitudes ($A>2$ mag in the V-band for Miras). We chose a lower limit of $A>2$ mag in amplitude when compared to the AAVSO definition ($A>2.5$ mag) in order to compensate for the effects of blending which can reduce the observed variability amplitudes in ASAS-SN.

We use APOGEE DR16 data \citep{2019arXiv191202905A} to study the spectroscopic and chemical properties of pulsating AGB stars. The abundances come from the APOGEE Stellar Parameters and Chemical Abundances Pipeline (ASPCAP; \citealt{2015AJ....150..173N}). The various subsets of Mira and SR variables discussed in this section are summarized in Table \ref{tab:srsubsets}. For comparison with the semi-regular variables, we selected a set of RGB stars (APOGEE/RGB: $\rm T_{eff} \leq 5200$ K, $2 \leq \rm \log(g) \leq 3.5$, $N{\sim}160,000$) and likely AGB stars (APOGEE/AGB: $\rm T_{eff} \leq 4000$ K, $\rm \log(g) \leq 1$, $N{\sim}15,000$). The sample of candidate AGB stars is likely contaminated with some lower mass stars on the upper RGB, and does not contain any ASAS-SN SR variables. For detailed studies of the APOGEE DR16 abundance ratios, we required the flag \verb"X_FE_FLAG=0", where X= Mg, Al, C, N or O. We calculate the $\rm [X/Mg]$ abundance ratios as $\rm [X/Mg]=[X/H]-[Mg/H]$. Following \citet{2019ApJ...874..102W}, we chose Mg as the reference element because it is produced almost exclusively by core-collapse supernovae and is thus a simpler tracer of chemical enrichment than Fe. 
 
\begin{table*}
	\centering
	\caption{Subsets of Mira and SR variables discussed in $\S$4.3}
	\label{tab:srsubsets}
\begin{tabular}{|p{2cm}|p{5cm}| p{5cm}|p{1cm}|p{3cm}|}
		\hline
		Group & Description & Criteria & Count & Sections \\
		\hline
ALLSPEC/SR & Spectroscopic cross-matches to ASAS-SN SR variables & $\rm Prob>0.95$ and parallaxes better than $50\%$ & 5,390 & $\S$ 4.3.1\\
		\hline
ASAS-SN/SR  & ASAS-SN SR variables & & &  \\

& & $\rm Prob>0.95$ and parallaxes better than $50\%$ & 96,724 & $\S$ 4.3.1\\
& & $\rm Prob>0.95$, $K_s$-band photometry available & 124,836 & $\S$ 4.3.2\\
		\hline
ASAS-SN/MIRA  & ASAS-SN MIRA variables & & &  \\
& & $\rm Prob>0.95$ and parallaxes better than $50\%$ & 958 & $\S$ 4.3.1\\
& & $\rm Prob>0.95$, $K_s$-band photometry available & 2,737 & $\S$ 4.3.2\\
		\hline
APOGEE/SR  & ASAS-SN SR variables in APOGEE  & & &  \\
& & $\rm Prob>0.95$ and parallaxes better than $50\%$ & 991 & $\S$ 4.3.1\\
& & $\rm Prob>0.95$ & 1,615 & $\S$ 4.3.2, 4.3.3, 4.3.4, 4.3.5\\

		\hline
APOGEE/GIANTS & Giants in APOGEE & $\log(g)<3.8$ & 238,472 & $\S$ 4.3.3\\
		\hline
APOGEE/AGB  & AGB candidates in APOGEE & $\rm T_{eff} \leq 4000$ K, and $\rm \log(g) \leq 1$ & 15,050 & $\S$ 4.3.3, 4.3.4, 4.3.5\\
		\hline
APOGEE/RGB  & RGB candidates in APOGEE & $\rm T_{eff} \leq 5200$ K, and $2 \leq \rm \log(g) \leq 3.5$& 159,042 & $\S$ 4.3.3, 4.3.4, 4.3.5\\
\hline
\end{tabular}
\end{table*}

In $\S4.3.1$, we study the various TP-AGB period-luminosity sequences. We identify an APOGEE temperature---$W_{RP}-W_{JK}$ color index fit to the oxygen-rich AGB stars in $\S4.3.2$. We examine the $\alpha$ enhancements of the SR variables in $\S4.3.3$. In $\S4.3.4$, we study the Aluminum abundances of the semi-regular variables and characterize the pulsation period dependent Al depletion as a result of mass loss and dust production. We look at the Nitrogen abundances of the semi-regular variables in $\S4.3.5$ and identify a sample of likely intermediate-mass AGB stars undergoing hot-bottom burning.

\subsubsection{Period-luminosity sequences}

\begin{figure*}
	\includegraphics[width=0.8\textwidth]{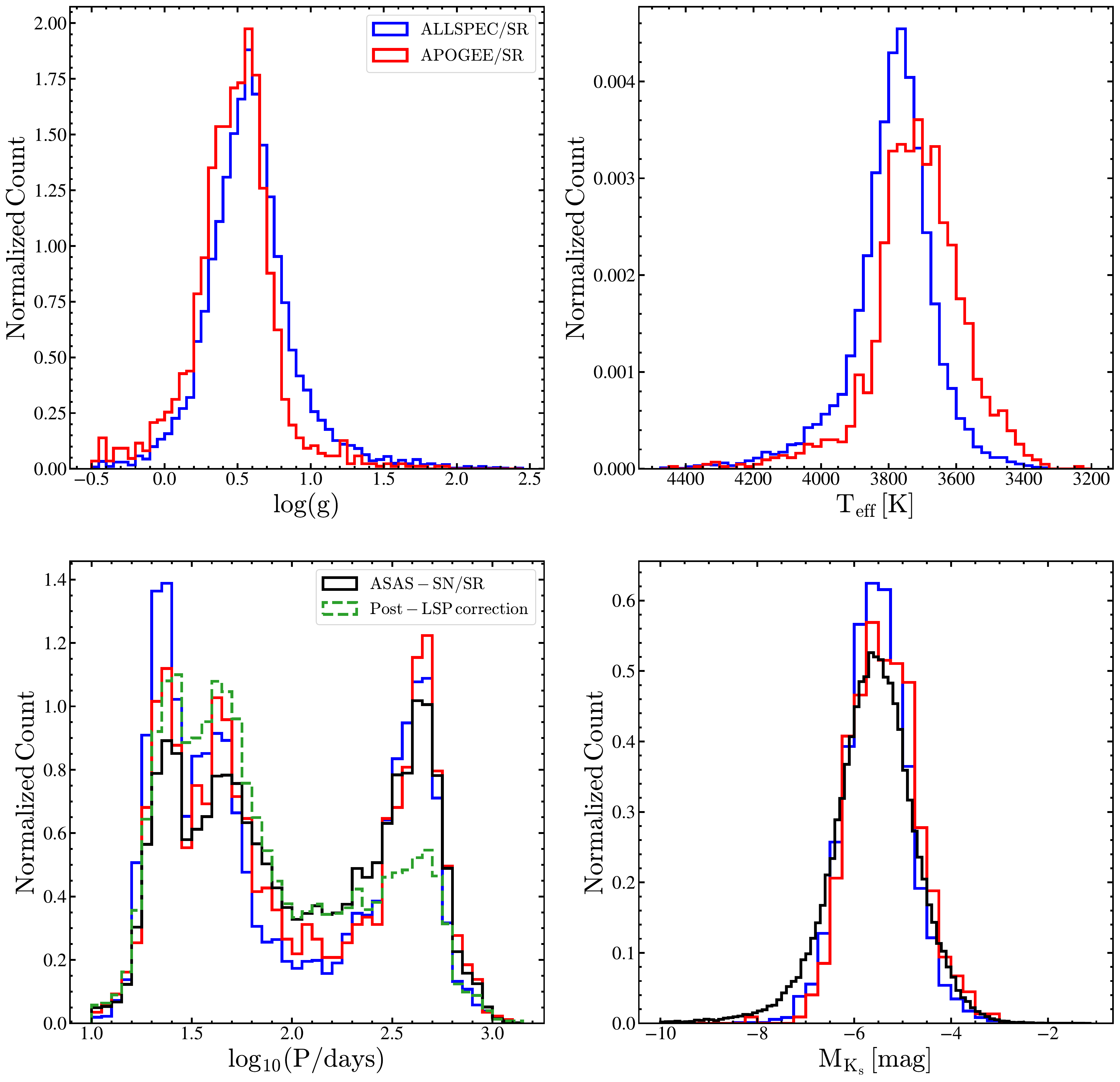}
    \caption{Distributions of the semi-regular variables for the complete spectroscopic sample (blue) and the APOGEE sample (red) in $\rm \log (g)$, $\rm T_{eff}$, $\rm \log_{10}(P/days)$ and $\rm M_{Ks}$. The $\rm \log_{10}(P/days)$ and $\rm M_{Ks}$ distributions of the semi-regular variables in the entire ASAS-SN sample are shown in black. The period distribution of the semi-regular variables after correcting for LSPs is shown in green.}
    
    \label{fig:fig17}
\end{figure*}

 Figure \ref{fig:fig17} shows the distributions in $\rm \log (g)$, $\rm T_{eff}$, $\rm \log_{10}(P/days)$ and $\rm M_{Ks}$ of the SR variables. The SR variables in APOGEE are shown separately. The spectroscopic parameters confirm the highly evolved nature of these stars, with the SR variables having a median surface gravity of $\rm \log (g){\sim}0.6$, and a median effective temperature of $\rm T_{eff}{\sim}3750$ K. The subset of SR variables in APOGEE DR16 are skewed towards cooler, more evolved SR variables when compared to the SR variables in other surveys. We will only consider the SR variables from APOGEE in the discussions below. 
 
 The \textit{Gaia} DR2 parallaxes of these luminous giants are poor, so we cannot directly assign them to the \citet{2000PASA...17...18W}'s A-E period-luminosity relations. However, we can use the periods to roughly group them. There are 3 distinct peaks in the period distribution shown in Figure \ref{fig:fig17} at $\rm \log_{10}(P/days){\sim}1.4,\,1.7,$ and $\,2.7$. SR variables with $1 \lesssim \rm \log_{10}(P/days) \lesssim 1.5$ (hereafter group I) fall on the \citet{2000PASA...17...18W} sequence A and have low variability amplitudes. Due to the overlap between the PLR sequences A and B, SR variables with $1.5 \lesssim \rm \log_{10}(P/days) \lesssim 1.9$ (hereafter group II) consist of sources on both sequences A and B. In addition, SR variables on sequence E also have periods in this range, even though they are less luminous than the variables on sequences A and B. The OGLE small amplitude variable red giants (OSARGs; \citealt{2004AcA....54..129S}) consist of both RGB and AGB stars, and follow a unique set of period-luminosity relations \citep{2007AcA....57..201S}. OSARGs have periods that span group I and have amplitudes as small as ${\sim}3$ mmag \citep{2007AcA....57..201S,2020arXiv200305459A}. Note that we are not considering the much lower variability amplitudes of ASAS-SN OSARGs used by \citet{2020arXiv200305459A} to determine the feasibility of ground-based asteroseismology of luminous giants. There is a paucity of SR variables with periods in the range $1.9 \lesssim \rm \log_{10}(P/days) \lesssim 2.3$ (hereafter group III) which spans the PLR sequence C corresponding to the Mira variables. It is thought that the PLR sequence C corresponds to the fundamental mode of pulsation, while sequences A and B correspond to higher order overtone modes \citep{2005AJ....129..768F}. SR variables on sequence E are less luminous, and are thought to be on the first ascent giant branch instead of the AGB. SR variables on sequence E have periods typical of contact binaries with a giant component \citep{2000PASA...17...18W} and will be included in groups II and III. SR variables with periods $\rm \log_{10}(P/days) \gtrsim 2.3$ (hereafter group IV) include sources on sequences C and D. 
 
A large fraction of the semi-regular variables with reported periods $\rm \log_{10}(P/days) \gtrsim 2.3$ (sequence D) are examples of semi-regular variables with long secondary periods (LSPs) of $\rm P_{LSP}{\sim}300-1500$ d that are ${\sim}10-15$ times longer than their actual pulsation period \citep{2004ApJ...604..800W}. The origin of the long secondary periods is debated \citep{2004ApJ...604..800W,2019MNRAS.487.5932P}. These LSPs can complicate our analysis of the pulsational properties of these variables. Thus, we re-derived the periods of the SR variables using the Generalized Lomb-Scargle (GLS, \citealt{2009A&A...496..577Z,1982ApJ...263..835S}) periodogram implemented in \verb"astrobase" \citep{2018zndo...1469822B}. From the 5 best periods in the periodogram, SR variables with LSPs were identified if there was a significantly shorter period $\rm P$ in the range $8 \leq \rm P_{LSP}/P \leq 30$ relative to the primary period. For SR variables identified as LSPs, we assign the shorter period as the pulsation period. Of the SR variables with $\rm \log_{10}(P/days) \gtrsim 2.3$, ${\sim}36\%$ were LSPs. This estimate of the LSP fraction at these periods agrees with the estimates (${\sim}25-50\%$) from previous studies (see for e.g., \citealt{2004ApJ...604..800W,2007ApJ...660.1486S,2009MNRAS.399.2063N}). Of the LSPs, ${\sim}60\%$ had pulsational periods in the range $1.5 \lesssim \rm \log_{10}(P/days) \lesssim 1.9$ and the LSP was on average ${\sim}11$ times longer than the actual pulsation period. This agrees with the results from \citet{2017ApJ...847..139T} who found that LSPs had pulsation periods between the PLR sequences B and C' ($1.5 \lesssim \rm \log_{10}(P/days) \lesssim 1.7$). The period distribution of the semi-regular variables after correcting for LSPs is shown in Figure \ref{fig:fig17}. We assign the LSPs into a separate group (group V) for comparison with the other SR variables.

\begin{figure*}
	\includegraphics[width=\textwidth]{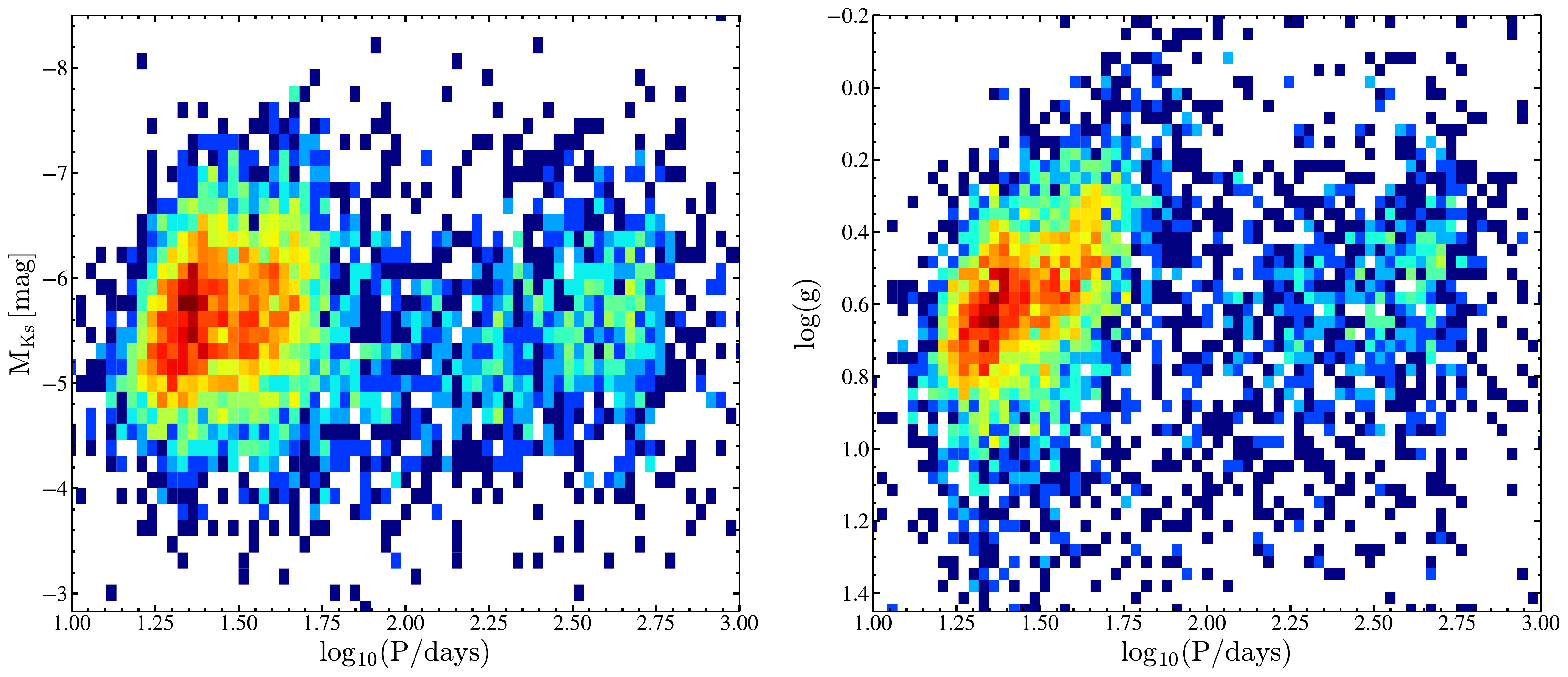}
    \caption{2-D histograms of $\rm M_{Ks}$ (left) and $\rm \log (g)$ (right) as a function of period for the SR variables in APOGEE DR16 after correcting for LSPs. The bins are colored by the number density.}
    \label{fig:fig18}
\end{figure*}

Figure \ref{fig:fig18} shows the distribution of the SR variables in $\rm M_{Ks}$ (left) and $\rm \log (g)$ with period. SR variables on the PLR sequences A, B and D are prominent, with sources on sequence C largely absent from this sample. The SR variables in these sequences overlap in both the period-luminosity and period-surface gravity spaces. It is difficult to disentangle the PLR sequences given the distance uncertainties --- only ${\sim}15\%$ (${\sim}9\%$) of these sources have \textit{Gaia} DR2 parallaxes better than $20\%$ ($10\%$). The various spectroscopic parameters and chemical abundance ratios from APOGEE DR16 for the five groups (I-V, defined above) are summarized in Table \ref{tab:srspec}. 

The median effective temperatures are similar for all the groups, however the median surface gravity for group I is larger than the other groups, suggesting that these sources might be less evolved. Sources in group I might be contaminated with some sources on the upper RGB or are relatively new to the TP-AGB phase. Sources in group III and IV have the largest median $V$-band amplitudes. This is not too surprising as these stars are fundamental-mode pulsators. Sources in group I have the smallest amplitudes, which is consistent with them being overtone pulsators. The metallicities of the sources in groups II, III and IV are largely similar. However, group I has a slightly different chemical profile, with median abundance ratios higher than for the other groups. However, the $\rm [X/Mg]$ abundance ratios are consistent across these groups given the reported dispersions. The biggest difference (${\sim}0.2$ dex) is seen in $\rm [Al/Mg]$, with the median abundance trending from $\rm [Al/Mg]{\sim}-0.1$ for $\rm \log_{10}(P/days)<1.5$ to  $\rm [Al/Mg]{\sim}-0.3$ for $\rm \log_{10}(P/days)>1.5$. The LSPs are less evolved than the SR variables in group IV with $ \rm \Delta \log(g) {\sim}0.1$, and are more similar to the SR variables in group I and II in their properties. This is not surprising given that the pulsation periods of the LSPs are consistent with the periods of SR variables in groups I and II \citep{2017ApJ...847..139T}. 

\begin{table*}
	\centering
	\caption{Distribution of the APOGEE DR16 spectroscopic parameters and chemical abundance ratios for the SR variables, binned by period. The SR variables are sorted into period bins as described in the text. The median, and standard deviation for each parameter is shown.}
	\label{tab:srspec}
\begin{tabular}{l|rrrrr}
		\hline
		 & Group I & Group II & Group III & Group IV & Group V (LSP)\\
		\hline		
		 & $1 \lesssim \rm \log_{10}(P/days) \lesssim 1.5$ & $1.5 \lesssim \rm \log_{10}(P/days) \lesssim 1.9$ & $1.9 \lesssim \rm \log_{10}(P/days) \lesssim 2.3$ & $\rm \log_{10}(P/days) \gtrsim 2.3$ & $\rm \log_{10}(P/days) \gtrsim 2.3$\\
\hline
$N$ & 407 & 475 & 169 & 334 & 300\\
$ \rm Amplitude\,[mag]$ & 0.11$\pm$0.04 & 0.20$\pm$0.14 & 0.29$\pm$0.28 & 0.30$\pm$0.27 & 0.16$\pm$0.14\\
\hline
$ \rm T_{eff}\,[K]$ & 3733$\pm$106 & 3695$\pm$128 & 3705$\pm$176 & 3705$\pm$149 & 3722$\pm$126\\
$ \rm \log(g)$ & 0.62$\pm$0.20 & 0.41$\pm$0.22 & 0.34$\pm$0.41 & 0.41$\pm$0.31 & 0.51$\pm$0.24\\
\hline
$ \rm [M/H]$ & $-0.42\pm$0.23 & $-0.63\pm$0.31 & $-0.71\pm$0.33 & $-0.66\pm$0.29 & $-0.57\pm$0.25\\
$ \rm [\alpha/M]$ & $0.10\pm$0.09 & $0.11\pm$0.10 & $0.10\pm$0.11 & $0.10\pm$0.11 & $0.10\pm$0.10\\
$ \rm [Fe/H]$ & $-0.44\pm$0.22 & $-0.64\pm$0.29 & $-0.71\pm$0.31 & $-0.66\pm$0.28 & $-0.59\pm$0.24\\
$ \rm [Mg/Fe]$ & $0.17\pm$0.11 & $0.18\pm$0.11 & $0.15\pm$0.13 & $0.17\pm$0.12 & $0.17\pm$0.11\\
\hline
$ \rm [C/Mg]$ & $-0.13\pm$0.08 & $-0.15\pm$0.15 & $-0.16\pm$0.19 & $-0.14\pm$0.14 & $-0.15\pm$0.11\\
$ \rm [N/Mg]$ & $0.05\pm$0.14 & $0.00\pm$0.18 & $-0.00\pm$0.20 & $0.01\pm$0.20 & $0.02\pm$0.15\\
$ \rm [O/Mg]$ & $-0.03\pm$0.04 & $-0.04\pm$0.05 & $-0.05\pm$0.06 & $-0.05\pm$0.05 & $-0.04\pm$0.04\\
$ \rm [Si/Mg]$ & $-0.11\pm$0.04 & $-0.12\pm$0.05 & $-0.12\pm$0.07 & $-0.12\pm$0.05 & $-0.12\pm$0.04\\
$ \rm [Ca/Mg]$ & $-0.08\pm$0.07 & $-0.07\pm$0.08 & $-0.07\pm$0.11 & $-0.08\pm$0.10 & $-0.08\pm$0.08\\
$ \rm [Al/Mg]$ & $-0.15\pm$0.09 & $-0.27\pm$0.16 & $-0.35\pm$0.15 & $-0.32\pm$0.14 & $-0.32\pm$0.14\\
$ \rm [Na/Mg]$ & $-0.07\pm$0.14 & $-0.09\pm$0.21 & $-0.10\pm$0.26 & $-0.09\pm$0.23 & $-0.10\pm$0.19\\
$ \rm [P/Mg]$ & $-0.02\pm$0.13 & $-0.06\pm$0.21 & $-0.04\pm$0.27 & $-0.03\pm$0.25 & $-0.04\pm$0.16\\
$ \rm [K/Mg]$ & $0.02\pm$0.11 & $0.01\pm$0.11 & $-0.00\pm$0.12 & $0.01\pm$0.11 & $0.01\pm$0.10\\
$ \rm [Cr/Mg]$ & $-0.17\pm$0.10 & $-0.16\pm$0.12 & $-0.14\pm$0.17 & $-0.15\pm$0.15 & $-0.13\pm$0.11\\
$ \rm [V/Mg]$ & $-0.06\pm$0.08 & $-0.05\pm$0.13 & $-0.07\pm$0.13 & $-0.06\pm$0.14 & $-0.05\pm$0.10\\
$ \rm [Co/Mg]$ & $-0.09\pm$0.08 & $-0.11\pm$0.09 & $-0.10\pm$0.13 & $-0.10\pm$0.11 & $-0.10\pm$0.10\\
$ \rm [Mn/Mg]$ & $-0.06\pm$0.16 & $-0.11\pm$0.18 & $-0.09\pm$0.19 & $-0.09\pm$0.18 & $-0.09\pm$0.16\\
$ \rm [Ni/Mg]$ & $-0.16\pm$0.08 & $-0.17\pm$0.08 & $-0.15\pm$0.11 & $-0.16\pm$0.10 & $-0.17\pm$0.08\\
$ \rm [Ti/Mg]$ & $-0.03\pm$0.13 & $-0.08\pm$0.14 & $-0.11\pm$0.15 & $-0.09\pm$0.13 & $-0.07\pm$0.13\\
\hline
$ \rm [C/N]$ & $-0.17\pm$0.14 & $-0.17\pm$0.23 & $-0.18\pm$0.27 & $-0.18\pm$0.24 & $-0.18\pm$0.17\\
$ \rm [C/O]$ & $-0.10\pm$0.06 & $-0.11\pm$0.12 & $-0.10\pm$0.17 & $-0.10\pm$0.13 & $-0.11\pm$0.09\\
$ \rm [N/O]$ & $0.07\pm$0.14 & $0.05\pm$0.17 & $0.06\pm$0.19 & $0.07\pm$0.19 & $0.07\pm$0.15\\
$ \rm [Al/O]$ & $-0.12\pm$0.08 & $-0.24\pm$0.14 & $-0.28\pm$0.15 & $-0.27\pm$0.13 & $-0.19\pm$0.12\\
		\hline
\end{tabular}
\end{table*}

\begin{figure*}
	\includegraphics[width=0.8\textwidth]{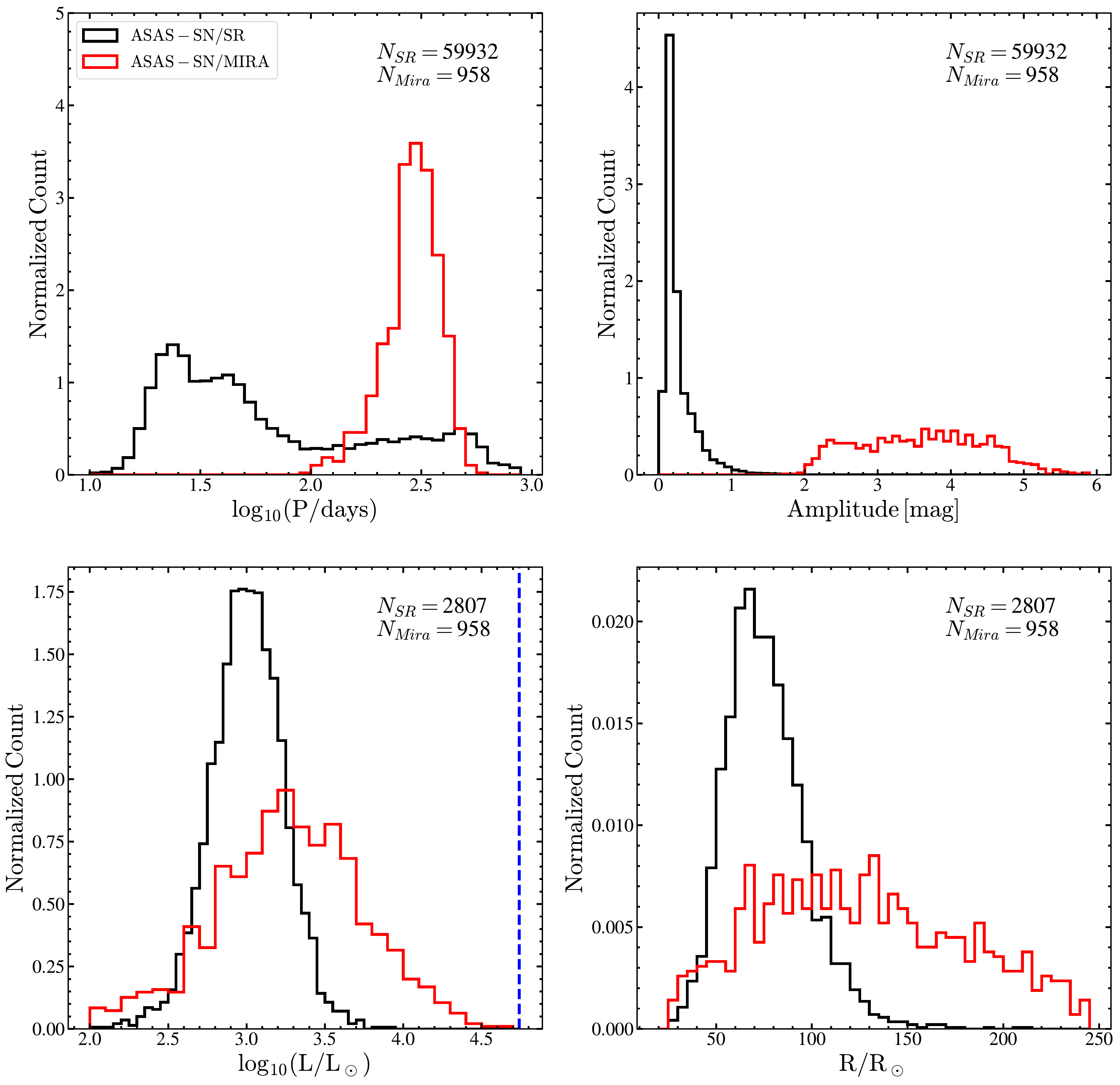}
    \caption{Distributions of the semi-regular variables after correcting for LSPs (black) and Mira variables (red) with $A_V<2$ mag and parallaxes better than $50\%$, in $\rm \log_{10}(P/days)$, $V$-band amplitude, $\log(L/L_\odot)$ and $R/R_\odot$. The blue dashed line corresponds to the maximum luminosity of AGB stars ($\log(L/L_\odot)=4.74$) as calculated by \citet{1970AcA....20...47P}.}
    \label{fig:fig19}
\end{figure*}

We compare the periods, amplitudes, bolometric luminosities and radii of the Mira and semi-regular variables with $A_V<2$ mag and parallaxes better than $50\%$ in Figure \ref{fig:fig19}. Very few Miras were in the spectroscopic sample. We calculate the bolometric luminosities as \begin{equation}
\log(L/L_\odot) = \rm 0.4\big(M_{bol,\odot}-M_{Ks}-BC_K\big),   
\end{equation} where $\rm BC_K=-6.75\log(\rm T_{eff}/9500)$ is the K-band bolometric correction for stars with $3300\leq \rm T_{eff}/K \leq 5000$  from \citet{2010MNRAS.403.1592B}. The $\rm M_{Ks}$ magnitudes were corrected for extinction using the SFD estimates \citep{2011ApJ...737..103S,1998ApJ...500..525S}. The radii are then calculated as  \begin{equation}
\bigg(\frac{R}{R_\odot}\bigg) = \bigg(\frac{T}{T_\odot} \bigg)^{-2} \bigg(\frac{L}{L_\odot} \bigg)^{1/2}. 
\end{equation} The bolometric corrections and radii of the SR variables were calculated using the spectroscopic temperatures, whereas we used temperature estimates from \textit{Gaia} DR2 for the Mira variables. The Mira variables have a median period of $\rm \log_{10}(P/days) {\sim} 2.5$ ($\rm P{\sim}320$ days). The amplitudes of the Mira variables range from ${\sim}2-6$ mag, whereas most SR variables have amplitudes $<1$ mag. The luminosity distributions of the Mira and SR variables are different, with the Mira variables being more luminous. The median luminosities of the SR variables and Mira variables are $\log(L/L_\odot){\sim}3.0$ and $\log(L/L_\odot){\sim}3.3$ respectively. Some Mira variables are very luminous, with $\log(L/L_\odot){\sim}4$. \citet{1970AcA....20...47P} calculated the maximum luminosity of AGB stars as $\log(L/L_\odot)=4.74$ based on the core-mass luminosity relationship derived for AGB stars that go through third dredge up. In our sample, all the Mira and SR variables had luminosities below this limit. The median radius of the Mira variables ($R{\sim}132\, R_\odot$) is almost twice that of the SR variables ($R{\sim}72\, R_\odot$), confirming the more evolved nature of the Mira variables when compared to the SR variables.

\subsubsection{A multi-band photometric calibration for the temperatures of oxygen-rich AGB stars}

 Most AGB stars lack spectroscopic information, but well calibrated temperatures are useful for studying these evolved stars. Here, we develop a temperature calibration based on the APOGEE temperatures, and a combination of \textit{Gaia} and 2MASS photometry. \citet{2018A&A...616L..13L} developed a multi-band approach using the Wesenheit magnitudes, $W_{RP}$ and $W_{JK}$, to distinguish between the various types of AGB stars. The intrinsic spread of red giants in the NIR colors is small, and the Wesenheit magnitudes adequately correct for the interstellar extinction. The visual colors of red giants are more sensitive to their surface temperatures and chemical compositions, resulting in a bigger spread in their optical colors. In their work, \citet{2018A&A...616L..13L} found that the color index $W_{RP}-W_{JK}$ traces the temperature and molecular features in the stellar spectra of AGB stars and that the $W_{RP}-W_{JK}$ index can be used to distinguish between carbon-rich and oxygen-rich AGB stars, with the carbon-rich (oxygen-rich) stars having $W_{RP}-W_{JK}\gtrsim 0.8$ mag ($W_{RP}-W_{JK}\lesssim 0.8$ mag). \citet{2020ApJS..247...44A} used the $W_{RP}-W_{JK}$ index and the 2MASS $J-Ks$ colors to study long period variables in the KELT survey \citep{2007PASP..119..923P}.

\begin{figure*}
	\includegraphics[width=\textwidth]{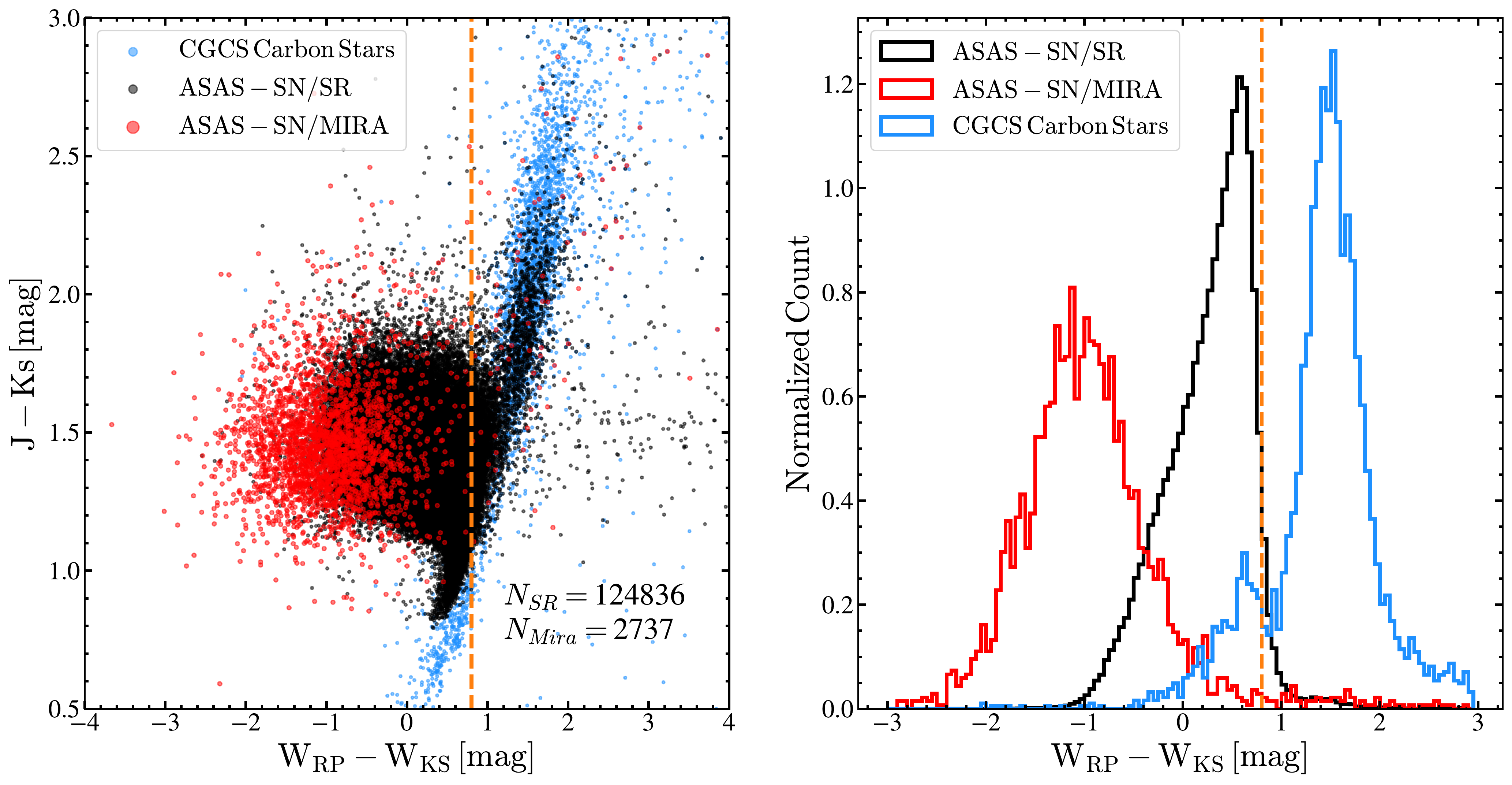}
    \caption{Distributions of the semi-regular variables (black) and Mira variables (red) in $W_{RP}-W_{JK}$ vs. $J-Ks$ (left) and $W_{RP}-W_{JK}$ (right). The sources from the Catalog of Galactic Carbon Stars \citep{2001BaltA..10....1A} (light blue) are also shown for reference. The orange dashed line corresponds to the $W_{RP}-W_{JK}$ index used to separate oxygen-rich ($W_{RP}-W_{JK}\lesssim 0.8$ mag) and carbon-rich ($W_{RP}-W_{JK}\gtrsim 0.8$ mag) AGB stars.}
    \label{fig:fig20}
\end{figure*}

We illustrate the distribution of ASAS-SN Mira and semi-regular variables in the ($W_{RP}-W_{JK}$)---($J-Ks$) plane in Figure \ref{fig:fig20}. For reference, we also show the sources from the Catalog of Galactic Carbon Stars \citep{2001BaltA..10....1A}. The carbon stars form a sharp, linear locus in this color-color space and have a median $W_{RP}-W_{JK}{\sim} 1.5 $ mag. In contrast to the carbon rich stars, the oxygen rich sources form a broader and bluer distribution in this plane. The vast majority of the ASAS-SN Mira and semi-regular variables have values of $W_{RP}-W_{JK}$ consistent with oxygen-rich stars. Of the Mira variables and semi-regular variables, ${\sim}97\%$ and ${\sim}95\%$ appear to be oxygen-rich AGB stars, respectively. The carbon-rich semi-regular variables follow the tight carbon-star locus in the $W_{RP}-W_{JK}$---$J-Ks$ plane at $1 \, {\rm mag} \lesssim W_{RP}-W_{JK} \lesssim 2 \,{\rm mag}$. In general, the carbon-rich semi-regular variables with $W_{RP}-W_{JK}\gtrsim 0.8$ mag have longer median periods ($\rm \log_{10}(P/days) {\sim} 2.1$) than the oxygen-rich stars ($\rm \log_{10}(P/days) {\sim} 1.9$). The Mira variables have a distinct distribution in $W_{RP}-W_{JK}$ that skews lower than that of the semi-regular variables.

\begin{figure*}
	\includegraphics[width=\textwidth]{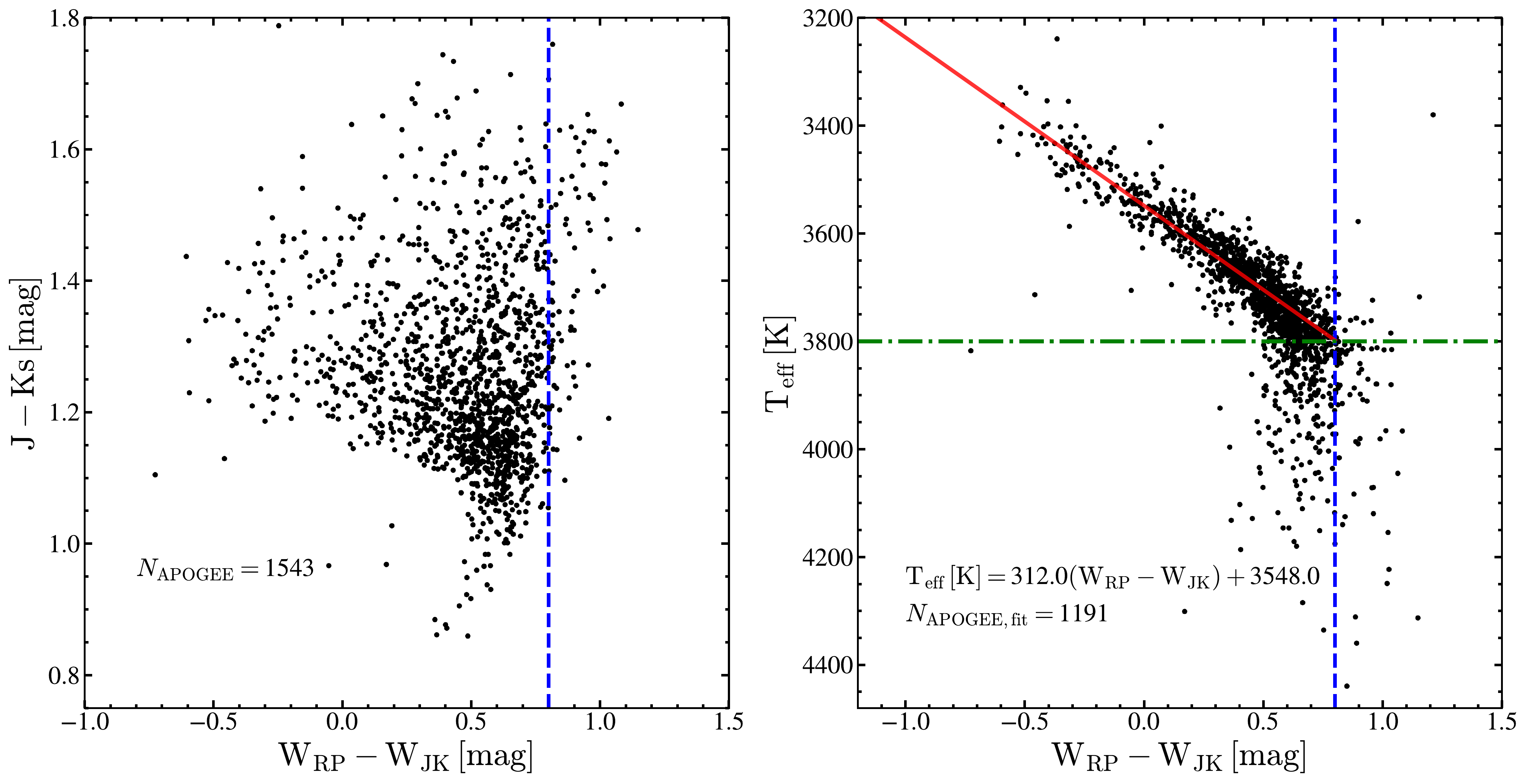}
    \caption{Distributions of the semi-regular variables in APOGEE DR16 in $W_{RP}-W_{JK}$ vs. $J-Ks$ (left) and $W_{RP}-W_{JK}$ vs. $\rm T_{eff}$ (right). The blue dashed line corresponds to the $W_{RP}-W_{JK}$ index ($0.8$ mag) used to separate oxygen-rich and carbon-rich AGB stars. The red line shows the linear fit to the sources with $\rm T_{eff}<3800$ K and  $W_{RP}-W_{JK}<0.8$. The green dashed line corresponds to the temperature cutoff of $\rm T_{eff}=3800$ K used in the fit. }
    \label{fig:fig21}
\end{figure*}

To investigate the dependence of temperature on the $W_{RP}-W_{JK}$ index, we show the distribution of 1639 semi-regular variables in APOGEE DR16 in the NIR $J-Ks$ color and $\rm T_{eff}$ against the $W_{RP}-W_{JK}$ index in Figure \ref{fig:fig21}. These are essentially all oxygen-rich semi-regular variables because there are few carbon-rich semi-regular variables in the APOGEE data. The temperatures of the oxygen-rich semi-regular variables are remarkably well correlated with the $W_{RP}-W_{JK}$ index for $\rm T_{eff}\lesssim3800$ K. The semi-regular variables with $\rm T_{eff}\gtrsim3800$ K have significantly more scatter. We fit a linear relationship of \begin{equation}
    \rm T_{eff} =3548(\pm2)K+\rm 312(\pm4)K\, \bigg(W_{RP}-W_{JK}\bigg)
	\label{eq:wrpkstemp}
\end{equation} to the 1191 sources with $\rm T_{eff}<3800 \rm \, K$ and $-0.7\, {\rm mag}<W_{RP}-W_{JK}<0.8 \, {\rm mag}$. On average, this fit returns temperatures that are within $\pm0.7\%$ of the APOGEE temperatures. The fit is only done for $-0.7 \, {\rm mag} \leq W_{RP}-W_{JK} \leq 0.8 \, {\rm mag}$ due to the lack of semi-regular variables with $W_{RP}-W_{JK}<-0.7$ mag.

As a test, we use this relation to estimate the temperatures of the Mira variables shown in Figure \ref{fig:fig20}. The majority of these sources have $W_{RP}-W_{JK}<-0.7$ mag, thus they lie outside the parameter space of the sources used in the fit. Applying equation \ref{eq:wrpkstemp} to these Mira variables, we find the median temperature to be $\rm T_{eff}{\sim}3230$ K (M5 spectral type), with the $1^{\rm st}$ percentile being $\rm T_{eff}{\sim}2800$ K (M8 spectral type) and the $99^{\rm th}$ percentile being $\rm T_{eff}{\sim}3687$ K (M1 spectral type). Mira variables have spectral temperatures that range from M0-M10 \citep{2017ApJS..232...16Y}, which is entirely consistent with the extrapolation. Furthermore, the median \textit{Gaia} DR2 temperature for these Mira variables is $\rm T_{eff}{\sim}3290$ K, which is very similar to the median temperature estimated using equation \ref{eq:wrpkstemp}.

\clearpage
\clearpage
\subsubsection{$\alpha$-enhancements on the TP-AGB}

During the TP-AGB, numerous nuclides are produced during H and He burning. The surfaces of AGB stars are chemically enriched primarily through the third dredge up (TDU) events that follow a thermal pulse. AGB nucleosynthesis depends on the efficiency of the TDU events, the minimum core mass at which TDU begins, the size of the convective envelope, and the mass of the He intershell \citep{2002PASA...19..515K}. The efficiency of TDU increases with increasing stellar mass and decreasing metallicity \citep{2002PASA...19..515K}. In particular, the surface abundances of CNO elements are enhanced by an order of magnitude or more through successive TDU episodes \citep{2004ApJ...602..377I}. The surfaces of AGB stars are also enriched in heavy elements that are produced by the s-process (e.g., Zr, Sr, Ba, etc., \citealt{1986ApJ...311..843S,1988ApJ...333..219S}). Hot-bottom burning (HBB) in the more massive AGB stars can also further enhance the surface N abundances \citep{1975ApJ...196..805S,2007MNRAS.378.1089M}. Thus, the surface abundances of intermediate-mass AGB stars are dependent on both the TDU and HBB processes. 

\begin{figure}
	\includegraphics[width=0.4\textwidth]{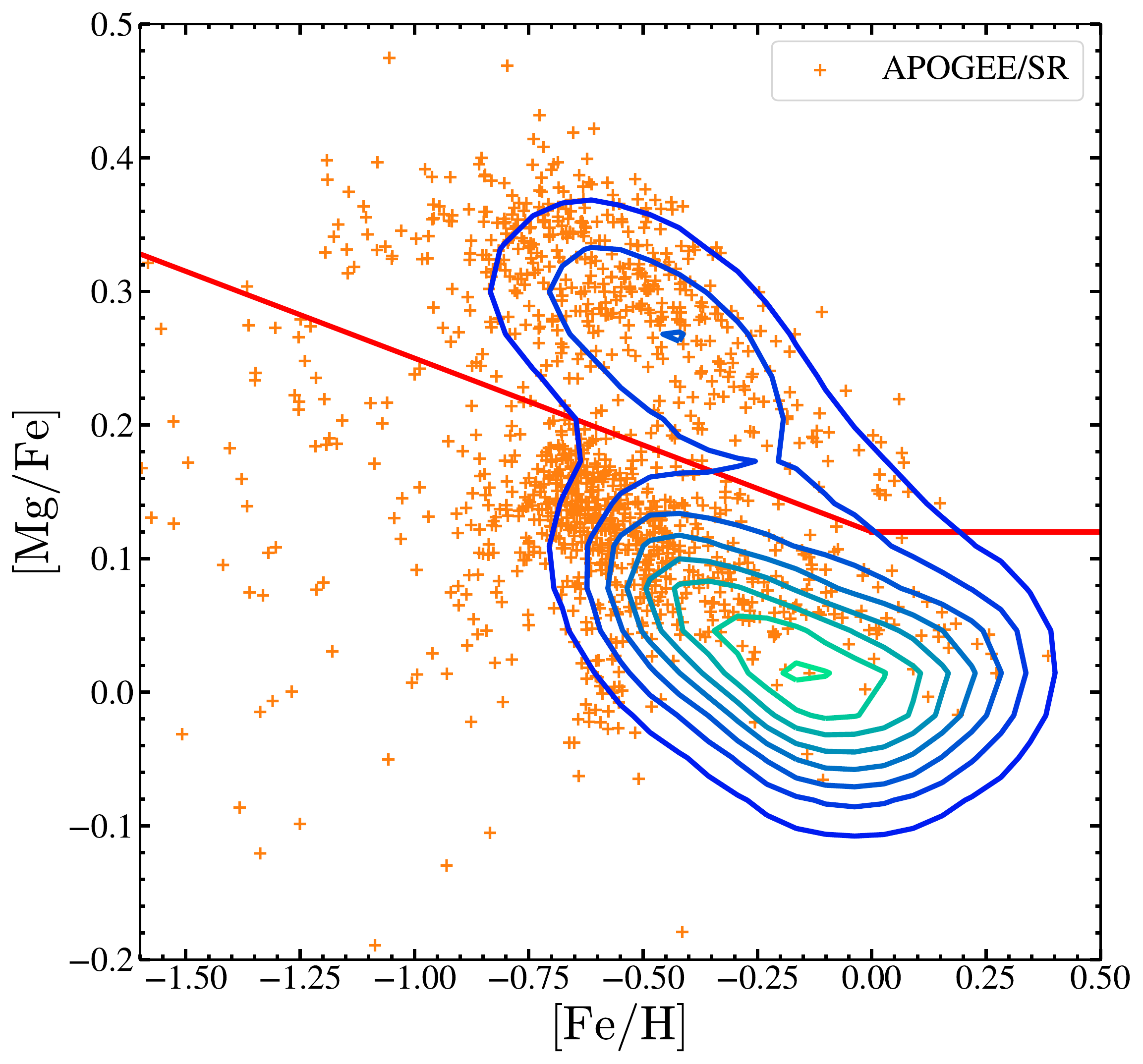}
    \caption{The distribution of the semi-regular variables in the APOGEE DR16 $\rm [Mg/Fe]$ vs. $\rm [Fe/H]$ plane. The contours show the distribution of a reference sample of APOGEE giants. The red line shows the division between the low-$\alpha$ and high-$\alpha$ populations from \citet{2019ApJ...874..102W}.}
    \label{fig:fig22}
\end{figure}

We show the distribution of the semi-regular variables in the $\rm [Mg/Fe]$---$\rm [Fe/H]$ plane as compared to the distribution of a reference sample of ${\sim}238,000$ giants in APOGEE DR16 with $\log(g)<3.8$ in Figure \ref{fig:fig22}. In the reference sample, ${\sim}27\%$ of the sources belong in the high-$\alpha$ sequence. The bimodality in $\rm [Mg/Fe]$ at sub-solar $\rm [Fe/H]$ is clearly seen for the semi-regular variables as well, but ${\sim}38\%$ of the semi-regular variables lie in the high-$\alpha$ sequence, which is an enhancement of $+41\%$ compared to the overall APOGEE sample. 

\begin{figure*}
	\includegraphics[width=\textwidth]{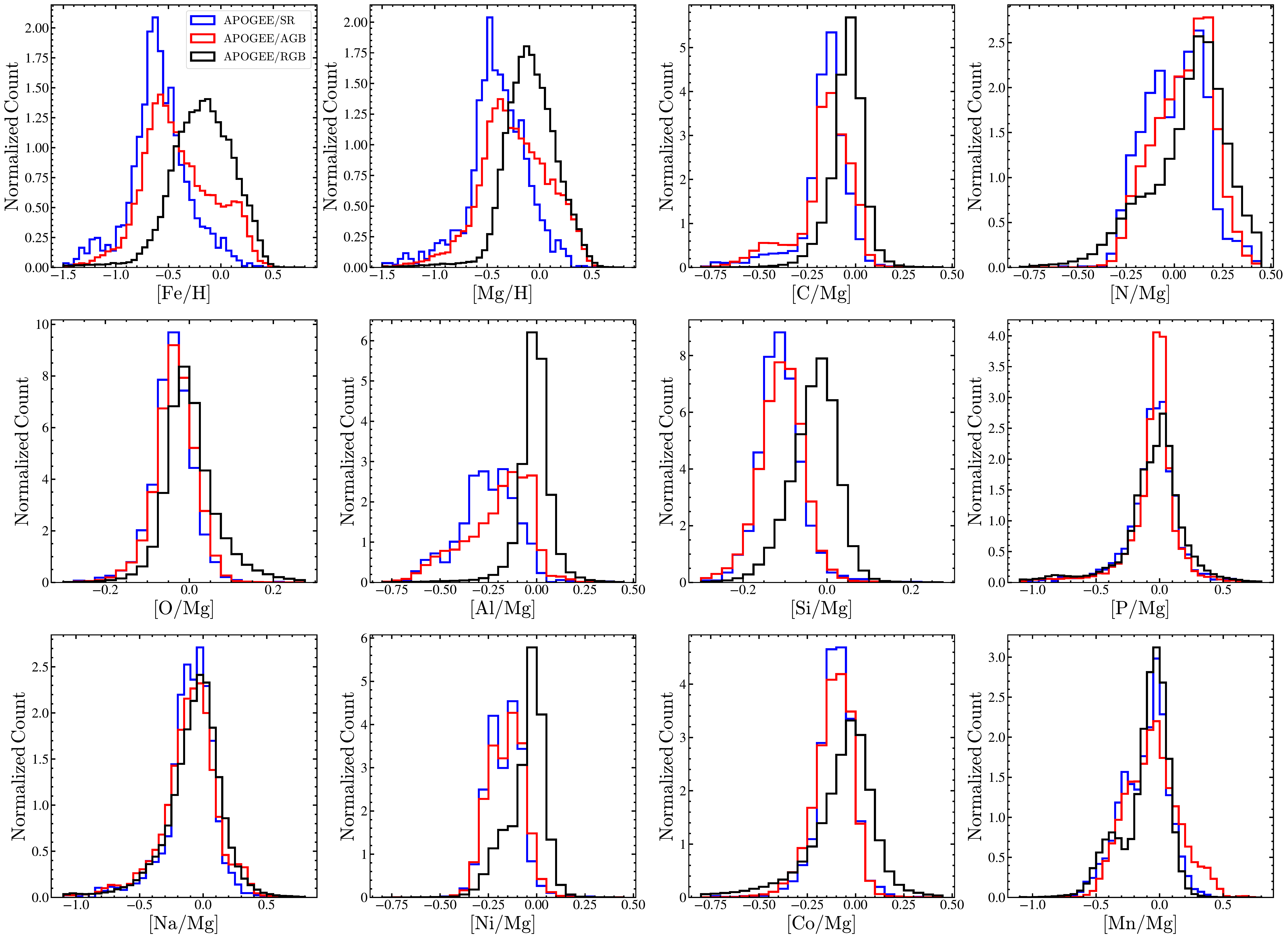}
    \caption{$\rm [Fe/H]$, $\rm [Mg/H]$ and $\rm [X/Mg]$ abundance ratios, where $X=$ C, N, O, Al, Si, P, Na, Ni, Co, and Mn, for the semi-regular variables (blue), APOGEE/AGB stars (red) and APOGEE/RGB stars (black).}
    \label{fig:fig24}
\end{figure*}

We also illustrate the distributions of various surface abundances of the semi-regular variables, APOGEE/RGB stars and APOGEE/AGB stars in Figure \ref{fig:fig24}. The semi-regular variables and the APOGEE/AGB stars are significantly metal-poor. Semi-regular variables with $\rm [Fe/H]>0$ or $\rm [Mg/H]>0$ are rare (${\sim}3\%$ and ${\sim}7\%$). The APOGEE/AGB stars populate the super-solar abundance bins significantly more than the semi-regular variables, but this maybe due to the contamination from metal-rich upper RGB stars. The distributions of $\rm [Na/Mg]$, $\rm [P/Mg]$, and $\rm [Mn/Mg]$ for the three samples are mostly similar. The semi-regular variables and the APOGEE/AGB stars have lower $\rm [C/Mg]$, $\rm [N/Mg]$ and $\rm [O/Mg]$ abundance ratios than the APOGEE/RGB stars. The biggest deviation from the APOGEE/RGB stars is in the distribution of $\rm [Al/Mg]$, with the semi-regular variables peaking ${\sim}-0.3$ dex below the peak of the APOGEE/RGB stars. There are virtually no APOGEE/RGB stars with $\rm [Al/Mg]<-0.25$. There are more semi-regular variables than APOGEE/AGB stars in these lower metallicity bins, perhaps suggesting a pulsational dependence to the $\rm [Al/Mg]$ abundance ratio. We will further investigate this phenomenon in $\S 4.3.4$. The semi-regular variables and APOGEE/AGB stars also show substantial shifts relative to the APOGEE/RGB stars in the $\rm [Si/Mg]$ ($\Delta{\sim}-0.1$ dex), $\rm [Ni/Mg]$ ($\Delta{\sim}-0.1$ dex) and $\rm [Co/Mg]$ ($\Delta{\sim}-0.1$ dex) abundance ratios.

\begin{figure*}
	\includegraphics[width=0.8\textwidth]{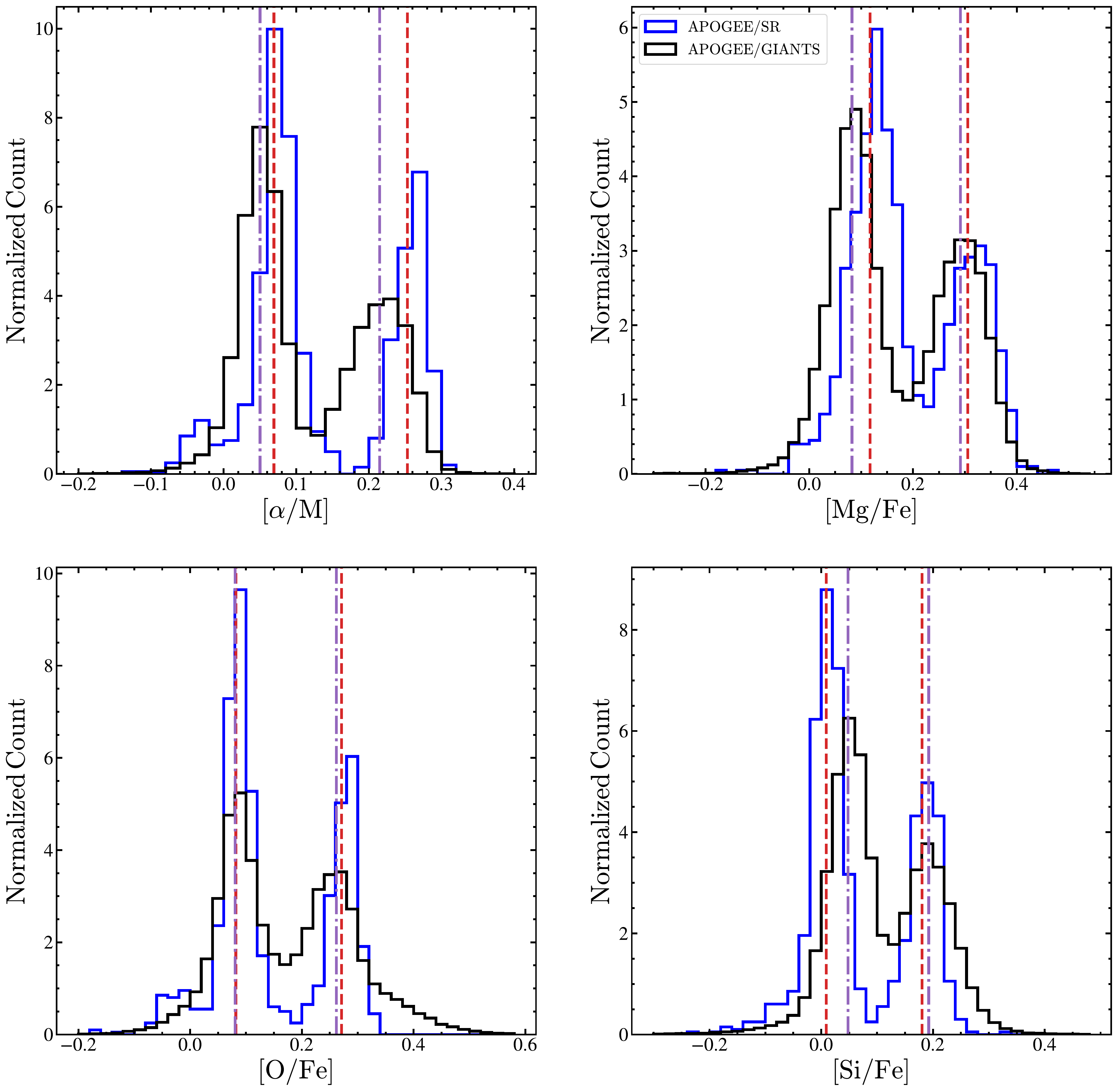}
    \caption{The distribution of the SR variables (blue) and giant stars from APOGEE DR16 in $\rm [\alpha/M]$, $\rm [Mg/Fe]$, $\rm [O/Fe]$ and $\rm [Si/Fe]$. The median abundance ratios of the high-$\alpha$ and low-$\alpha$ SR variables (APOGEE giants) are shown as red (purple) dashed (dot-dashed) lines.}
    \label{fig:fig25}
\end{figure*}

While the morphology of the SR distribution in Figure \ref{fig:fig22} is similar to the APOGEE giants, the actual metallicity and abundance distributions are shifted, as shown in Figure \ref{fig:fig24} and \ref{fig:fig25}. SR variables tend to be more metal-poor (peak $\rm [Fe/H]{\sim}-0.6$ instead of $\rm [Fe/H]{\sim}-0.2$) and $\alpha$-rich (median $\rm [Mg/Fe]{\sim}0.16$ instead of $\rm [Mg/Fe]{\sim}0.07$). The low-$\alpha$ SR variables peak at $\rm [Mg/Fe]{\sim}+0.12$, compared to $\rm [Mg/Fe]{\sim}+0.03$ for the low-$\alpha$ APOGEE/GIANTS. We illustrate the shifts in the distribution of the $\rm [\alpha/M]$, $\rm [Mg/Fe]$, $\rm [O/Fe]$ and $\rm [Si/Fe]$ abundance ratios in Figure \ref{fig:fig25}. We see significant shifts for both the low-$\alpha$ and high-$\alpha$ stars in all but the $\rm [Si/Fe]$ abundance ratio. 

To distinguish between the contribution of (i) the initial abundances at star formation and (ii) dredge-up effects, to the $\rm [\alpha/M]$, $\rm [Mg/Fe]$, $\rm [O/Fe]$ and $\rm [Si/Fe]$ abundance ratios (Figure \ref{fig:fig25}), we re-calculate the peak shifts by limiting the sample of SR variables and APOGEE giants to $\rm -1 \leq [Fe/H] \leq -0.25$. In this metallicity-limited sample of SR variables and APOGEE giants, we see that the magnitude of the abundance shifts are smaller, as summarized in Table \ref{tab:alphashift}. For example, the peak shift of the low-$\alpha$ SR variables in $\rm [Mg/Fe]$ is now ${\sim}+0.04$, compared to ${\sim}+0.09$ for the full sample. Furthermore, the low-$\alpha$ SR variables appear to have significantly lower median $\rm [Si/Fe]$ abundance ratios than the APOGEE giants. This is suggestive of Si depletion through dust-formation in the winds of these AGB stars \citep{2018A&A...611A..29M}. 

These enhancements are suggestive of surface Mg enhancement during TDU episodes. During the AGB, isotopes of Mg are primarily created during thermal pulses through neutron capture processes, and the Ne-Mg reactions \citep{2011ApJS..197...17C}. Models of AGB stars have shown that the surface abundance of Mg is enhanced at low metallicities \citep{2011ApJS..197...17C,2015ApJS..219...40C}. For example, taking a model of a $1.5\rm M_\odot$ AGB star, the final surface $\rm [Mg/Fe]$ abundance ratio is 0.01 at $Z=6\times10^{-3}$, and 0.43 at $Z=1\times10^{-3}$ \citep{2011ApJS..197...17C}. The efficiency of dredge up also increases with decreasing metallicity \citep{2002PASA...19..515K,2007A&A...469..239M}. Considering these theoretical expectations, it is not surprising to see Mg enhancements in the atmospheres of these metal-poor pulsating AGB stars.

In summary, we find that the metallicities and abundance ratios of the SR variables are distinct from the RGB stars. There are more $\alpha$-rich SR variables than $\alpha$-rich RGB stars. The semi-regular variables are more metal-poor with a metallicity peak at $\rm [Fe/H]{\sim}-0.6$. Even after considering the initial abundances at star formation, we find some evidence of $\alpha$/Mg enrichment in the atmospheres of the SR variables, which is likely the result of the third dredge up during the AGB.

\begin{table*}
	\centering
	\caption{The shifts in the median abundance ratios of various $\alpha$ elements between the SR variables and APOGEE/GIANTS. The median abundance error is ${\sim}0.01$ dex.}
	\label{tab:alphashift}
\begin{tabular}{llrrr}
		\hline
		Abundance & low-$\alpha$ shift & high-$\alpha$ shift & low-$\alpha$ shift & high-$\alpha$ shift  \\
		 & \textit{All} & \textit{All} & \textit{Metallicity-limited} & \textit{Metallicity-limited}  \\		
		\hline
$\rm [\alpha/M]$ & +0.06 & +0.06 & +0.02 & +0.04 \\
$\rm [Mg/Fe]$ & +0.09 & +0.04 & +0.04 & +0.01 \\
$\rm [O/Fe]$ & +0.08 & +0.05 & +0.0 & +0.01 \\
$\rm [Si/Fe]$ & +0.01 & +0.01 & -0.04 & -0.01 \\
		\hline
\end{tabular}
\end{table*}

\clearpage
\clearpage

\subsubsection{Surface Aluminium depletion on the TP-AGB}

Al is a light, odd $Z$ element that is almost entirely produced due to core-collapse supernovae (see for e.g., \citealt{2019ApJ...874..102W,2019ApJ...886...84G}). The oxygen-rich atmospheres of most AGB stars are conducive to the formation of dust seeds close to the stellar surface, including $\rm Al_2O_3$, $\rm SiO_2$ and $\rm TiO_2$ \citep{2013A&A...555A.119G,2017A&A...608A..55D}. The presence of alumina ($\rm Al_2O_3$) dust around oxygen-rich AGB stars is well known \citep{1989A&A...218..169O}. \citet{2012ApJ...753L..20B} reported the near-IR detection of several AlO bands in the wavelength range $1.0-1.35\mu \rm m$ and rotational transitions of this molecule have also been reported \citep{2017A&A...598A..53D, 2017A&A...608A..55D}. It is thought that AlO is efficiently depleted from the gas around oxygen-rich AGB stars to form alumina dust seeds \citep{2017A&A...598A..53D}. AGB stars lose mass through a slow wind (${\sim}10\,\rm km/s$) driven by radiation pressure on dust grains \citep{1983ARA&A..21..271I,2015A&A...577A.114K}. Understanding the formation of dust seeds like alumina and silicate is a crucial consideration when studying mass loss in AGB stars.

\begin{figure*}
	\includegraphics[width=\textwidth]{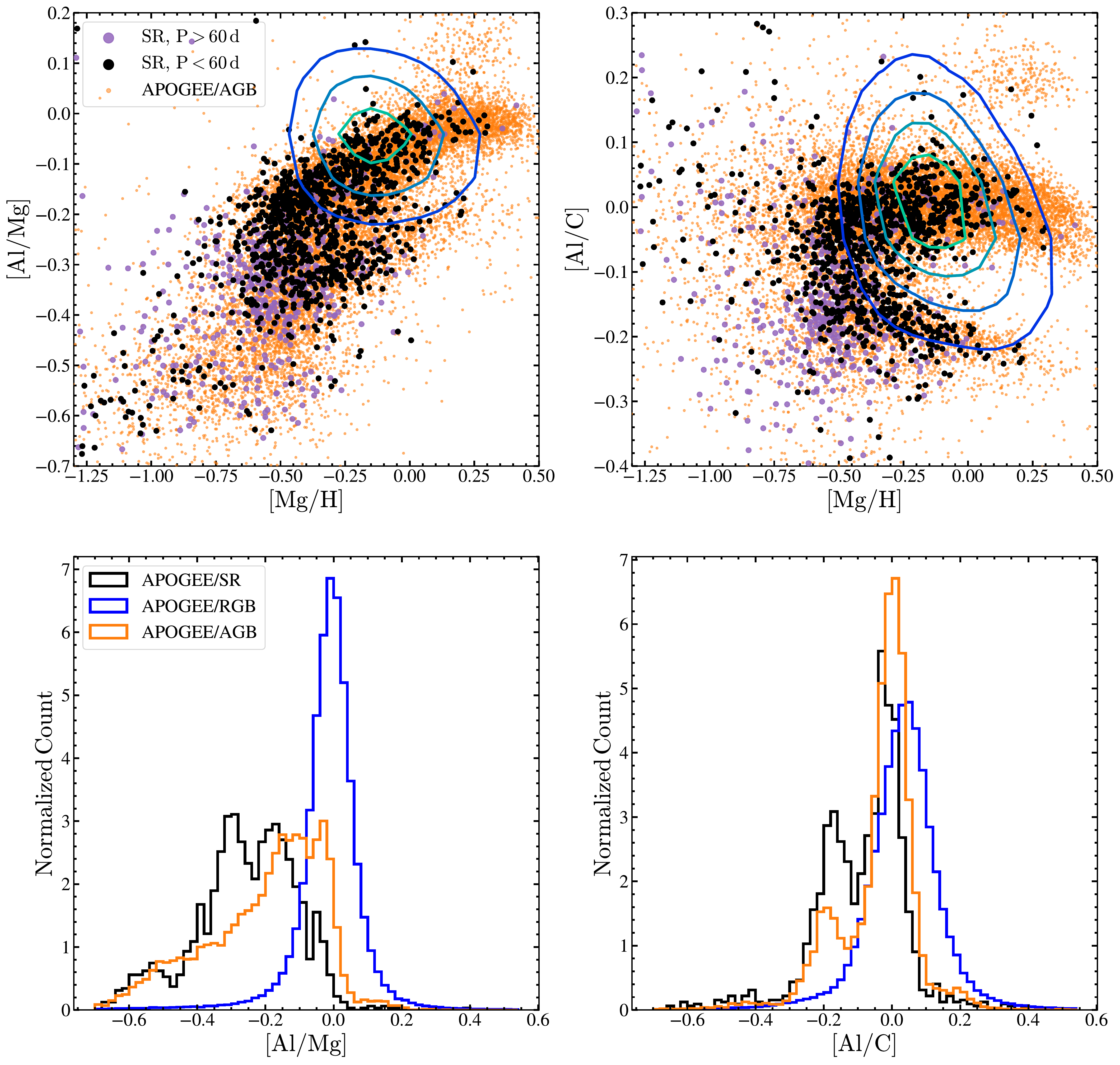}
    \caption{$\rm [Al/Mg]$ (top left) and $\rm [Al/C]$ (top right) vs. $\rm [Mg/H]$ for the semi-regular variables ($\rm P<60$d: black, $\rm P>60$d: purple) and APOGEE AGB stars (orange). The distributions of the APOGEE RGB stars in these planes are shown as contours. The distributions of $\rm [Al/Mg]$ (bottom left) and $\rm [Al/C]$ (bottom right) for the semi-regular variables (black) APOGEE AGB (orange) and RGB (blue) stars are shown as histograms.}
    \label{fig:fig26}
\end{figure*}

In $\S4.3.3$, we noted that the semi-regular variables and the sample of APOGEE/AGB stars in APOGEE DR16 had lower $\rm [Al/Mg]$ abundance ratios than the APOGEE/RGB stars. The median $\rm[Al/Mg]$ abundance ratios for the semi-regular variables, APOGEE/AGB stars and APOGEE/RGB stars are $-0.25$, $-0.16$ and $-0.01$ respectively. We investigate the $\rm [Al/Mg]$ and $\rm [Al/C]$ abundance ratios for the semi-regular variables, APOGEE/AGB and APOGEE/RGB stars in Figure \ref{fig:fig26}. In the $\rm[Al/Mg]$---$\rm [Mg/H]$ plane, both the semi-regular variables and APOGEE/AGB stars form two distinct sequences of increasing $\rm[Al/Mg]$ with increasing $\rm [Mg/H]$. In general, the APOGEE/AGB stars have higher $\rm [Al/Mg]$ abundance ratios than the semi-regular variables. The $\rm [Al/Mg]$ abundance ratios of the APOGEE/RGB stars appear to follow a Gaussian distribution centered at $\rm [Al/Mg]=0$. Of the APOGEE/RGB stars, only ${\sim}3\%$ have $\rm [Al/Mg]<-0.2$, whereas ${\sim}61\%$ of the semi-regular variables and ${\sim}42\%$ of the APOGEE/AGB stars have $[\rm Al/Mg]<-0.2$. This suggests that Al is being depleted on the AGB, with the pulsating TP-AGB stars depleting more Al than other AGB stars. 

The two Al sequences (hereafter the ``high-Al'' and ``low-Al'' sequences) are even more distinct in the $\rm [Al/C]$---$\rm [Mg/H]$ plane (Figure \ref{fig:fig26}). The ``high-Al'' and ``low-Al'' populations are centered on $\rm [Al/C]{\sim}0$ and $\rm [Al/C]{\sim}-0.15$. The APOGEE/RGB stars again appear to follow a Gaussian distribution in $ \rm [Al/C]$, centered at $ \rm [Al/C]{\sim}+0.05$. However, both the semi-regular variables and APOGEE/AGB stars have a bi-modal distribution in $\rm [Al/C]$. The semi-regular variables populate the ``low-Al'' sequence relatively more than the APOGEE/AGB stars, with ${\sim}40\%$ of the semi-regular variables having $[\rm Al/C]<-0.1$ compared to only ${\sim}23\%$ of the APOGEE/AGB stars. This again hints at a correlation between Al depletion and pulsations in AGB stars.

\begin{figure*}
	\includegraphics[width=\textwidth]{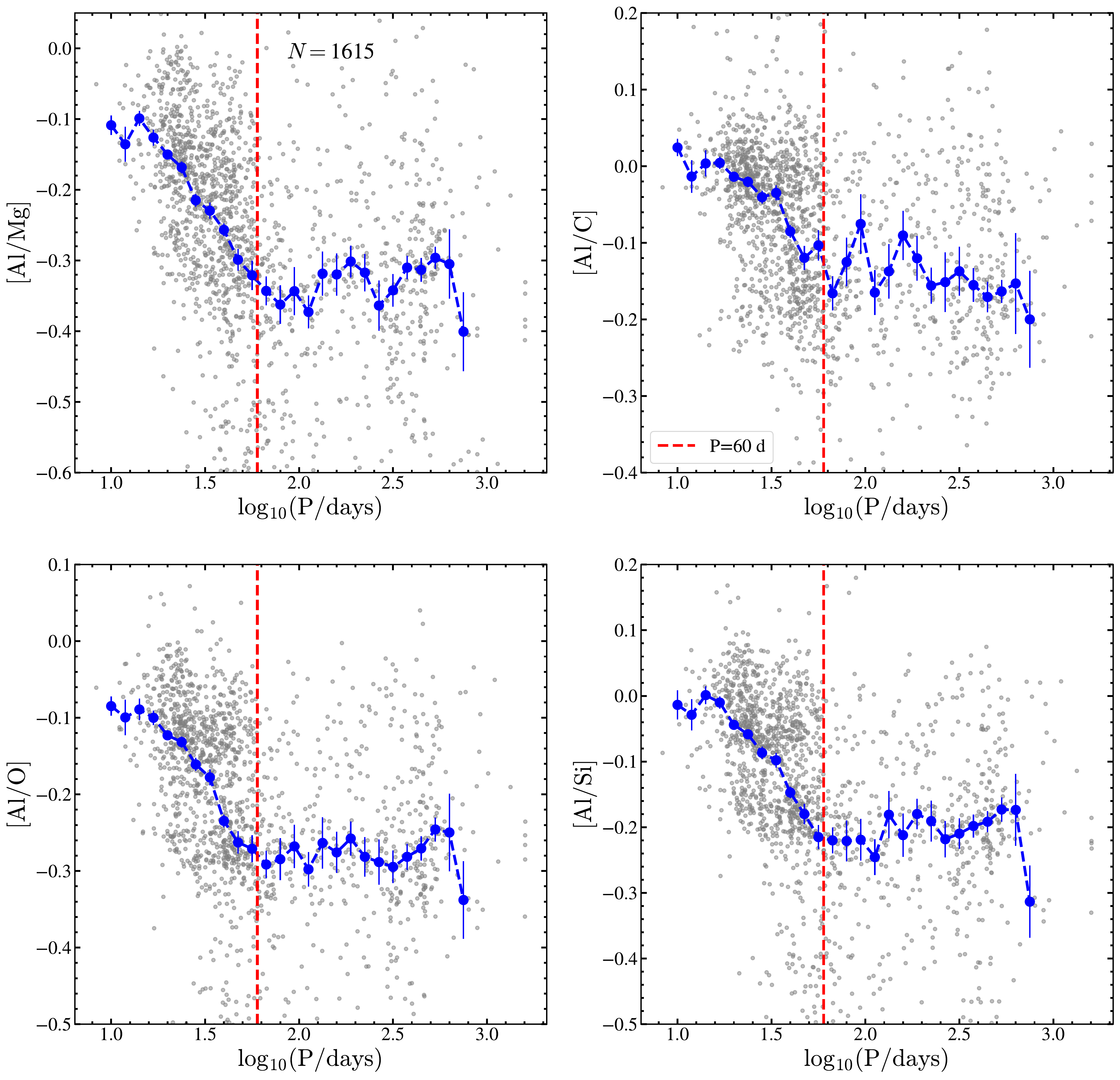}
    \caption{$\rm [Al/Mg]$, $\rm [Al/C]$, $\rm [Al/Si]$ and $\rm [Al/O]$ vs. $\rm \log_{10}(P/days)$ for the semi-regular variables. The binned median abundance ratios are shown in blue. The red dashed line shows the period at which increased dust formation first occurs for Galactic AGB stars ($\rm P{\gtrsim}60$ days). These correlations are far less clear if we use the LSP periods.}
    \label{fig:fig27}
\end{figure*}

We investigate the dependence of $\rm [Al/X]$, where $\rm X=$ Mg, C, Si, O, with the pulsation period in Figure \ref{fig:fig27}. The reference elements C, Si and O, were chosen due to their significant presence in dust grains. Previous studies have established that strong mass loss and increased dust formation first occurs for pulsation periods of $\rm P{\gtrsim}60$ days for Galactic AGB stars \citep{2009MNRAS.395L..11G,2018MNRAS.481.4984M}. The $\rm [Al/X]$ abundance ratios of the semi-regular variables are strongly correlated with the pulsational period. On average, the $\rm [Al/X]$ abundance ratios decrease with period up until $\rm P{\sim}60$ days and then flatten. The most striking dependence is seen in the distribution of $\rm [Al/O]$ with period. The $\rm [Al/O]$ trend largely plateaus at $\rm [Al/O]{\sim}-0.3$ beyond $P{\sim}60$ days. The fraction of sources with $\rm [Al/O]<-0.22$ varies between the PLR groups (see $\S 4.3.1$), with ${\sim}16\%$ in group I, ${\sim}54\%$ in group II, ${\sim}74\%$ in group III and ${\sim}69\%$ in group IV. The fraction of sources with $\rm [Al/C]<-0.1$ follow a similar trend, with ${\sim}14\%$ in group I, ${\sim}45\%$ in group II, ${\sim}62\%$ in group III and ${\sim}63\%$ in group IV. Of the LSPs, ${\sim}44\%$ and ${\sim}41\%$ had $\rm [Al/O]<-0.22$ and $\rm [Al/C]<-0.1$, respectively. These period trends were far less obvious when we assigned stars their LSPs, which strongly supports the argument that the shorter periods are more physically important. 

\begin{figure*}
	\includegraphics[width=\textwidth]{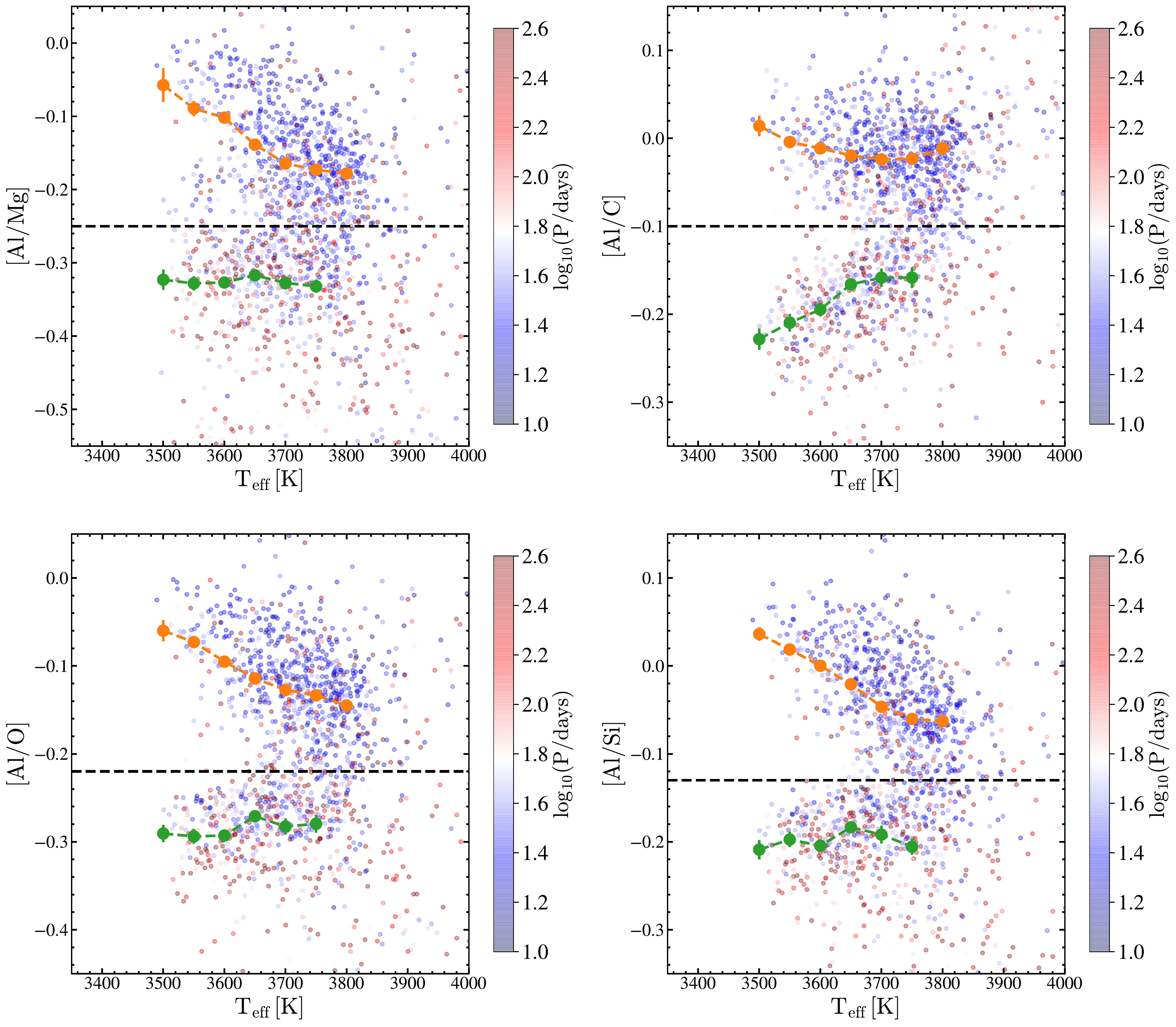}
    \caption{The dependence of $\rm [Al/Mg]$, $\rm [Al/C]$, $\rm [Al/Si]$ and $\rm [Al/O]$ on $\rm T_{eff}$ for the semi-regular variables. The points are colored by $\rm \log_{10}(P/days)$ with the changes in hue at periods near 60 days. The trends in the Al abundance ratios with $\rm T_{eff}$ for the high-Al and low-Al sequences are shown in orange and green respectively. The black dashed line shows the suggested abundance cuts ($\rm [Al/Mg]=-0.25$, $\rm [Al/C]=-0.1$, $\rm [Al/O]=-0.22$ or $\rm [Al/Si]=-0.13$) for separating the high-Al and low-Al sequences.}
    \label{fig:fig28}
\end{figure*}

Figure \ref{fig:fig28} shows the dependence of $\rm [Al/X]$, where $\rm X=$ Mg, C, Si, and O, with effective temperature for the semi-regular variables. The high-Al and low-Al sequences that were identified in the $\rm[Al/Mg]$---$\rm [Mg/H]$ and $\rm[Al/C]$---$\rm [Mg/H]$ planes are also apparent in temperature---abundance space. Furthermore, we see evidence of the period dependence on the Al abundance ratios, with short period sources falling into the high-Al sequence. Cuts of $\rm [Al/Mg]=-0.25$, $\rm [Al/C]=-0.1$, $\rm [Al/O]=-0.22$ or $\rm [Al/Si]=-0.13$ roughly separate the high-Al and low-Al sequences. The Al abundance ratios of the high-Al stars decrease with increasing temperature, and is best seen in the trends of the $\rm[Al/O]$ and $\rm[Al/Si]$ abundance ratios with $\rm T_{eff}$. In contrast, the Al abundances of the low-Al stars are not correlated with temperature. 

Separating based on their $\rm[Al/C]$ abundance ratio, we find that the high-Al variables have a median period of $\rm \log_{10}(P/days) {\sim} 1.5\pm0.4$, whereas the low-Al variables have a median period of $\rm \log_{10}(P/days) {\sim} 1.8\pm0.5$. In general, the high-Al variables have periods that fall below the period at which dust formation and mass loss is significant for AGB stars ($\rm \log_{10}(P/days) {\sim} 1.75$). Most low-Al stars have periods above the canonical threshold ($\rm P{\gtrsim}60$ days) for dust formation in Galactic AGB stars (Figure \ref{fig:fig27}). This suggests that the low-Al sequence could be the result of significant Al depletion through the formation of dust grains containing Al (for example, alumina) in the stellar winds of AGB stars. 

The various APOGEE DR16 spectroscopic parameters and chemical abundance ratios for the high-Al and low-Al sequences are summarized in Table \ref{tab:srspeclowhigh} for these two groups. The median $V$-band amplitudes of the low-Al stars are more than twice that of the high-Al stars. The high-Al stars have $\rm [\alpha/M]$ abundance ratios that are consistent with being on the low-$\alpha$ sequence, whereas the low-Al stars have $\rm [\alpha/M]$ consistent with being high-$\alpha$ stars. Based on the division in \citet{2019ApJ...874..102W}, we find that ${\sim}54\%$ of the low-Al stars are high-$\alpha$ stars, whereas only ${\sim}34\%$ of the high-Al variables are on the high-$\alpha$ sequence. 

The high-Al stars have enhanced N, Na, P, Cr, Mn, Ni and Ti abundances when compared to the low-Al stars. The low-Al stars tend to have larger Mg abundances ($\rm [Mg/Fe]_{~low~Al}=0.24\pm0.12$) and lower metallicities ($\rm [Fe/H]_{~low~Al}=-0.69\pm0.28$) when compared to the high-Al stars ($\rm [Mg/Fe]_{~high~Al}=0.14\pm0.10$, $\rm [Fe/H]_{~high~Al}=-0.52\pm0.25$). The $\rm [Mg/Fe]$ abundance ratios (at star formation) for both the high-$\alpha$ and low-$\alpha$ sequences tend to increase with decreasing metallicity. In addition, models of both low-mass and intermediate-mass AGB stars show that surface $\rm[Mg/Fe]$ is enhanced at low metallicities due to dredge up processes \citep{2011ApJS..197...17C,2015ApJS..219...40C}. In $\S4.3.3$, we showed that the SR variables could be enhanced in Mg through dredge up. Therefore, the larger Mg abundances observed for the low-Al stars can be explained by their birth $\rm [Mg/Fe]$ distribution and possibly with a contribution from dredge up processes in AGB stars.

In summary, an analysis of the $\rm [Al/X]$ abundance ratios of the SR variables show that the Al abundances are correlated with the pulsation period. We identified two distinct sequences amongst the SR variables corresponding to SR variables with high and low Al abundances. The trends of the Al abundances with pulsation period and temperature suggest that the low-Al sequence is likely the result of significant Al depletion through the formation of dust grains containing Al in the stellar winds of AGB stars. The low-Al sequence likely consists of pulsating AGB stars that lose mass through dust-driven winds.

\begin{table*}
	\centering
	\caption{Distribution of the APOGEE DR16 spectroscopic parameters and chemical abundances for the SR variables grouped by their abundances. The SR variables are sorted into these groups based on their $\rm [Al/O]$ and $\rm [N/O]$ abundances. The median, and standard deviation for each parameter is shown.}
	\label{tab:srspeclowhigh}
\begin{tabular}{l|rrrrr}
		\hline
		 & High-Al & Low-Al & High-N & Low-N& HBB\\
		\hline		
		 & $\rm [Al/O]>-0.22$ & $\rm [Al/O]<-0.22$ & $\rm 0.02 \leq[N/O]\leq0.3$ & $\rm [N/O]<0.02$ & $\rm [N/O]>0.30$\\
\hline
$N$ & 874 & 812 & 872 & 728 & 86\\
$ \rm \log_{10}(P/days)$ & 1.5$\pm$0.4 & 1.8$\pm$0.5 & 1.6$\pm$0.5 & 1.6$\pm$0.5 & 2.0$\pm$0.6\\
$ \rm Amplitude\,[mag]$ & 0.13$\pm$0.13 & 0.26$\pm$0.23 & 0.15$\pm$0.19 & 0.20$\pm$0.21 & 0.17$\pm$0.21\\
\hline
$ \rm T_{eff}\,[K]$ & 3730$\pm$126 & 3683$\pm$139 & 3727$\pm$126 & 3677$\pm$120 & 3822$\pm$194\\
$ \rm \log(g)$ & 0.54$\pm$0.29 & 0.42$\pm$0.25 & 0.46$\pm$0.28 & 0.55$\pm$0.21 & 0.07$\pm$0.55\\
\hline
$ \rm [M/H]$ & -0.50$\pm$0.26 & -0.68$\pm$0.29 & -0.57$\pm$0.31 & -0.63$\pm$0.28 & -0.52$\pm$0.31\\
$ \rm [\alpha/M]$ & 0.08$\pm$0.09 & 0.22$\pm$0.11 & 0.07$\pm$0.05 & 0.25$\pm$0.06 & -0.03$\pm$0.07\\
$ \rm [Fe/H]$ & -0.52$\pm$0.25 & -0.69$\pm$0.28 & -0.59$\pm$0.30 & -0.64$\pm$0.27 & -0.56$\pm$0.30\\
$ \rm [Mg/Fe]$ & 0.14$\pm$0.10 & 0.24$\pm$0.12 & 0.13$\pm$0.07 & 0.30$\pm$0.08 & 0.03$\pm$0.09\\
\hline
$ \rm [C/Mg]$ & $-0.14\pm$0.09 & $-0.15\pm$0.16 & $-0.14\pm$0.13 & $-0.15\pm$0.14 & $-0.22\pm$0.18\\
$ \rm [N/Mg]$ & $0.08\pm$0.16 & $-0.07\pm$0.18 & $0.10\pm$0.08 & $-0.13\pm$0.09 & $0.36\pm$0.19\\
$ \rm [O/Mg]$ & $-0.03\pm$0.04 & $-0.04\pm$0.05 & $-0.04\pm$0.05 & $-0.04\pm$0.05 & $-0.05\pm$0.05\\
$ \rm [Si/Mg]$ & $-0.11\pm$0.05 & $-0.12\pm$0.06 & $-0.11\pm$0.05 & $-0.12\pm$0.05 & $-0.08\pm$0.06\\
$ \rm [Ca/Mg]$ & $-0.07\pm$0.07 & $-0.10\pm$0.10 & $-0.06\pm$0.06 & $-0.14\pm$0.08 & $0.05\pm$0.08\\
$ \rm [Al/Mg]$ & $-0.16\pm$0.09 & $-0.34\pm$0.11 & $-0.20\pm$0.14 & $-0.31\pm$0.14 & $-0.31\pm$0.14\\
$ \rm [Na/Mg]$ & $-0.04\pm$0.15 & $-0.14\pm$0.24 & $-0.02\pm$0.16 & $-0.17\pm$0.20 & $0.10\pm$0.17\\
$ \rm [P/Mg]$ & $-0.01\pm$0.15 & $-0.07\pm$0.25 & $0.01\pm$0.17 & $-0.11\pm$0.20 & $0.18\pm$0.23\\
$ \rm [K/Mg]$ & $0.02\pm$0.10 & $-0.01\pm$0.11 & $0.03\pm$0.10 & $-0.02\pm$0.11 & $0.09\pm$0.11\\
$ \rm [Cr/Mg]$ & $-0.11\pm$0.11 & $-0.21\pm$0.15 & $-0.08\pm$0.11 & $-0.25\pm$0.09 & $0.02\pm$0.12\\
$ \rm [V/Mg]$ & $-0.04\pm$0.09 & $-0.07\pm$0.13 & $-0.03\pm$0.12 & $-0.09\pm$0.09 & $-0.05\pm$0.13\\
$ \rm [Co/Mg]$ & $-0.08\pm$0.09 & $-0.12\pm$0.11 & $-0.06\pm$0.10 & $-0.14\pm$0.08 & $-0.09\pm$0.12\\
$ \rm [Mn/Mg]$ & $-0.04\pm$0.16 & $-0.17\pm$0.18 & $-0.02\pm$0.14 & $-0.24\pm$0.14 & $0.02\pm$0.16\\
$ \rm [Ni/Mg]$ & $-0.13\pm$0.08 & $-0.20\pm$0.09 & $-0.11\pm$0.07 & $-0.23\pm$0.06 & $-0.12\pm$0.10\\
$ \rm [Ti/Mg]$ & $-0.06\pm$0.14 & $-0.10\pm$0.12 & $-0.07\pm$0.16 & $-0.09\pm$0.11 & $-0.01\pm$0.11\\
\hline
$ \rm [C/N]$ & $-0.20\pm$0.18 & $-0.11\pm$0.24 & $-0.23\pm$0.12 & $-0.02\pm$0.14 & $-0.60\pm$0.32\\
$ \rm [C/O]$ & $-0.10\pm$0.07 & $-0.11\pm$0.14 & $-0.10\pm$0.11 & $-0.11\pm$0.11 & $-0.17\pm$0.17\\
$ \rm [N/O]$ & $0.11\pm$0.15 & $-0.03\pm$0.17 & $0.13\pm$0.06 & $-0.09\pm$0.07 & $0.41\pm$0.19\\
$ \rm [Al/O]$ & $-0.12\pm$0.07 & $-0.29\pm$0.10 & $-0.14\pm$0.12 & $-0.27\pm$0.12 & $-0.27\pm$0.20\\
		\hline
\end{tabular}
\end{table*}
\clearpage

\clearpage

\subsubsection{Nitrogen Abundances on the TP-AGB}
Nitrogen production during the AGB phase is dominated by hot bottom burning (HBB) in intermediate mass stars with $M\gtrsim 3 M_\odot$ \citep{1975ApJ...196..805S,2007MNRAS.378.1089M}. During hot-bottom burning, proton-capture nucelosynthesis occcurs at the base of the outer envelope, favoring the conversion of C to N through the CN cycle \citep{1992ApJ...393L..21B} where the $^{12}\rm C$,$^{15}\rm N$, $^{16}\rm O$ and $^{18}\rm O$ isotopes are destroyed to produce $^{14}\rm N$ \citep{1996MmSAI..67..729L}.This can only happen if the temperature at the bottom of the envelope exceeds ${\sim}50\times10^6$ K \citep{2013A&A...555L...3G}. In these intermediate mass stars, the combination of third dredge-up and hot-bottom burning results in increased N abundances \citep{2007MNRAS.378.1089M}. At the onset of the thermal pulses, hot-bottom burning reduces the $^{12}\rm C$ abundance to ${\sim}1/15$ of the MS value and increases the $^{14}\rm N$ abundance to ${\sim}5-6$ times the MS value \citep{2004agbs.book...23L}. TDU episodes can continue after HBB ceases to operate, allowing the C abundances to again increase \citep{1996ApJ...473..383F}.

\begin{figure*}
	\includegraphics[width=\textwidth]{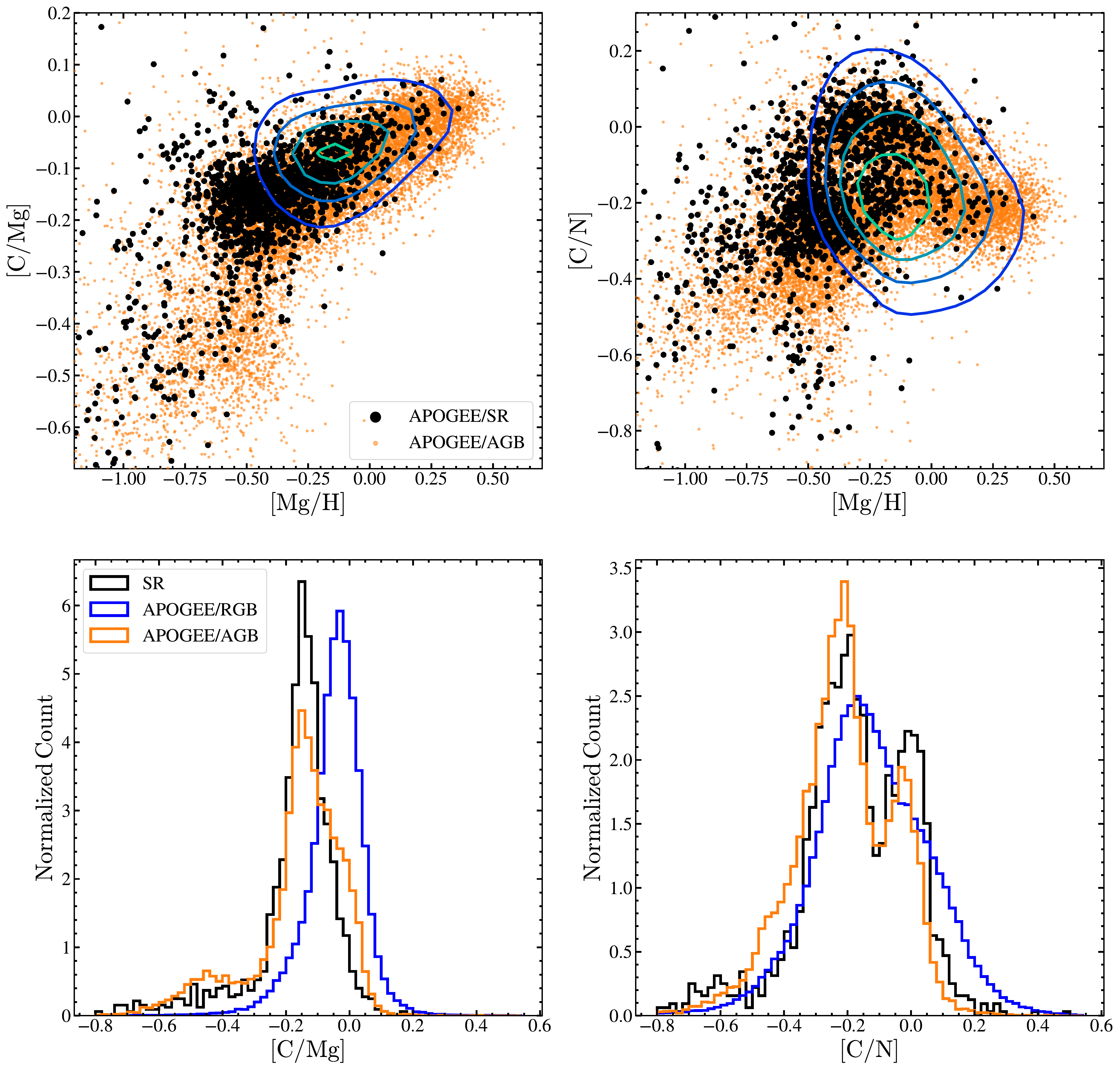}
    \caption{$\rm [C/Mg]$ (top left) and $\rm [C/N]$ (top right) vs. $\rm [Mg/H]$ for the semi-regular variables (black) and APOGEE AGB stars (orange). The distributions of the APOGEE RGB stars in these planes are shown as contours. The distributions of $\rm [C/Mg]$ (bottom left) and $\rm [C/N]$ (bottom right) for the semi-regular variables (black) APOGEE AGB (orange) and RGB (blue) stars are shown as histograms.}
    \label{fig:fig29}
\end{figure*}

We investigate the $\rm [C/Mg]$ and $\rm [C/N]$ abundance ratios for the semi-regular variables, APOGEE AGB and RGB stars in Figure \ref{fig:fig29}. The $\rm[C/Mg]$ abundance ratios for both the semi-regular variables and APOGEE/AGB stars are positively correlated with $\rm [Mg/H]$. The APOGEE/AGB stars and semi-regular variables tend to be carbon-poor when compared to the APOGEE/RGB stars. Only ${\sim}20\%$ of the APOGEE/RGB stars have $\rm [C/Mg]{\lesssim}-0.1$, whereas ${\sim}75\%$ of the semi-regular variables and ${\sim}64\%$ of the APOGEE/AGB stars have $\rm [C/Mg]{\lesssim}-0.1$. The APOGEE/AGB stars form three distinct clusters in the $\rm [C/N]$---$\rm [Mg/H]$ plane at ($-0.4,-0.25$), ($-0.25,0$) and ($0.25,-0.25$). However, the vast majority of the semi-regular variables only populate the two clusters with $\rm [Mg/H]<0$. This is not too surprising as semi-regular variables with $\rm [Mg/H]>0$ are rare in our catalog. The distribution of the APOGEE/RGB stars overlap the metal-rich cluster of APOGEE/AGB stars at ($0.25,-0.25$) that is devoid of semi-regular variables, suggesting that the sample of APOGEE/AGB stars is likely contaminated with stars on the RGB. The distribution of $\rm[C/N]$ is strongly bi-modal for both the semi-regular variables and the APOGEE/AGB stars ($\rm [C/N]{\sim}-0.2$ and $\rm [C/N]{\sim}0$), whereas the APOGEE/RGB stars form a continuous distribution in $\rm[C/N]$, peaking at $\rm [C/N]{\sim}-0.2$. Compared to the APOGEE/RGB stars, there is a small enhancement in the distribution of APOGEE/AGB stars and semi-regular variables with $\rm [C/N]{\lesssim}-0.5$. These are potentially sources with significant N enhancement due to HBB.

\begin{figure*}
	\includegraphics[width=\textwidth]{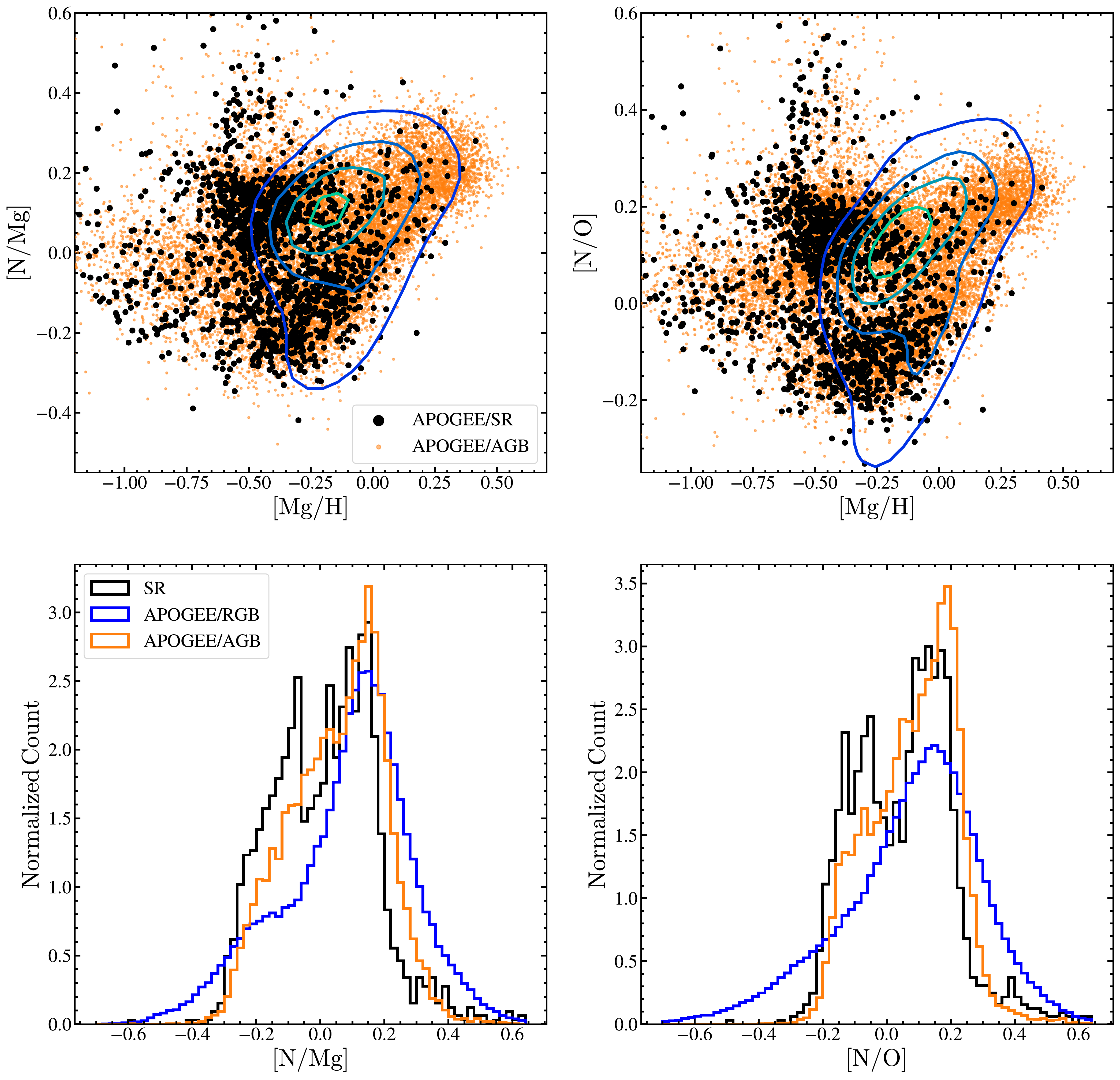}
    \caption{$\rm [N/Mg]$ (top left) and $\rm [N/O]$ (top right) vs. $\rm [Mg/H]$ for the semi-regular variables (black) and APOGEE AGB stars (orange). The distributions of the APOGEE RGB stars in these planes are shown as contours. The distributions of $\rm [N/Mg]$ (bottom left) and $\rm [N/O]$ (bottom right) for the semi-regular variables (black) APOGEE AGB (orange) and RGB (blue) stars are shown as histograms.}
    \label{fig:fig30}
\end{figure*}

In the $\rm [N/Mg]$---$\rm [Mg/H]$ plane (Figure \ref{fig:fig30}), the APOGEE/AGB stars and semi-regular variables form clusters at ($-0.4,0.1$) and ($-0.25,-0.15$). Similarly, in the $\rm [Mg/H]$---$\rm [N/O]$ plane, the APOGEE/AGB stars form three distinct clusters at ($-0.25,-0.1$), ($-0.4,0.15$) and ($0.25,0.15$), with the semi-regular variables populating the two clusters with $\rm [Mg/H]<0$. The distribution of APOGEE/RGB stars in $\rm [N/Mg]$ appears to be bi-modal, with peaks that are similar to those seen for the semi-regular variables and APOGEE/AGB stars. Unlike the APOGEE/RGB stars, the distribution in $\rm [N/O]$ is strongly bi-modal for both the semi-regular variables and APOGEE/AGB stars. We also note the presence of some stars with large $\rm [N/Mg]$ ($\rm [N/Mg]\gtrsim0.25$) and $\rm [N/O]$ ($\rm [N/O]{\gtrsim}0.3$) abundance ratios. These N-rich sources are candidates for intermediate-mass AGB stars where the surface abundances of N are greatly enhanced due to HBB.

Semi-regular variables and APOGEE/AGB stars with $\rm [N/Mg]{\lesssim}-0.3$ are very rare. Similarly, there is a sharp cutoff in $\rm [N/O]$ for both the semi-regular variables and APOGEE/AGB stars at $\rm [N/O]{\sim}-0.3$. Only ${\sim}0.7\%$ of the semi-regular variables and ${\sim}0.5\%$ of the APOGEE/AGB stars have $\rm [N/O]{\lesssim}-0.3$, whereas ${\sim}4.7\%$ of the APOGEE/RGB stars have $\rm [N/O]{\lesssim}-0.3$. While the N abundances are enhanced and the O abundances are depleted due to the CN cycle, the Mg-Al chain also operates during HBB, destroying the Mg isotopes $\rm^{24}Mg$ and $\rm ^{25}Mg$. In a scenario where HBB operates in an AGB star, the surface abundances of $\rm [N/Mg]$ and $\rm [N/O]$ will increase due to N enrichment and Mg/O depletion. Thus, the observed cutoffs in the $\rm [N/Mg]$ and $\rm [N/O]$ abundance ratios could potentially be related to the nucelosynthetic processes that occur during HBB. 

Based on their $\rm [N/O]$ abundance ratios, we group the semi-regular variables into the ``high-N'' group ($\rm 0.02 \leq[N/O]\leq0.3$), ``low-N'' group ($\rm [N/O]<0.02$) and intermediate-mass AGB stars that are candidates for HBB ($\rm [N/O]>0.30$). The various APOGEE DR16 spectroscopic parameters and chemical abundance ratios for these groups are summarized in Table \ref{tab:srspeclowhigh}. 

The HBB candidates with $\rm [N/O]>0.30$ are distinct from the high-N and low-N stars. They have significantly lower $ \rm \log(g){\sim}0.07$, when compared to the high-N ($ \rm \log(g){\sim}0.46$) and low-N ($ \rm \log(g){\sim}0.55$) groups. The HBB candidates are generally hotter than the high-N and low-N stars, with a median temperature of $ \rm T_{eff}\,{\sim}3822\pm194$ K. The HBB candidates have a median period of $ \rm \log_{10}(P/days){\sim}2$, which is consistent with these stars pulsating in the fundamental mode \citep{2017AstL...43..602F}. Much like the high-N stars, the HBB candidates also have low $\rm [\alpha/M]$ abundance ratios ($\rm [\alpha/M]{\sim}-0.04$) consistent with the low-$\alpha$ sequence. The HBB stars have average $\rm [Al/Mg]$ and $\rm [Al/O]$ abundance ratios ($\rm [Al/Mg]{\sim}-0.3$, $\rm [Al/O]{\sim}-0.3$) that are consistent with the low-Al sequence identified in $\S 4.3.4$. Compared to both the high-N and low-N groups, the HBB candidates are poor in Mg and C and they have enhanced N, Ca, Na, P, K, Cr, Mn and Ti abundances compared to both the high-N and low-N groups. The $\rm [C/N]$ abundance ratios of the HBB candidates are significantly different--- these HBB candidates have $\rm [C/N]{\simeq}-0.6\pm0.3$, compared to $\rm [C/N]{\simeq}-0.2\pm0.1$ for the high-N and $\rm [C/N]{\simeq}0.0\pm0.1$ for the low-N stars. 

The high-N and low-N stars have similar periods, temperatures and $ \rm \log(g)$. The slight differences in these measures($ \rm \Delta Teff {\sim}50$ K, $ \rm \Delta \log(g) {\sim}0.1$) might suggest that the high-N stars are somewhat more evolved and hotter than the low-N stars. The high-N stars have $\rm [\alpha/M]$ abundance ratios that are consistent with them being on the low-$\alpha$ sequence, whereas the low-N stars have $\rm [\alpha/M]$ consistent with them being high-$\alpha$ stars. The low-N stars have larger Mg abundances ($\rm \Delta [Mg/Fe] {\sim}0.2$ dex) than the high-N stars. The $\rm [Al/Mg]$ and $\rm [Al/O]$ abundance ratios of the low-N stars are lower than that of the high-N stars and possibly indicate that these stars are undergoing mass loss. The high-N stars have enhanced N, Ca, Na, P, K, Cr, V, Co, Mn, and Ni abundances when compared to the low-N stars.

In summary, we identified a ``high-N'' group ($\rm 0.02 \leq[N/O]\leq0.3$), and a ``low-N'' group ($\rm [N/O]<0.02$) amongst the SR variables. The high-N stars differ in their chemical profiles when compared to the low-N stars. We also identified a sample of N-enhanced, likely intermediate-mass AGB stars ($\rm [N/O]>0.30$) going through significant hot-bottom burning (HBB). We also find that the HBB candidates are carbon poor with $\rm [C/N]{\sim}-0.6$.
   

\clearpage
\clearpage
\section{Conclusions}

We systematically searched for variable sources in ASAS-SN with $V<17$ mag in the $V$-band light curves of ${\sim}61.5$ million sources. Through our search which first began in 2018, we identified ${\sim}426,000$ variable sources, of which ${\sim}219,000$ are new discoveries. The V-band light curves of all the ${\sim}61.5$M sources studied in this work are available online at the ASAS-SN Photometry Database (\url{https://asas-sn.osu.edu/photometry}). The $V$-band light curves and other information on the variable stars identified in our work are available on the ASAS-SN variable stars database (\url{https://asas-sn.osu.edu/variables}). ASAS-SN has significantly improved the census of semi-regular/irregular variables ($+235\%$), $\delta$ Scuti variables ($+81\%$), rotational variables ($+116\%$) and detached eclipsing binaries ($+90\%$) with $V<17$ mag. Most (${\sim}74\%$) of our discoveries were in the Southern hemisphere. 

Due to the overlap between ASAS-SN and modern major wide-field spectroscopic surveys, we are able to utilize spectroscopic information to closely study various variable types. We cross-matched our catalog of variables with the APOGEE DR16 catalog, the RAVE-on catalog, the LAMOST DR5 v4 catalog and the GALAH DR2 catalog, and identified ${\sim}39,000$ unique cross-matches. We find that data from the LAMOST and RAVE surveys are best suited to the characterization of pulsators, eclipsing binaries and rotational variables. APOGEE data is excellent for the characterization of the cooler semi-regular and irregular variables. Our main results for eclipsing binaries, rotational variables and semi-regular variables are summarized below.\newline

\textbf{Eclipsing binaries:}
\begin{enumerate}
  \item EW type contact binaries are significantly cooler ($\rm T_{eff}{\sim}5900$ K) than both the semi-detached EB binaries ($\rm T_{eff}{\sim}6700$ K) and the detached EA systems ($\rm T_{eff}{\sim}6300$ K). 
  \item There is significant overlap in period---temperature space between the early-type EW and EB binaries. This overlap is consistent with the predictions of the thermal relaxation oscillation (TRO) models for contact binaries, where systems can transition between the semi-detached and contact phases.
  \item Most (${\sim}63\%$) eclipsing binaries have metallicities $\rm-0.5<[Fe/H]<0$. This is consistent with recent findings of higher binary fractions at lower metallicity \citep{2018ApJ...854..147B,2019ApJ...875...61M,2019MNRAS.482L.139E}.
  \item The period---temperature distributions depend on metallicity, with lower metallicity binaries having shorter periods at fixed temperature.
  \item Contact binaries have an observed period-temperature relationship that falls below that of the semi-detached and detached binaries at any given temperature and tracks Roche expectations.

\end{enumerate}

   %ROT

\textbf{Rotational variables:}
\begin{enumerate}   
   \item We find rotational variables on the MS/pre-MS ($\rm \log (g){\sim}4.5$, $\rm M_{Ks}{\sim}4$ mag), the base of the red-giant branch ($\rm \log (g){\sim}3.5$, $\rm M_{Ks}{\sim}1$ mag) and the red clump ($\rm \log (g){\sim}2.6$, $\rm M_{Ks}{\sim}-1.5$ mag).
   \item ${\sim}81\%$ of the rotating giants with spectroscopic $v\sin(i)$ were rapid rotators with $v\sin(i)>10 \rm \, km/s$.    
   \item ${\sim}80\%$ of the giants have $v_{\rm rot}>10 \rm \, km/s$, consistent with the estimate of rapid rotators with spectroscopic $v\sin(i)$. 
   \item ${\sim}98\%$ of the rotational variables on the red clump have $v_{\rm rot}>10 \rm \, km/s$. 
   \item ${\sim}30\%$ of the rapidly rotating red clump stars are metal-poor with $\rm [Fe/H]<-0.5$, whereas only ${\sim}8\%$ are metal-rich with $\rm [Fe/H]>0$. 
   \item ${\sim}87\%$ of the rotating giants had NUV excesses from \citet{2020arXiv200500577D} consistent with $v\sin(i)>10 \rm \, km/s$. This is consistent with our other estimates of the fraction of sources with rapid rotation ($v\sin(i)>10 \rm \, km/s$). 
   \item Evolved rotating giants have larger amplitudes than main-sequence rotational variables at any given period, with the variability amplitudes peaking at $\rm \log(g){\sim}3.2$.    
   \item The NUV excesses of these rotating giants follow a similar trend with period to the work of \citet{2020arXiv200500577D}. 
   \item The rotational variables in APOGEE are low-$\alpha$ stars strongly clustered towards $\rm [Mg/Fe]{\sim}-0.1$, and $\rm [Fe/H]{\sim}-0.1$, distinct from the typical giant stars or the SR variables and AGB stars. These abundances are unusual and suggestive of systematic issues in the ASPCAP pipeline when dealing with rapidly rotating stars.

\end{enumerate}   
   %SR

\textbf{Semi-regular variables and Miras:}  
\begin{enumerate}   
   \item Semi-regular variables tend to be lower metallicity ($\rm [Fe/H]{\sim}-0.5$) than most giant stars.
   \item Of the semi-regular variables with $\rm \log_{10}(P/days) \gtrsim 2.3$, ${\sim}36\%$ were LSPs. ${\sim}60\%$ of the LSPs had pulsational periods in the range $1.5 \lesssim \rm \log_{10}(P/days) \lesssim 1.9$. The LSP was on average ${\sim}11$ times longer than the actual pulsation period.
   \item Many semi-regular variables have LSPs, and while it seems clear that the shorter period rather than the LSP is more closely related to the physical properties of the star, we could identify no spectroscopic property ($\rm T_{eff}$, $ \rm \log(g)$, abundances etc.) which distinguished SR variables with LSPs from those without.
   \item The vast majority of the ASAS-SN Mira (${\sim}97\%$) and semi-regular variables (${\sim}95\%$) are oxygen-rich AGB stars.
   \item We fit a temperature---$W_{RP}-W_{JK}$ relation that returns temperatures that are, on average, within $\pm0.7\%$ of the APOGEE temperatures for the oxygen-rich SR variables with $\rm T_{eff}<3800$ K and $-0.7 {\, \rm mag} \leq W_{RP}-W_{JK}\leq 0.8 {\, \rm mag}$. We tested an extrapolation of this temperature---$W_{RP}-W_{JK}$ relation to estimate the temperatures of the Mira variables with $W_{RP}-W_{JK}<-0.7$ mag. We find the results of our extrapolation consistent with the expected temperatures of Miras \citep{2017ApJS..232...16Y}.
   \item  The peak shifts in the $\rm[Mg/Fe]$, and $\rm[\alpha/M]$ abundance ratios for the high-$\alpha$ and low-$\alpha$ stars relative to the APOGEE giants indicate possible $\alpha$-enhancements due to the effects of the third dredge up in AGB stars.
   \item The $\rm [Al/X]$ abundance ratios of the semi-regular variables are correlated with the pulsation period. These abundance ratios decrease with period and plateau beyond the canonical period of $P{\sim}60$ days where dust formation first occurs. 
   \item We identified a ``high-Al'' and a ``low-Al'' sequence amongst the SR variables. The Al depleted ``low-Al'' sequence likely corresponds to AGB stars with significant dust production and mass loss. Cuts of $\rm [Al/Mg]=-0.25$, $\rm [Al/C]=-0.1$, $\rm [Al/O]=-0.25$ or $\rm [Al/Si]=-0.13$ roughly separate the high-Al and low-Al sequences.
   \item We identified a ``high-N'' group ($\rm 0.02 \leq[N/O]\leq0.3$), and a ``low-N'' group ($\rm [N/O]<0.02$) amongst the SR variables. The high-N stars have enhanced Ca, N, Na, P, K, Cr, V, Co, Mn, and Ni abundances when compared to the low-N stars.
   \item We identified a sample of likely intermediate-mass AGB stars going through significant hot-bottom burning (HBB) using the cutoff $\rm [N/O]>0.30$. The HBB candidates are poor in Mg and C ($\rm [C/N]{\sim}-0.6$), but have enhanced N, Ca, Na, P, K, Cr, Mn and Ti abundances compared to both the high-N and low-N groups.

\end{enumerate}
\vspace{20 mm}
This work is a first exploration of the powerful synergy between wide-field photometric and spectroscopic surveys towards deciphering the various properties of variable stars. As the number of stars with spectroscopic measurements increases by orders of magnitude over the coming years, it will be possible to study much larger samples of variable stars in greater detail. While we have not examined it here, larger samples with models of selection effects can be used to examine the detailed fractions of variable types as a function of their spectroscopic properties, similar to the recent studies of the dependence of binary fraction on metallicity \citep{2018ApJ...854..147B,2019ApJ...875...61M,2019MNRAS.482L.139E}. 

\section*{Acknowledgements}

This work is dedicated to the memory of our friend and colleague David Will, whose enduring support enabled the success of the ASAS-SN project.

We thank the anonymous referee for the very useful comments that improved our presentation of this work. We thank Dr. Jamie Tayar for her help with interpreting the APOGEE data used in this analysis. We thank the Las Cumbres Observatory and its staff for its
continuing support of the ASAS-SN project. We also thank the Ohio State University College of Arts and Sciences Technology Services for helping us set up and maintain the ASAS-SN variable stars and photometry databases.

ASAS-SN is supported by the Gordon and Betty Moore
Foundation through grant GBMF5490 to the Ohio State
University, and NSF grants AST-1515927 and AST-1908570. Development of
ASAS-SN has been supported by NSF grant AST-0908816,
the Mt. Cuba Astronomical Foundation, the Center for Cosmology 
and AstroParticle Physics at the Ohio State University, 
the Chinese Academy of Sciences South America Center
for Astronomy (CAS- SACA), the Villum Foundation, and
George Skestos. 

KZS and CSK are supported by NSF grants AST-1515927, AST-1814440, and 
AST-1908570. BJS is supported by NSF grants AST-1908952, AST-1920392, 
and AST-1911074. TAT acknowledges support from a Simons Foundation 
Fellowship and from an IBM Einstein Fellowship from the Institute for 
Advanced Study, Princeton. Support for JLP is provided in part by the
Ministry of Economy, Development, and Tourism's Millennium Science 
Initiative through grant IC120009, awarded to The Millennium Institute 
of Astrophysics, MAS. Support for MP and OP has been provided by 
INTER-EXCELLENCE grant LTAUSA18093 from the Czech Ministry of 
Education, Youth, and Sports. The research of OP has also been 
supported by Horizon 2020 ERC Starting Grant ``Cat-In-hAT'' 
(grant agreement \#803158) and PRIMUS/SCI/17 award from Charles 
University. This work was partly supported by NSFC 11721303.

Funding for the Sloan Digital Sky Survey IV has been provided by the Alfred P. Sloan Foundation, the U.S. Department of Energy Office of Science, and the Participating Institutions. SDSS-IV acknowledges
support and resources from the Center for High-Performance Computing at
the University of Utah. The SDSS web site is www.sdss.org.

SDSS-IV is managed by the Astrophysical Research Consortium for the 
Participating Institutions of the SDSS Collaboration including the 
Brazilian Participation Group, the Carnegie Institution for Science, 
Carnegie Mellon University, the Chilean Participation Group, the French Participation Group, Harvard-Smithsonian Center for Astrophysics, 
Instituto de Astrof\'isica de Canarias, The Johns Hopkins University, Kavli Institute for the Physics and Mathematics of the Universe (IPMU) / 
University of Tokyo, the Korean Participation Group, Lawrence Berkeley National Laboratory, 
Leibniz Institut f\"ur Astrophysik Potsdam (AIP),  
Max-Planck-Institut f\"ur Astronomie (MPIA Heidelberg), 
Max-Planck-Institut f\"ur Astrophysik (MPA Garching), 
Max-Planck-Institut f\"ur Extraterrestrische Physik (MPE), 
National Astronomical Observatories of China, New Mexico State University, 
New York University, University of Notre Dame, 
Observat\'ario Nacional / MCTI, The Ohio State University, 
Pennsylvania State University, Shanghai Astronomical Observatory, 
United Kingdom Participation Group,
Universidad Nacional Aut\'onoma de M\'exico, University of Arizona, 
University of Colorado Boulder, University of Oxford, University of Portsmouth, 
University of Utah, University of Virginia, University of Washington, University of Wisconsin, 
Vanderbilt University, and Yale University.

This work has made use of data from the European Space Agency (ESA)
mission {\it Gaia} (\url{https://www.cosmos.esa.int/gaia}), processed by
the {\it Gaia} Data Processing and Analysis Consortium. This publication makes 
use of data products from the Two Micron All Sky Survey, as well as
data products from the Wide-field Infrared Survey Explorer.
This research was also made possible through the use of the AAVSO Photometric 
All-Sky Survey (APASS), funded by the Robert Martin Ayers Sciences Fund.

This research has made use of the VizieR catalogue access tool, CDS, Strasbourg, France. 
This research also made use of Astropy, a community-developed core Python package for 
Astronomy (Astropy Collaboration, 2013).

\section*{Data Availability}

The ASAS-SN photometric data underlying this article are available on the ASAS-SN Photometry Database (\url{https://asas-sn.osu.edu/photometry}) and the ASAS-SN variable stars database (\url{https://asas-sn.osu.edu/variables}). The external photometric data underlying this article were accessed from sources in the public domain: {\it Gaia} (\url{https://www.cosmos.esa.int/gaia}), 2MASS (\url{https://old.ipac.caltech.edu/2mass/overview/access.html}), AllWISE (\url{http://wise2.ipac.caltech.edu/docs/release/allwise/}) and \textit{GALEX} (\url{https://archive.stsci.edu/missions-and-data/galex-1/}). The spectroscopic datasets underlying this article were accessed from sources in the public domain: APOGEE (\url{https://www.sdss.org/dr16/}), LAMOST (\url{http://dr5.lamost.org/}), GALAH (\url{https://galah-survey.org/publicdata/}) and RAVE (\url{https://www.rave-survey.org/}).

%%%%%%%%%%%%%%%%%%%%%%%%%%%%%%%%%%%%%%%%%%%%%%%%%%



\bibliographystyle{mnras}
\bibliography{ref} % if your bibtex file is called example.bib

\begin{thebibliography}{}
\makeatletter
\relax
\def\mn@urlcharsother{\let\do\@makeother \do\$\do\&\do\#\do\^\do\_\do\%\do\~}
\def\mn@doi{\begingroup\mn@urlcharsother \@ifnextchar [ {\mn@doi@}
  {\mn@doi@[]}}
\def\mn@doi@[#1]#2{\def\@tempa{#1}\ifx\@tempa\@empty \href
  {http://dx.doi.org/#2} {doi:#2}\else \href {http://dx.doi.org/#2} {#1}\fi
  \endgroup}
\def\mn@eprint#1#2{\mn@eprint@#1:#2::\@nil}
\def\mn@eprint@arXiv#1{\href {http://arxiv.org/abs/#1} {{\tt arXiv:#1}}}
\def\mn@eprint@dblp#1{\href {http://dblp.uni-trier.de/rec/bibtex/#1.xml}
  {dblp:#1}}
\def\mn@eprint@#1:#2:#3:#4\@nil{\def\@tempa {#1}\def\@tempb {#2}\def\@tempc
  {#3}\ifx \@tempc \@empty \let \@tempc \@tempb \let \@tempb \@tempa \fi \ifx
  \@tempb \@empty \def\@tempb {arXiv}\fi \@ifundefined
  {mn@eprint@\@tempb}{\@tempb:\@tempc}{\expandafter \expandafter \csname
  mn@eprint@\@tempb\endcsname \expandafter{\@tempc}}}

\bibitem[\protect\citeauthoryear{{Ahumada} et~al.,}{{Ahumada}
  et~al.}{2019}]{2019arXiv191202905A}
{Ahumada} R.,  et~al., 2019, arXiv e-prints, \href
  {https://ui.adsabs.harvard.edu/abs/2019arXiv191202905A} {p. arXiv:1912.02905}

\bibitem[\protect\citeauthoryear{{Alard}}{{Alard}}{2000}]{2000A&AS..144..363A}
{Alard} C.,  2000, \mn@doi [\aaps] {10.1051/aas:2000214}, \href
  {https://ui.adsabs.harvard.edu/abs/2000A&AS..144..363A} {144, 363}

\bibitem[\protect\citeauthoryear{{Alard} \& {Lupton}}{{Alard} \&
  {Lupton}}{1998}]{1998ApJ...503..325A}
{Alard} C.,  {Lupton} R.~H.,  1998, \mn@doi [\apj] {10.1086/305984}, \href
  {https://ui.adsabs.harvard.edu/abs/1998ApJ...503..325A} {503, 325}

\bibitem[\protect\citeauthoryear{{Alcock} et~al.,}{{Alcock}
  et~al.}{1997}]{1997ApJ...486..697A}
{Alcock} C.,  et~al., 1997, \mn@doi [\apj] {10.1086/304535}, \href
  {https://ui.adsabs.harvard.edu/abs/1997ApJ...486..697A} {486, 697}

\bibitem[\protect\citeauthoryear{{Alksnis}, {Balklavs}, {Dzervitis}, {Eglitis},
  {Paupers}  \& {Pundure}}{{Alksnis} et~al.}{2001}]{2001BaltA..10....1A}
{Alksnis} A.,  {Balklavs} A.,  {Dzervitis} U.,  {Eglitis} I.,  {Paupers} O.,
  {Pundure} I.,  2001, \mn@doi [Baltic Astronomy] {10.1515/astro-2001-1-202},
  \href {https://ui.adsabs.harvard.edu/abs/2001BaltA..10....1A} {10, 1}

\bibitem[\protect\citeauthoryear{{Arnold} et~al.,}{{Arnold}
  et~al.}{2020}]{2020ApJS..247...44A}
{Arnold} R.~A.,  et~al., 2020, \mn@doi [\apjs] {10.3847/1538-4365/ab6bdb},
  \href {https://ui.adsabs.harvard.edu/abs/2020ApJS..247...44A} {247, 44}

\bibitem[\protect\citeauthoryear{{Auge} et~al.,}{{Auge}
  et~al.}{2020}]{2020arXiv200305459A}
{Auge} C.,  et~al., 2020, arXiv e-prints, \href
  {https://ui.adsabs.harvard.edu/abs/2020arXiv200305459A} {p. arXiv:2003.05459}

\bibitem[\protect\citeauthoryear{{Badenes} et~al.,}{{Badenes}
  et~al.}{2018}]{2018ApJ...854..147B}
{Badenes} C.,  et~al., 2018, \mn@doi [\apj] {10.3847/1538-4357/aaa765}, \href
  {https://ui.adsabs.harvard.edu/abs/2018ApJ...854..147B} {854, 147}

\bibitem[\protect\citeauthoryear{{Bailer-Jones}, {Rybizki}, {Fouesneau},
  {Mantelet}  \& {Andrae}}{{Bailer-Jones} et~al.}{2018}]{2018AJ....156...58B}
{Bailer-Jones} C.~A.~L.,  {Rybizki} J.,  {Fouesneau} M.,  {Mantelet} G.,
  {Andrae} R.,  2018, \mn@doi [\aj] {10.3847/1538-3881/aacb21}, \href
  {https://ui.adsabs.harvard.edu/abs/2018AJ....156...58B} {156, 58}

\bibitem[\protect\citeauthoryear{{Banerjee}, {Varricatt}, {Mathew}, {Launila}
  \& {Ashok}}{{Banerjee} et~al.}{2012}]{2012ApJ...753L..20B}
{Banerjee} D.~P.~K.,  {Varricatt} W.~P.,  {Mathew} B.,  {Launila} O.,   {Ashok}
  N.~M.,  2012, \mn@doi [\apjl] {10.1088/2041-8205/753/1/L20}, \href
  {https://ui.adsabs.harvard.edu/abs/2012ApJ...753L..20B} {753, L20}

\bibitem[\protect\citeauthoryear{{Beaton} et~al.,}{{Beaton}
  et~al.}{2018}]{2018SSRv..214..113B}
{Beaton} R.~L.,  et~al., 2018, \mn@doi [\ssr] {10.1007/s11214-018-0542-1},
  \href {https://ui.adsabs.harvard.edu/abs/2018SSRv..214..113B} {214, 113}

\bibitem[\protect\citeauthoryear{{Bellm} et~al.,}{{Bellm}
  et~al.}{2019}]{2019PASP..131a8002B}
{Bellm} E.~C.,  et~al., 2019, \mn@doi [\pasp] {10.1088/1538-3873/aaecbe}, \href
  {https://ui.adsabs.harvard.edu/abs/2019PASP..131a8002B} {131, 018002}

\bibitem[\protect\citeauthoryear{{Bhatti}, {Bouma}, {Joshua}, {John}  \&
  {Price-Whelan}}{{Bhatti} et~al.}{2018}]{2018zndo...1469822B}
{Bhatti} W.,  {Bouma} L.,  {Joshua} {John}  {Price-Whelan} A.,  2018,
  {Waqasbhatti/Astrobase: Astrobase V0.3.20}, \mn@doi{10.5281/zenodo.1469822}

\bibitem[\protect\citeauthoryear{{Bianchi}, {Shiao}  \& {Thilker}}{{Bianchi}
  et~al.}{2017}]{2017ApJS..230...24B}
{Bianchi} L.,  {Shiao} B.,   {Thilker} D.,  2017, \mn@doi [\apjs]
  {10.3847/1538-4365/aa7053}, \href
  {https://ui.adsabs.harvard.edu/abs/2017ApJS..230...24B} {230, 24}

\bibitem[\protect\citeauthoryear{{Blanton} et~al.,}{{Blanton}
  et~al.}{2017}]{2017AJ....154...28B}
{Blanton} M.~R.,  et~al., 2017, \mn@doi [\aj] {10.3847/1538-3881/aa7567}, \href
  {https://ui.adsabs.harvard.edu/abs/2017AJ....154...28B} {154, 28}

\bibitem[\protect\citeauthoryear{{Boothroyd} \& {Sackmann}}{{Boothroyd} \&
  {Sackmann}}{1992}]{1992ApJ...393L..21B}
{Boothroyd} A.~I.,  {Sackmann} I.~J.,  1992, \mn@doi [\apjl] {10.1086/186441},
  \href {https://ui.adsabs.harvard.edu/abs/1992ApJ...393L..21B} {393, L21}

\bibitem[\protect\citeauthoryear{{Bredall} et~al.,}{{Bredall}
  et~al.}{2020}]{2020arXiv200514201B}
{Bredall} J.~W.,  et~al., 2020, arXiv e-prints, \href
  {https://ui.adsabs.harvard.edu/abs/2020arXiv200514201B} {p. arXiv:2005.14201}

\bibitem[\protect\citeauthoryear{{Brown} et~al.,}{{Brown}
  et~al.}{2013}]{2013PASP..125.1031B}
{Brown} T.~M.,  et~al., 2013, \mn@doi [\pasp] {10.1086/673168}, \href
  {https://ui.adsabs.harvard.edu/abs/2013PASP..125.1031B} {125, 1031}

\bibitem[\protect\citeauthoryear{{Buder} et~al.,}{{Buder}
  et~al.}{2018}]{2018MNRAS.478.4513B}
{Buder} S.,  et~al., 2018, \mn@doi [\mnras] {10.1093/mnras/sty1281}, \href
  {https://ui.adsabs.harvard.edu/abs/2018MNRAS.478.4513B} {478, 4513}

\bibitem[\protect\citeauthoryear{{Buzzoni}, {Patelli}, {Bellazzini}, {Pecci}
  \& {Oliva}}{{Buzzoni} et~al.}{2010}]{2010MNRAS.403.1592B}
{Buzzoni} A.,  {Patelli} L.,  {Bellazzini} M.,  {Pecci} F.~F.,   {Oliva} E.,
  2010, \mn@doi [\mnras] {10.1111/j.1365-2966.2009.16223.x}, \href
  {https://ui.adsabs.harvard.edu/abs/2010MNRAS.403.1592B} {403, 1592}

\bibitem[\protect\citeauthoryear{{Casey} et~al.,}{{Casey}
  et~al.}{2017}]{2017ApJ...840...59C}
{Casey} A.~R.,  et~al., 2017, \mn@doi [\apj] {10.3847/1538-4357/aa69c2}, \href
  {https://ui.adsabs.harvard.edu/abs/2017ApJ...840...59C} {840, 59}

\bibitem[\protect\citeauthoryear{{Ceillier} et~al.,}{{Ceillier}
  et~al.}{2017}]{2017A&A...605A.111C}
{Ceillier} T.,  et~al., 2017, \mn@doi [\aap] {10.1051/0004-6361/201629884},
  \href {https://ui.adsabs.harvard.edu/abs/2017A&A...605A.111C} {605, A111}

\bibitem[\protect\citeauthoryear{{Chen}, {Wang}, {Deng}, {de Grijs}  \&
  {Yang}}{{Chen} et~al.}{2018}]{2018ApJS..237...28C}
{Chen} X.,  {Wang} S.,  {Deng} L.,  {de Grijs} R.,   {Yang} M.,  2018, \mn@doi
  [\apjs] {10.3847/1538-4365/aad32b}, \href
  {https://ui.adsabs.harvard.edu/abs/2018ApJS..237...28C} {237, 28}

\bibitem[\protect\citeauthoryear{{Chen}, {Wang}, {Deng}, {de Grijs}, {Yang}  \&
  {Tian}}{{Chen} et~al.}{2020}]{2020arXiv200508662C}
{Chen} X.,  {Wang} S.,  {Deng} L.,  {de Grijs} R.,  {Yang} M.,   {Tian} H.,
  2020, arXiv e-prints, \href
  {https://ui.adsabs.harvard.edu/abs/2020arXiv200508662C} {p. arXiv:2005.08662}

\bibitem[\protect\citeauthoryear{{Choi}, {Dotter}, {Conroy}, {Cantiello},
  {Paxton}  \& {Johnson}}{{Choi} et~al.}{2016}]{2016ApJ...823..102C}
{Choi} J.,  {Dotter} A.,  {Conroy} C.,  {Cantiello} M.,  {Paxton} B.,
  {Johnson} B.~D.,  2016, \mn@doi [\apj] {10.3847/0004-637X/823/2/102}, \href
  {https://ui.adsabs.harvard.edu/abs/2016ApJ...823..102C} {823, 102}

\bibitem[\protect\citeauthoryear{{Cristallo} et~al.,}{{Cristallo}
  et~al.}{2011}]{2011ApJS..197...17C}
{Cristallo} S.,  et~al., 2011, \mn@doi [\apjs] {10.1088/0067-0049/197/2/17},
  \href {https://ui.adsabs.harvard.edu/abs/2011ApJS..197...17C} {197, 17}

\bibitem[\protect\citeauthoryear{{Cristallo}, {Straniero}, {Piersanti}  \&
  {Gobrecht}}{{Cristallo} et~al.}{2015}]{2015ApJS..219...40C}
{Cristallo} S.,  {Straniero} O.,  {Piersanti} L.,   {Gobrecht} D.,  2015,
  \mn@doi [\apjs] {10.1088/0067-0049/219/2/40}, \href
  {https://ui.adsabs.harvard.edu/abs/2015ApJS..219...40C} {219, 40}

\bibitem[\protect\citeauthoryear{{Cui} et~al.,}{{Cui}
  et~al.}{2012}]{2012RAA....12.1197C}
{Cui} X.-Q.,  et~al., 2012, \mn@doi [Research in Astronomy and Astrophysics]
  {10.1088/1674-4527/12/9/003}, \href
  {https://ui.adsabs.harvard.edu/abs/2012RAA....12.1197C} {12, 1197}

\bibitem[\protect\citeauthoryear{{Cutri} \& {et al.}}{{Cutri} \& {et
  al.}}{2013}]{2013yCat.2328....0C}
{Cutri} R.~M.,  {et al.} 2013, VizieR Online Data Catalog, \href
  {https://ui.adsabs.harvard.edu/abs/2013yCat.2328....0C} {p. II/328}

\bibitem[\protect\citeauthoryear{{De Beck}, {Decin}, {Ramstedt}, {Olofsson},
  {Menten}, {Patel}  \& {Vlemmings}}{{De Beck}
  et~al.}{2017}]{2017A&A...598A..53D}
{De Beck} E.,  {Decin} L.,  {Ramstedt} S.,  {Olofsson} H.,  {Menten} K.~M.,
  {Patel} N.~A.,   {Vlemmings} W.~H.~T.,  2017, \mn@doi [\aap]
  {10.1051/0004-6361/201628928}, \href
  {https://ui.adsabs.harvard.edu/abs/2017A&A...598A..53D} {598, A53}

\bibitem[\protect\citeauthoryear{{De Silva} et~al.,}{{De Silva}
  et~al.}{2015}]{2015MNRAS.449.2604D}
{De Silva} G.~M.,  et~al., 2015, \mn@doi [\mnras] {10.1093/mnras/stv327}, \href
  {https://ui.adsabs.harvard.edu/abs/2015MNRAS.449.2604D} {449, 2604}

\bibitem[\protect\citeauthoryear{{Decin} et~al.,}{{Decin}
  et~al.}{2017}]{2017A&A...608A..55D}
{Decin} L.,  et~al., 2017, \mn@doi [\aap] {10.1051/0004-6361/201730782}, \href
  {https://ui.adsabs.harvard.edu/abs/2017A&A...608A..55D} {608, A55}

\bibitem[\protect\citeauthoryear{{Derue} et~al.,}{{Derue}
  et~al.}{2002}]{2002A&A...389..149D}
{Derue} F.,  et~al., 2002, \mn@doi [\aap] {10.1051/0004-6361:20020570}, \href
  {https://ui.adsabs.harvard.edu/abs/2002A&A...389..149D} {389, 149}

\bibitem[\protect\citeauthoryear{{Dixon}, {Tayar}  \& {Stassun}}{{Dixon}
  et~al.}{2020}]{2020arXiv200500577D}
{Dixon} D.,  {Tayar} J.,   {Stassun} K.~G.,  2020, arXiv e-prints, \href
  {https://ui.adsabs.harvard.edu/abs/2020arXiv200500577D} {p. arXiv:2005.00577}

\bibitem[\protect\citeauthoryear{{Dotter}}{{Dotter}}{2016}]{2016ApJS..222....8D}
{Dotter} A.,  2016, \mn@doi [\apjs] {10.3847/0067-0049/222/1/8}, \href
  {https://ui.adsabs.harvard.edu/abs/2016ApJS..222....8D} {222, 8}

\bibitem[\protect\citeauthoryear{{Drake} et~al.,}{{Drake}
  et~al.}{2014}]{2014ApJS..213....9D}
{Drake} A.~J.,  et~al., 2014, \mn@doi [\apjs] {10.1088/0067-0049/213/1/9},
  \href {https://ui.adsabs.harvard.edu/abs/2014ApJS..213....9D} {213, 9}

\bibitem[\protect\citeauthoryear{{Eggleton}}{{Eggleton}}{1983}]{1983ApJ...268..368E}
{Eggleton} P.~P.,  1983, \mn@doi [\apj] {10.1086/160960}, \href
  {https://ui.adsabs.harvard.edu/abs/1983ApJ...268..368E} {268, 368}

\bibitem[\protect\citeauthoryear{{El-Badry} \& {Rix}}{{El-Badry} \&
  {Rix}}{2019}]{2019MNRAS.482L.139E}
{El-Badry} K.,  {Rix} H.-W.,  2019, \mn@doi [\mnras] {10.1093/mnrasl/sly206},
  \href {https://ui.adsabs.harvard.edu/abs/2019MNRAS.482L.139E} {482, L139}

\bibitem[\protect\citeauthoryear{{Fadeyev}}{{Fadeyev}}{2017}]{2017AstL...43..602F}
{Fadeyev} Y.~A.,  2017, \mn@doi [Astronomy Letters]
  {10.1134/S1063773717080059}, \href
  {https://ui.adsabs.harvard.edu/abs/2017AstL...43..602F} {43, 602}

\bibitem[\protect\citeauthoryear{{Feast} \& {Whitelock}}{{Feast} \&
  {Whitelock}}{2014}]{2014IAUS..298...40F}
{Feast} M.,  {Whitelock} P.~A.,  2014, in {Feltzing} S.,  {Zhao} G.,  {Walton}
  N.~A.,   {Whitelock} P.,  eds,  IAU Symposium Vol. 298, Setting the scene for
  Gaia and LAMOST. pp 40--52 (\mn@eprint {arXiv} {1310.3928}),
  \mn@doi{10.1017/S1743921313006182}

\bibitem[\protect\citeauthoryear{{Flannery}}{{Flannery}}{1976}]{1976ApJ...205..217F}
{Flannery} B.~P.,  1976, \mn@doi [\apj] {10.1086/154266}, \href
  {https://ui.adsabs.harvard.edu/abs/1976ApJ...205..217F} {205, 217}

\bibitem[\protect\citeauthoryear{{Fraser}, {Hawley}, {Cook}  \&
  {Keller}}{{Fraser} et~al.}{2005}]{2005AJ....129..768F}
{Fraser} O.~J.,  {Hawley} S.~L.,  {Cook} K.~H.,   {Keller} S.~C.,  2005,
  \mn@doi [\aj] {10.1086/426749}, \href
  {https://ui.adsabs.harvard.edu/abs/2005AJ....129..768F} {129, 768}

\bibitem[\protect\citeauthoryear{{Frost} \& {Lattanzio}}{{Frost} \&
  {Lattanzio}}{1996}]{1996ApJ...473..383F}
{Frost} C.~A.,  {Lattanzio} J.~C.,  1996, \mn@doi [\apj] {10.1086/178152},
  \href {https://ui.adsabs.harvard.edu/abs/1996ApJ...473..383F} {473, 383}

\bibitem[\protect\citeauthoryear{{Fuhrmann}}{{Fuhrmann}}{1998}]{1998A&A...338..161F}
{Fuhrmann} K.,  1998, \aap, \href
  {https://ui.adsabs.harvard.edu/abs/1998A&A...338..161F} {338, 161}

\bibitem[\protect\citeauthoryear{{Gaia Collaboration} et~al.,}{{Gaia
  Collaboration} et~al.}{2018}]{2018A&A...616A...1G}
{Gaia Collaboration} et~al., 2018, \mn@doi [\aap]
  {10.1051/0004-6361/201833051}, \href
  {https://ui.adsabs.harvard.edu/abs/2018A&A...616A...1G} {616, A1}

\bibitem[\protect\citeauthoryear{{Gaia Collaboration} et~al.,}{{Gaia
  Collaboration} et~al.}{2019}]{2019A&A...623A.110G}
{Gaia Collaboration} et~al., 2019, \mn@doi [\aap]
  {10.1051/0004-6361/201833304}, \href
  {https://ui.adsabs.harvard.edu/abs/2019A&A...623A.110G} {623, A110}

\bibitem[\protect\citeauthoryear{{Gail}, {Wetzel}, {Pucci}  \&
  {Tamanai}}{{Gail} et~al.}{2013}]{2013A&A...555A.119G}
{Gail} H.~P.,  {Wetzel} S.,  {Pucci} A.,   {Tamanai} A.,  2013, \mn@doi [\aap]
  {10.1051/0004-6361/201321807}, \href
  {https://ui.adsabs.harvard.edu/abs/2013A&A...555A.119G} {555, A119}

\bibitem[\protect\citeauthoryear{{Garc{\'\i}a-Hern{\'a}ndez}, {Zamora},
  {Yag{\"u}e}, {Uttenthaler}, {Karakas}, {Lugaro}, {Ventura}  \&
  {Lambert}}{{Garc{\'\i}a-Hern{\'a}ndez} et~al.}{2013}]{2013A&A...555L...3G}
{Garc{\'\i}a-Hern{\'a}ndez} D.~A.,  {Zamora} O.,  {Yag{\"u}e} A.,
  {Uttenthaler} S.,  {Karakas} A.~I.,  {Lugaro} M.,  {Ventura} P.,   {Lambert}
  D.~L.,  2013, \mn@doi [\aap] {10.1051/0004-6361/201321818}, \href
  {https://ui.adsabs.harvard.edu/abs/2013A&A...555L...3G} {555, L3}

\bibitem[\protect\citeauthoryear{{Garc{\'\i}a P{\'e}rez} et~al.,}{{Garc{\'\i}a
  P{\'e}rez} et~al.}{2016}]{2016AJ....151..144G}
{Garc{\'\i}a P{\'e}rez} A.~E.,  et~al., 2016, \mn@doi [\aj]
  {10.3847/0004-6256/151/6/144}, \href
  {https://ui.adsabs.harvard.edu/abs/2016AJ....151..144G} {151, 144}

\bibitem[\protect\citeauthoryear{{Glass} \& {Evans}}{{Glass} \&
  {Evans}}{1981}]{1981Natur.291..303G}
{Glass} I.~S.,  {Evans} T.~L.,  1981, \mn@doi [\nat] {10.1038/291303a0}, \href
  {https://ui.adsabs.harvard.edu/abs/1981Natur.291..303G} {291, 303}

\bibitem[\protect\citeauthoryear{{Glass}, {Schultheis}, {Blommaert}, {Sahai},
  {Stute}  \& {Uttenthaler}}{{Glass} et~al.}{2009}]{2009MNRAS.395L..11G}
{Glass} I.~S.,  {Schultheis} M.,  {Blommaert} J.~A.~D.~L.,  {Sahai} R.,
  {Stute} M.,   {Uttenthaler} S.,  2009, \mn@doi [\mnras]
  {10.1111/j.1745-3933.2009.00628.x}, \href
  {https://ui.adsabs.harvard.edu/abs/2009MNRAS.395L..11G} {395, L11}

\bibitem[\protect\citeauthoryear{{Griffith}, {Johnson}  \&
  {Weinberg}}{{Griffith} et~al.}{2019}]{2019ApJ...886...84G}
{Griffith} E.,  {Johnson} J.~A.,   {Weinberg} D.~H.,  2019, \mn@doi [\apj]
  {10.3847/1538-4357/ab4b5d}, \href
  {https://ui.adsabs.harvard.edu/abs/2019ApJ...886...84G} {886, 84}

\bibitem[\protect\citeauthoryear{{Gunn} et~al.,}{{Gunn}
  et~al.}{2006}]{2006AJ....131.2332G}
{Gunn} J.~E.,  et~al., 2006, \mn@doi [\aj] {10.1086/500975}, \href
  {https://ui.adsabs.harvard.edu/abs/2006AJ....131.2332G} {131, 2332}

\bibitem[\protect\citeauthoryear{{Hawkins}, {Leistedt}, {Bovy}  \&
  {Hogg}}{{Hawkins} et~al.}{2017}]{2017MNRAS.471..722H}
{Hawkins} K.,  {Leistedt} B.,  {Bovy} J.,   {Hogg} D.~W.,  2017, \mn@doi
  [\mnras] {10.1093/mnras/stx1655}, \href
  {https://ui.adsabs.harvard.edu/abs/2017MNRAS.471..722H} {471, 722}

\bibitem[\protect\citeauthoryear{{Heinze} et~al.,}{{Heinze}
  et~al.}{2018}]{2018AJ....156..241H}
{Heinze} A.~N.,  et~al., 2018, \mn@doi [\aj] {10.3847/1538-3881/aae47f}, \href
  {https://ui.adsabs.harvard.edu/abs/2018AJ....156..241H} {156, 241}

\bibitem[\protect\citeauthoryear{{Henden}, {Levine}, {Terrell}  \&
  {Welch}}{{Henden} et~al.}{2015}]{2015AAS...22533616H}
{Henden} A.~A.,  {Levine} S.,  {Terrell} D.,   {Welch} D.~L.,  2015, in
  American Astronomical Society Meeting Abstracts \#225. p. 336.16

\bibitem[\protect\citeauthoryear{{Herwig}}{{Herwig}}{2005}]{2005ARA&A..43..435H}
{Herwig} F.,  2005, \mn@doi [\araa] {10.1146/annurev.astro.43.072103.150600},
  \href {https://ui.adsabs.harvard.edu/abs/2005ARA&A..43..435H} {43, 435}

\bibitem[\protect\citeauthoryear{{Holl} et~al.,}{{Holl}
  et~al.}{2018}]{2018A&A...618A..30H}
{Holl} B.,  et~al., 2018, \mn@doi [\aap] {10.1051/0004-6361/201832892}, \href
  {https://ui.adsabs.harvard.edu/abs/2018A&A...618A..30H} {618, A30}

\bibitem[\protect\citeauthoryear{{Holoien} et~al.,}{{Holoien}
  et~al.}{2014}]{2014MNRAS.445.3263H}
{Holoien} T.~W.~S.,  et~al., 2014, \mn@doi [\mnras] {10.1093/mnras/stu1922},
  \href {https://ui.adsabs.harvard.edu/abs/2014MNRAS.445.3263H} {445, 3263}

\bibitem[\protect\citeauthoryear{{Holoien} et~al.,}{{Holoien}
  et~al.}{2016}]{2016MNRAS.455.2918H}
{Holoien} T.~W.~S.,  et~al., 2016, \mn@doi [\mnras] {10.1093/mnras/stv2486},
  \href {https://ui.adsabs.harvard.edu/abs/2016MNRAS.455.2918H} {455, 2918}

\bibitem[\protect\citeauthoryear{{Holoien} et~al.,}{{Holoien}
  et~al.}{2017}]{2017MNRAS.471.4966H}
{Holoien} T.~W.~S.,  et~al., 2017, \mn@doi [\mnras] {10.1093/mnras/stx1544},
  \href {https://ui.adsabs.harvard.edu/abs/2017MNRAS.471.4966H} {471, 4966}

\bibitem[\protect\citeauthoryear{{Holtzman} et~al.,}{{Holtzman}
  et~al.}{2015}]{2015AJ....150..148H}
{Holtzman} J.~A.,  et~al., 2015, \mn@doi [\aj] {10.1088/0004-6256/150/5/148},
  \href {https://ui.adsabs.harvard.edu/abs/2015AJ....150..148H} {150, 148}

\bibitem[\protect\citeauthoryear{{Iben} \& {Renzini}}{{Iben} \&
  {Renzini}}{1983}]{1983ARA&A..21..271I}
{Iben} I. J.,  {Renzini} A.,  1983, \mn@doi [\araa]
  {10.1146/annurev.aa.21.090183.001415}, \href
  {https://ui.adsabs.harvard.edu/abs/1983ARA&A..21..271I} {21, 271}

\bibitem[\protect\citeauthoryear{{Ivanova} et~al.,}{{Ivanova}
  et~al.}{2013}]{2013A&ARv..21...59I}
{Ivanova} N.,  et~al., 2013, \mn@doi [\aapr] {10.1007/s00159-013-0059-2}, \href
  {https://ui.adsabs.harvard.edu/abs/2013A&ARv..21...59I} {21, 59}

\bibitem[\protect\citeauthoryear{{Iwamoto}, {Kajino}, {Mathews}, {Fujimoto}  \&
  {Aoki}}{{Iwamoto} et~al.}{2004}]{2004ApJ...602..377I}
{Iwamoto} N.,  {Kajino} T.,  {Mathews} G.~J.,  {Fujimoto} M.~Y.,   {Aoki} W.,
  2004, \mn@doi [\apj] {10.1086/380989}, \href
  {https://ui.adsabs.harvard.edu/abs/2004ApJ...602..377I} {602, 377}

\bibitem[\protect\citeauthoryear{{Iwanek} et~al.,}{{Iwanek}
  et~al.}{2019}]{2019ApJ...879..114I}
{Iwanek} P.,  et~al., 2019, \mn@doi [\apj] {10.3847/1538-4357/ab23f6}, \href
  {https://ui.adsabs.harvard.edu/abs/2019ApJ...879..114I} {879, 114}

\bibitem[\protect\citeauthoryear{{Jayasinghe} et~al.,}{{Jayasinghe}
  et~al.}{2018}]{Jayasinghe2018}
{Jayasinghe} T.,  et~al., 2018, \mn@doi [\mnras] {10.1093/mnras/sty838}, \href
  {https://ui.adsabs.harvard.edu/abs/2018MNRAS.477.3145J} {477, 3145}

\bibitem[\protect\citeauthoryear{{Jayasinghe} et~al.,}{{Jayasinghe}
  et~al.}{2019a}]{Jayasinghe2019b}
{Jayasinghe} T.,  et~al., 2019a, \mn@doi [\mnras] {10.1093/mnras/stz444}, \href
  {https://ui.adsabs.harvard.edu/abs/2019MNRAS.485..961J} {485, 961}

\bibitem[\protect\citeauthoryear{{Jayasinghe} et~al.,}{{Jayasinghe}
  et~al.}{2019b}]{Jayasinghe2019a}
{Jayasinghe} T.,  et~al., 2019b, \mn@doi [\mnras] {10.1093/mnras/stz844}, \href
  {https://ui.adsabs.harvard.edu/abs/2019MNRAS.486.1907J} {486, 1907}

\bibitem[\protect\citeauthoryear{{Jayasinghe} et~al.,}{{Jayasinghe}
  et~al.}{2019c}]{Jayasinghe2019c}
{Jayasinghe} T.,  et~al., 2019c, \mn@doi [Monthly Notices of the Royal
  Astronomical Society] {10.1093/mnras/stz2711}, 491, 13

\bibitem[\protect\citeauthoryear{{Jayasinghe} et~al.,}{{Jayasinghe}
  et~al.}{2020a}]{Jayasinghe2020a}
{Jayasinghe} T.,  et~al., 2020a, \mn@doi [\mnras] {10.1093/mnras/staa518},
  \href {https://ui.adsabs.harvard.edu/abs/2020MNRAS.493.4045J} {493, 4045}

\bibitem[\protect\citeauthoryear{{Jayasinghe} et~al.,}{{Jayasinghe}
  et~al.}{2020b}]{Jayasinghe2020b}
{Jayasinghe} T.,  et~al., 2020b, \mn@doi [\mnras] {10.1093/mnras/staa499},
  \href {https://ui.adsabs.harvard.edu/abs/2020MNRAS.493.4186J} {493, 4186}

\bibitem[\protect\citeauthoryear{{J{\"o}nsson} et~al.,}{{J{\"o}nsson}
  et~al.}{2020}]{2020arXiv200705537J}
{J{\"o}nsson} H.,  et~al., 2020, arXiv e-prints, \href
  {https://ui.adsabs.harvard.edu/abs/2020arXiv200705537J} {p. arXiv:2007.05537}

\bibitem[\protect\citeauthoryear{{Karakas}, {Lattanzio}  \& {Pols}}{{Karakas}
  et~al.}{2002}]{2002PASA...19..515K}
{Karakas} A.~I.,  {Lattanzio} J.~C.,   {Pols} O.~R.,  2002, \mn@doi [\pasa]
  {10.1071/AS02013}, \href
  {https://ui.adsabs.harvard.edu/abs/2002PASA...19..515K} {19, 515}

\bibitem[\protect\citeauthoryear{{Khouri}, {Waters}, {de Koter}, {Decin},
  {Min}, {de Vries}, {Lombaert}  \& {Cox}}{{Khouri}
  et~al.}{2015}]{2015A&A...577A.114K}
{Khouri} T.,  {Waters} L.~B.~F.~M.,  {de Koter} A.,  {Decin} L.,  {Min} M.,
  {de Vries} B.~L.,  {Lombaert} R.,   {Cox} N.~L.~J.,  2015, \mn@doi [\aap]
  {10.1051/0004-6361/201425092}, \href
  {https://ui.adsabs.harvard.edu/abs/2015A&A...577A.114K} {577, A114}

\bibitem[\protect\citeauthoryear{{Kiss}, {Szatm{\'a}ry}, {Cadmus}  \&
  {Mattei}}{{Kiss} et~al.}{1999}]{1999A&A...346..542K}
{Kiss} L.~L.,  {Szatm{\'a}ry} K.,  {Cadmus} R.~R. J.,   {Mattei} J.~A.,  1999,
  \aap, \href {https://ui.adsabs.harvard.edu/abs/1999A&A...346..542K} {346,
  542}

\bibitem[\protect\citeauthoryear{{Kochanek} et~al.,}{{Kochanek}
  et~al.}{2017}]{2017PASP..129j4502K}
{Kochanek} C.~S.,  et~al., 2017, \mn@doi [\pasp] {10.1088/1538-3873/aa80d9},
  \href {https://ui.adsabs.harvard.edu/abs/2017PASP..129j4502K} {129, 104502}

\bibitem[\protect\citeauthoryear{{Kraft}}{{Kraft}}{1967}]{1967ApJ...150..551K}
{Kraft} R.~P.,  1967, \mn@doi [\apj] {10.1086/149359}, \href
  {https://ui.adsabs.harvard.edu/abs/1967ApJ...150..551K} {150, 551}

\bibitem[\protect\citeauthoryear{{Lattanzio} \& {Wood}}{{Lattanzio} \&
  {Wood}}{2004}]{2004agbs.book...23L}
{Lattanzio} J.~C.,  {Wood} P.~R.,  2004, {Evolution, Nucleosynthesis, and
  Pulsation of AGB Stars}.
pp 23--104, \mn@doi{10.1007/978-1-4757-3876-6_2}

\bibitem[\protect\citeauthoryear{{Lattanzio}, {Frost}, {Cannon}  \&
  {Wood}}{{Lattanzio} et~al.}{1996}]{1996MmSAI..67..729L}
{Lattanzio} J.,  {Frost} C.,  {Cannon} R.,   {Wood} P.~R.,  1996, \memsai,
  \href {https://ui.adsabs.harvard.edu/abs/1996MmSAI..67..729L} {67, 729}

\bibitem[\protect\citeauthoryear{{Leavitt}}{{Leavitt}}{1908}]{1908AnHar..60...87L}
{Leavitt} H.~S.,  1908, Annals of Harvard College Observatory, \href
  {https://ui.adsabs.harvard.edu/abs/1908AnHar..60...87L} {60, 87}

\bibitem[\protect\citeauthoryear{{Lebzelter}, {Mowlavi}, {Marigo},
  {Pastorelli}, {Trabucchi}, {Wood}  \& {Lecoeur-Ta{\"\i}bi}}{{Lebzelter}
  et~al.}{2018}]{2018A&A...616L..13L}
{Lebzelter} T.,  {Mowlavi} N.,  {Marigo} P.,  {Pastorelli} G.,  {Trabucchi} M.,
   {Wood} P.~R.,   {Lecoeur-Ta{\"\i}bi} I.,  2018, \mn@doi [\aap]
  {10.1051/0004-6361/201833615}, \href
  {https://ui.adsabs.harvard.edu/abs/2018A&A...616L..13L} {616, L13}

\bibitem[\protect\citeauthoryear{{Lucy}}{{Lucy}}{1976}]{1976ApJ...205..208L}
{Lucy} L.~B.,  1976, \mn@doi [\apj] {10.1086/154265}, \href
  {https://ui.adsabs.harvard.edu/abs/1976ApJ...205..208L} {205, 208}

\bibitem[\protect\citeauthoryear{{Madore}}{{Madore}}{1982}]{1982ApJ...253..575M}
{Madore} B.~F.,  1982, \mn@doi [\apj] {10.1086/159659}, \href
  {https://ui.adsabs.harvard.edu/abs/1982ApJ...253..575M} {253, 575}

\bibitem[\protect\citeauthoryear{{Marigo} \& {Girardi}}{{Marigo} \&
  {Girardi}}{2007}]{2007A&A...469..239M}
{Marigo} P.,  {Girardi} L.,  2007, \mn@doi [\aap] {10.1051/0004-6361:20066772},
  \href {https://ui.adsabs.harvard.edu/abs/2007A&A...469..239M} {469, 239}

\bibitem[\protect\citeauthoryear{{Marrese}, {Marinoni}, {Fabrizio}  \&
  {Altavilla}}{{Marrese} et~al.}{2019}]{2019A&A...621A.144M}
{Marrese} P.~M.,  {Marinoni} S.,  {Fabrizio} M.,   {Altavilla} G.,  2019,
  \mn@doi [\aap] {10.1051/0004-6361/201834142}, \href
  {https://ui.adsabs.harvard.edu/abs/2019A&A...621A.144M} {621, A144}

\bibitem[\protect\citeauthoryear{{Massalkhi} et~al.,}{{Massalkhi}
  et~al.}{2018}]{2018A&A...611A..29M}
{Massalkhi} S.,  et~al., 2018, \mn@doi [\aap] {10.1051/0004-6361/201732038},
  \href {https://ui.adsabs.harvard.edu/abs/2018A&A...611A..29M} {611, A29}

\bibitem[\protect\citeauthoryear{{Mateu} \& {Vivas}}{{Mateu} \&
  {Vivas}}{2018}]{2018MNRAS.479..211M}
{Mateu} C.,  {Vivas} A.~K.,  2018, \mn@doi [\mnras] {10.1093/mnras/sty1373},
  \href {https://ui.adsabs.harvard.edu/abs/2018MNRAS.479..211M} {479, 211}

\bibitem[\protect\citeauthoryear{{Matsunaga}}{{Matsunaga}}{2018}]{2018IAUS..334...57M}
{Matsunaga} N.,  2018, in {Chiappini} C.,  {Minchev} I.,  {Starkenburg} E.,
  {Valentini} M.,  eds,  IAU Symposium Vol. 334, Rediscovering Our Galaxy. pp
  57--64 (\mn@eprint {arXiv} {1804.04942}), \mn@doi{10.1017/S1743921317007943}

\bibitem[\protect\citeauthoryear{{Matsunaga} et~al.,}{{Matsunaga}
  et~al.}{2006}]{2006MNRAS.370.1979M}
{Matsunaga} N.,  et~al., 2006, \mn@doi [\mnras]
  {10.1111/j.1365-2966.2006.10620.x}, \href
  {https://ui.adsabs.harvard.edu/abs/2006MNRAS.370.1979M} {370, 1979}

\bibitem[\protect\citeauthoryear{{McDonald}, {De Beck}, {Zijlstra}  \&
  {Lagadec}}{{McDonald} et~al.}{2018}]{2018MNRAS.481.4984M}
{McDonald} I.,  {De Beck} E.,  {Zijlstra} A.~A.,   {Lagadec} E.,  2018, \mn@doi
  [\mnras] {10.1093/mnras/sty2607}, \href
  {https://ui.adsabs.harvard.edu/abs/2018MNRAS.481.4984M} {481, 4984}

\bibitem[\protect\citeauthoryear{{McQuillan}, {Mazeh}  \&
  {Aigrain}}{{McQuillan} et~al.}{2013}]{2013ApJ...775L..11M}
{McQuillan} A.,  {Mazeh} T.,   {Aigrain} S.,  2013, \mn@doi [\apjl]
  {10.1088/2041-8205/775/1/L11}, \href
  {https://ui.adsabs.harvard.edu/abs/2013ApJ...775L..11M} {775, L11}

\bibitem[\protect\citeauthoryear{{McQuillan}, {Mazeh}  \&
  {Aigrain}}{{McQuillan} et~al.}{2014}]{2014ApJS..211...24M}
{McQuillan} A.,  {Mazeh} T.,   {Aigrain} S.,  2014, \mn@doi [\apjs]
  {10.1088/0067-0049/211/2/24}, \href
  {https://ui.adsabs.harvard.edu/abs/2014ApJS..211...24M} {211, 24}

\bibitem[\protect\citeauthoryear{{McSaveney}, {Wood}, {Scholz}, {Lattanzio}  \&
  {Hinkle}}{{McSaveney} et~al.}{2007}]{2007MNRAS.378.1089M}
{McSaveney} J.~A.,  {Wood} P.~R.,  {Scholz} M.,  {Lattanzio} J.~C.,   {Hinkle}
  K.~H.,  2007, \mn@doi [\mnras] {10.1111/j.1365-2966.2007.11845.x}, \href
  {https://ui.adsabs.harvard.edu/abs/2007MNRAS.378.1089M} {378, 1089}

\bibitem[\protect\citeauthoryear{{Moe}, {Kratter}  \& {Badenes}}{{Moe}
  et~al.}{2019}]{2019ApJ...875...61M}
{Moe} M.,  {Kratter} K.~M.,   {Badenes} C.,  2019, \mn@doi [\apj]
  {10.3847/1538-4357/ab0d88}, \href
  {https://ui.adsabs.harvard.edu/abs/2019ApJ...875...61M} {875, 61}

\bibitem[\protect\citeauthoryear{{Nicholls}, {Wood}, {Cioni}  \&
  {Soszy{\'n}ski}}{{Nicholls} et~al.}{2009}]{2009MNRAS.399.2063N}
{Nicholls} C.~P.,  {Wood} P.~R.,  {Cioni} M. R.~L.,   {Soszy{\'n}ski} I.,
  2009, \mn@doi [\mnras] {10.1111/j.1365-2966.2009.15401.x}, \href
  {https://ui.adsabs.harvard.edu/abs/2009MNRAS.399.2063N} {399, 2063}

\bibitem[\protect\citeauthoryear{{Nidever} et~al.,}{{Nidever}
  et~al.}{2015}]{2015AJ....150..173N}
{Nidever} D.~L.,  et~al., 2015, \mn@doi [\aj] {10.1088/0004-6256/150/6/173},
  \href {https://ui.adsabs.harvard.edu/abs/2015AJ....150..173N} {150, 173}

\bibitem[\protect\citeauthoryear{{Onaka}, {de Jong}  \& {Willems}}{{Onaka}
  et~al.}{1989}]{1989A&A...218..169O}
{Onaka} T.,  {de Jong} T.,   {Willems} F.~J.,  1989, \aap, \href
  {https://ui.adsabs.harvard.edu/abs/1989A&A...218..169O} {218, 169}

\bibitem[\protect\citeauthoryear{{Paczy{\'n}ski}}{{Paczy{\'n}ski}}{1970}]{1970AcA....20...47P}
{Paczy{\'n}ski} B.,  1970, \actaa, \href
  {https://ui.adsabs.harvard.edu/abs/1970AcA....20...47P} {20, 47}

\bibitem[\protect\citeauthoryear{{Paczy{\'n}ski}, {Szczygie{\l}}, {Pilecki}  \&
  {Pojma{\'n}ski}}{{Paczy{\'n}ski} et~al.}{2006}]{2006MNRAS.368.1311P}
{Paczy{\'n}ski} B.,  {Szczygie{\l}} D.~M.,  {Pilecki} B.,   {Pojma{\'n}ski} G.,
   2006, \mn@doi [\mnras] {10.1111/j.1365-2966.2006.10223.x}, \href
  {https://ui.adsabs.harvard.edu/abs/2006MNRAS.368.1311P} {368, 1311}

\bibitem[\protect\citeauthoryear{{Pawlak} et~al.,}{{Pawlak}
  et~al.}{2019}]{2019MNRAS.487.5932P}
{Pawlak} M.,  et~al., 2019, \mn@doi [\mnras] {10.1093/mnras/stz1681}, \href
  {https://ui.adsabs.harvard.edu/abs/2019MNRAS.487.5932P} {487, 5932}

\bibitem[\protect\citeauthoryear{{Pepper} et~al.,}{{Pepper}
  et~al.}{2007}]{2007PASP..119..923P}
{Pepper} J.,  et~al., 2007, \mn@doi [\pasp] {10.1086/521836}, \href
  {https://ui.adsabs.harvard.edu/abs/2007PASP..119..923P} {119, 923}

\bibitem[\protect\citeauthoryear{{Percy} \& {Colivas}}{{Percy} \&
  {Colivas}}{1999}]{1999PASP..111...94P}
{Percy} J.~R.,  {Colivas} T.,  1999, \mn@doi [\pasp] {10.1086/316290}, \href
  {https://ui.adsabs.harvard.edu/abs/1999PASP..111...94P} {111, 94}

\bibitem[\protect\citeauthoryear{{Pojmanski}}{{Pojmanski}}{2002}]{2002AcA....52..397P}
{Pojmanski} G.,  2002, \actaa, \href
  {https://ui.adsabs.harvard.edu/abs/2002AcA....52..397P} {52, 397}

\bibitem[\protect\citeauthoryear{{Price-Whelan} et~al.,}{{Price-Whelan}
  et~al.}{2020}]{2020arXiv200200014P}
{Price-Whelan} A.~M.,  et~al., 2020, arXiv e-prints, \href
  {https://ui.adsabs.harvard.edu/abs/2020arXiv200200014P} {p. arXiv:2002.00014}

\bibitem[\protect\citeauthoryear{{Prochaska}, {Naumov}, {Carney}, {McWilliam}
  \& {Wolfe}}{{Prochaska} et~al.}{2000}]{2000AJ....120.2513P}
{Prochaska} J.~X.,  {Naumov} S.~O.,  {Carney} B.~W.,  {McWilliam} A.,   {Wolfe}
  A.~M.,  2000, \mn@doi [\aj] {10.1086/316818}, \href
  {https://ui.adsabs.harvard.edu/abs/2000AJ....120.2513P} {120, 2513}

\bibitem[\protect\citeauthoryear{{Ricker} et~al.,}{{Ricker}
  et~al.}{2015}]{2015JATIS...1a4003R}
{Ricker} G.~R.,  et~al., 2015, \mn@doi [Journal of Astronomical Telescopes,
  Instruments, and Systems] {10.1117/1.JATIS.1.1.014003}, \href
  {https://ui.adsabs.harvard.edu/abs/2015JATIS...1a4003R} {1, 014003}

\bibitem[\protect\citeauthoryear{{Salaris}, {Pietrinferni}, {Piersimoni}  \&
  {Cassisi}}{{Salaris} et~al.}{2015}]{2015A&A...583A..87S}
{Salaris} M.,  {Pietrinferni} A.,  {Piersimoni} A.~M.,   {Cassisi} S.,  2015,
  \mn@doi [\aap] {10.1051/0004-6361/201526951}, \href
  {https://ui.adsabs.harvard.edu/abs/2015A&A...583A..87S} {583, A87}

\bibitem[\protect\citeauthoryear{{Scalo}, {Despain}  \& {Ulrich}}{{Scalo}
  et~al.}{1975}]{1975ApJ...196..805S}
{Scalo} J.~M.,  {Despain} K.~H.,   {Ulrich} R.~K.,  1975, \mn@doi [\apj]
  {10.1086/153471}, \href
  {https://ui.adsabs.harvard.edu/abs/1975ApJ...196..805S} {196, 805}

\bibitem[\protect\citeauthoryear{{Scargle}}{{Scargle}}{1982}]{1982ApJ...263..835S}
{Scargle} J.~D.,  1982, \mn@doi [\apj] {10.1086/160554}, \href
  {https://ui.adsabs.harvard.edu/abs/1982ApJ...263..835S} {263, 835}

\bibitem[\protect\citeauthoryear{{Schlafly} \& {Finkbeiner}}{{Schlafly} \&
  {Finkbeiner}}{2011}]{2011ApJ...737..103S}
{Schlafly} E.~F.,  {Finkbeiner} D.~P.,  2011, \mn@doi [\apj]
  {10.1088/0004-637X/737/2/103}, \href
  {https://ui.adsabs.harvard.edu/abs/2011ApJ...737..103S} {737, 103}

\bibitem[\protect\citeauthoryear{{Schlegel}, {Finkbeiner}  \&
  {Davis}}{{Schlegel} et~al.}{1998}]{1998ApJ...500..525S}
{Schlegel} D.~J.,  {Finkbeiner} D.~P.,   {Davis} M.,  1998, \mn@doi [\apj]
  {10.1086/305772}, \href
  {https://ui.adsabs.harvard.edu/abs/1998ApJ...500..525S} {500, 525}

\bibitem[\protect\citeauthoryear{{Shappee} et~al.,}{{Shappee}
  et~al.}{2014}]{2014ApJ...788...48S}
{Shappee} B.~J.,  et~al., 2014, \mn@doi [\apj] {10.1088/0004-637X/788/1/48},
  \href {https://ui.adsabs.harvard.edu/abs/2014ApJ...788...48S} {788, 48}

\bibitem[\protect\citeauthoryear{{Shetrone} et~al.,}{{Shetrone}
  et~al.}{2019}]{2019ApJ...872..137S}
{Shetrone} M.,  et~al., 2019, \mn@doi [\apj] {10.3847/1538-4357/aaff66}, \href
  {https://ui.adsabs.harvard.edu/abs/2019ApJ...872..137S} {872, 137}

\bibitem[\protect\citeauthoryear{{Skrutskie} et~al.,}{{Skrutskie}
  et~al.}{2006}]{2006AJ....131.1163S}
{Skrutskie} M.~F.,  et~al., 2006, \mn@doi [\aj] {10.1086/498708}, \href
  {https://ui.adsabs.harvard.edu/abs/2006AJ....131.1163S} {131, 1163}

\bibitem[\protect\citeauthoryear{{Smith} \& {Lambert}}{{Smith} \&
  {Lambert}}{1986}]{1986ApJ...311..843S}
{Smith} V.~V.,  {Lambert} D.~L.,  1986, \mn@doi [\apj] {10.1086/164823}, \href
  {https://ui.adsabs.harvard.edu/abs/1986ApJ...311..843S} {311, 843}

\bibitem[\protect\citeauthoryear{{Smith} \& {Lambert}}{{Smith} \&
  {Lambert}}{1988}]{1988ApJ...333..219S}
{Smith} V.~V.,  {Lambert} D.~L.,  1988, \mn@doi [\apj] {10.1086/166738}, \href
  {https://ui.adsabs.harvard.edu/abs/1988ApJ...333..219S} {333, 219}

\bibitem[\protect\citeauthoryear{{Solano} \& {Fernley}}{{Solano} \&
  {Fernley}}{1997}]{1997A&AS..122..131S}
{Solano} E.,  {Fernley} J.,  1997, \mn@doi [\aaps] {10.1051/aas:1997329}, \href
  {https://ui.adsabs.harvard.edu/abs/1997A&AS..122..131S} {122, 131}

\bibitem[\protect\citeauthoryear{{Soszy{\'n}ski}}{{Soszy{\'n}ski}}{2007}]{2007ApJ...660.1486S}
{Soszy{\'n}ski} I.,  2007, \mn@doi [\apj] {10.1086/513012}, \href
  {https://ui.adsabs.harvard.edu/abs/2007ApJ...660.1486S} {660, 1486}

\bibitem[\protect\citeauthoryear{{Soszynski}, {Udalski}, {Kubiak}, {Szymanski},
  {Pietrzynski}, {Zebrun}, {Szewczyk}  \& {Wyrzykowski}}{{Soszynski}
  et~al.}{2004}]{2004AcA....54..129S}
{Soszynski} I.,  {Udalski} A.,  {Kubiak} M.,  {Szymanski} M.,  {Pietrzynski}
  G.,  {Zebrun} K.,  {Szewczyk} O.,   {Wyrzykowski} L.,  2004, \actaa, \href
  {https://ui.adsabs.harvard.edu/abs/2004AcA....54..129S} {54, 129}

\bibitem[\protect\citeauthoryear{{Soszynski} et~al.,}{{Soszynski}
  et~al.}{2007}]{2007AcA....57..201S}
{Soszynski} I.,  et~al., 2007, \actaa, \href
  {https://ui.adsabs.harvard.edu/abs/2007AcA....57..201S} {57, 201}

\bibitem[\protect\citeauthoryear{{Soszy{\'n}ski} et~al.,}{{Soszy{\'n}ski}
  et~al.}{2009}]{2009AcA....59..239S}
{Soszy{\'n}ski} I.,  et~al., 2009, \actaa, \href
  {https://ui.adsabs.harvard.edu/abs/2009AcA....59..239S} {59, 239}

\bibitem[\protect\citeauthoryear{{Tayar} et~al.,}{{Tayar}
  et~al.}{2015}]{2015ApJ...807...82T}
{Tayar} J.,  et~al., 2015, \mn@doi [\apj] {10.1088/0004-637X/807/1/82}, \href
  {https://ui.adsabs.harvard.edu/abs/2015ApJ...807...82T} {807, 82}

\bibitem[\protect\citeauthoryear{{Taylor}}{{Taylor}}{2005}]{2005ASPC..347...29T}
{Taylor} M.~B.,  2005, {TOPCAT \&amp; STIL: Starlink Table/VOTable Processing
  Software}.
p.~29

\bibitem[\protect\citeauthoryear{{Thompson} et~al.,}{{Thompson}
  et~al.}{2019}]{2019Sci...366..637T}
{Thompson} T.~A.,  et~al., 2019, \mn@doi [Science] {10.1126/science.aau4005},
  \href {https://ui.adsabs.harvard.edu/abs/2019Sci...366..637T} {366, 637}

\bibitem[\protect\citeauthoryear{{Thompson} et~al.,}{{Thompson}
  et~al.}{2020}]{2020arXiv200507653T}
{Thompson} T.~A.,  et~al., 2020, arXiv e-prints, \href
  {https://ui.adsabs.harvard.edu/abs/2020arXiv200507653T} {p. arXiv:2005.07653}

\bibitem[\protect\citeauthoryear{{Tonry} et~al.,}{{Tonry}
  et~al.}{2018a}]{2018PASP..130f4505T}
{Tonry} J.~L.,  et~al., 2018a, \mn@doi [\pasp] {10.1088/1538-3873/aabadf},
  \href {https://ui.adsabs.harvard.edu/abs/2018PASP..130f4505T} {130, 064505}

\bibitem[\protect\citeauthoryear{{Tonry} et~al.,}{{Tonry}
  et~al.}{2018b}]{2018ApJ...867..105T}
{Tonry} J.~L.,  et~al., 2018b, \mn@doi [\apj] {10.3847/1538-4357/aae386}, \href
  {https://ui.adsabs.harvard.edu/abs/2018ApJ...867..105T} {867, 105}

\bibitem[\protect\citeauthoryear{{Torres}, {Andersen}  \&
  {Gim{\'e}nez}}{{Torres} et~al.}{2010}]{2010A&ARv..18...67T}
{Torres} G.,  {Andersen} J.,   {Gim{\'e}nez} A.,  2010, \mn@doi [\aapr]
  {10.1007/s00159-009-0025-1}, \href
  {https://ui.adsabs.harvard.edu/abs/2010A&ARv..18...67T} {18, 67}

\bibitem[\protect\citeauthoryear{{Trabucchi}, {Wood}, {Montalb{\'a}n},
  {Marigo}, {Pastorelli}  \& {Girardi}}{{Trabucchi}
  et~al.}{2017}]{2017ApJ...847..139T}
{Trabucchi} M.,  {Wood} P.~R.,  {Montalb{\'a}n} J.,  {Marigo} P.,  {Pastorelli}
  G.,   {Girardi} L.,  2017, \mn@doi [\apj] {10.3847/1538-4357/aa8998}, \href
  {https://ui.adsabs.harvard.edu/abs/2017ApJ...847..139T} {847, 139}

\bibitem[\protect\citeauthoryear{{Udalski}}{{Udalski}}{2003}]{2003AcA....53..291U}
{Udalski} A.,  2003, \actaa, \href
  {https://ui.adsabs.harvard.edu/abs/2003AcA....53..291U} {53, 291}

\bibitem[\protect\citeauthoryear{{Walker} \& {Terndrup}}{{Walker} \&
  {Terndrup}}{1991}]{1991ApJ...378..119W}
{Walker} A.~R.,  {Terndrup} D.~M.,  1991, \mn@doi [\apj] {10.1086/170411},
  \href {https://ui.adsabs.harvard.edu/abs/1991ApJ...378..119W} {378, 119}

\bibitem[\protect\citeauthoryear{{Watson}, {Henden}  \& {Price}}{{Watson}
  et~al.}{2006}]{2006SASS...25...47W}
{Watson} C.~L.,  {Henden} A.~A.,   {Price} A.,  2006, Society for Astronomical
  Sciences Annual Symposium, \href
  {https://ui.adsabs.harvard.edu/abs/2006SASS...25...47W} {25, 47}

\bibitem[\protect\citeauthoryear{{Webbink}}{{Webbink}}{2003}]{2003ASPC..293...76W}
{Webbink} R.~F.,  2003, {Contact Binaries}.
p.~76

\bibitem[\protect\citeauthoryear{{Weinberg} et~al.,}{{Weinberg}
  et~al.}{2019}]{2019ApJ...874..102W}
{Weinberg} D.~H.,  et~al., 2019, \mn@doi [\apj] {10.3847/1538-4357/ab07c7},
  \href {https://ui.adsabs.harvard.edu/abs/2019ApJ...874..102W} {874, 102}

\bibitem[\protect\citeauthoryear{{Whitelock}, {Feast}  \& {Van
  Leeuwen}}{{Whitelock} et~al.}{2008}]{2008MNRAS.386..313W}
{Whitelock} P.~A.,  {Feast} M.~W.,   {Van Leeuwen} F.,  2008, \mn@doi [\mnras]
  {10.1111/j.1365-2966.2008.13032.x}, \href
  {https://ui.adsabs.harvard.edu/abs/2008MNRAS.386..313W} {386, 313}

\bibitem[\protect\citeauthoryear{{Wilson} et~al.,}{{Wilson}
  et~al.}{2019}]{2019PASP..131e5001W}
{Wilson} J.~C.,  et~al., 2019, \mn@doi [\pasp] {10.1088/1538-3873/ab0075},
  \href {https://ui.adsabs.harvard.edu/abs/2019PASP..131e5001W} {131, 055001}

\bibitem[\protect\citeauthoryear{{Wood}}{{Wood}}{2000}]{2000PASA...17...18W}
{Wood} P.~R.,  2000, \mn@doi [\pasa] {10.1071/AS00018}, \href
  {https://ui.adsabs.harvard.edu/abs/2000PASA...17...18W} {17, 18}

\bibitem[\protect\citeauthoryear{{Wood} et~al.,}{{Wood}
  et~al.}{1999}]{1999IAUS..191..151W}
{Wood} P.~R.,  et~al., 1999, in {Le Bertre} T.,  {Lebre} A.,   {Waelkens} C.,
  eds,  IAU Symposium Vol. 191, Asymptotic Giant Branch Stars. p.~151

\bibitem[\protect\citeauthoryear{{Wood}, {Olivier}  \& {Kawaler}}{{Wood}
  et~al.}{2004}]{2004ApJ...604..800W}
{Wood} P.~R.,  {Olivier} E.~A.,   {Kawaler} S.~D.,  2004, \mn@doi [\apj]
  {10.1086/382123}, \href
  {https://ui.adsabs.harvard.edu/abs/2004ApJ...604..800W} {604, 800}

\bibitem[\protect\citeauthoryear{{Wo{\'z}niak} et~al.,}{{Wo{\'z}niak}
  et~al.}{2004}]{2004AJ....127.2436W}
{Wo{\'z}niak} P.~R.,  et~al., 2004, \mn@doi [\aj] {10.1086/382719}, \href
  {https://ui.adsabs.harvard.edu/abs/2004AJ....127.2436W} {127, 2436}

\bibitem[\protect\citeauthoryear{{Wright} et~al.,}{{Wright}
  et~al.}{2010}]{2010AJ....140.1868W}
{Wright} E.~L.,  et~al., 2010, \mn@doi [\aj] {10.1088/0004-6256/140/6/1868},
  \href {https://ui.adsabs.harvard.edu/abs/2010AJ....140.1868W} {140, 1868}

\bibitem[\protect\citeauthoryear{{Yakut} \& {Eggleton}}{{Yakut} \&
  {Eggleton}}{2005}]{2005ApJ...629.1055Y}
{Yakut} K.,  {Eggleton} P.~P.,  2005, \mn@doi [\apj] {10.1086/431300}, \href
  {https://ui.adsabs.harvard.edu/abs/2005ApJ...629.1055Y} {629, 1055}

\bibitem[\protect\citeauthoryear{{Yao}, {Liu}, {Deng}, {de Grijs}  \&
  {Matsunaga}}{{Yao} et~al.}{2017}]{2017ApJS..232...16Y}
{Yao} Y.,  {Liu} C.,  {Deng} L.,  {de Grijs} R.,   {Matsunaga} N.,  2017,
  \mn@doi [\apjs] {10.3847/1538-4365/aa88a9}, \href
  {https://ui.adsabs.harvard.edu/abs/2017ApJS..232...16Y} {232, 16}

\bibitem[\protect\citeauthoryear{{Y{\i}ld{\i}z}}{{Y{\i}ld{\i}z}}{2014}]{2014MNRAS.437..185Y}
{Y{\i}ld{\i}z} M.,  2014, \mn@doi [\mnras] {10.1093/mnras/stt1874}, \href
  {https://ui.adsabs.harvard.edu/abs/2014MNRAS.437..185Y} {437, 185}

\bibitem[\protect\citeauthoryear{{Yu}, {Huber}, {Bedding}, {Stello}, {Hon},
  {Murphy}  \& {Khanna}}{{Yu} et~al.}{2018}]{2018ApJS..236...42Y}
{Yu} J.,  {Huber} D.,  {Bedding} T.~R.,  {Stello} D.,  {Hon} M.,  {Murphy}
  S.~J.,   {Khanna} S.,  2018, \mn@doi [\apjs] {10.3847/1538-4365/aaaf74},
  \href {https://ui.adsabs.harvard.edu/abs/2018ApJS..236...42Y} {236, 42}

\bibitem[\protect\citeauthoryear{{Zechmeister} \& {K{\"u}rster}}{{Zechmeister}
  \& {K{\"u}rster}}{2009}]{2009A&A...496..577Z}
{Zechmeister} M.,  {K{\"u}rster} M.,  2009, \mn@doi [\aap]
  {10.1051/0004-6361:200811296}, \href
  {https://ui.adsabs.harvard.edu/abs/2009A&A...496..577Z} {496, 577}

\makeatother
\end{thebibliography}

%\begin{thebibliography}{99}
%\end{thebibliography}

%%%%%%%%%%%%%%%%%%%%%%%%%%%%%%%%%%%%%%%%%%%%%%%%%%

%%%%%%%%%%%%%%%%% APPENDICES %%%%%%%%%%%%%%%%%%%%%

%%%%%%%%%%%%%%%%%%%%%%%%%%%%%%%%%%%%%%%%%%%%%%%%%%

% Don't change these lines
\bsp	% typesetting comment
\label{lastpage}
\end{document}